\newif\ifincludefigs
\includefigstrue 

\newif\iftypeDzero
\typeDzerotrue 

\newif\ifcross
\crosstrue 

\documentclass[prd,superscriptaddress,nofootinbib,floatfix,notitlepage]{revtex4-1}
\pdfoutput=1

\usepackage{amsmath,amssymb,epsfig}
\usepackage{hyperref}
\usepackage{natbib,ifthen}
\usepackage[toc,page]{appendix}

\usepackage{mathrsfs}
\usepackage{simplewick}
\usepackage[lofdepth,lotdepth,caption=false]{subfig}
\usepackage{color}
\usepackage{bbold}
\usepackage{array}

\usepackage{multirow}

\begin{document}

\newcommand{\jcap}{JCAP}
\newcommand{\apjl}{APJL~}
\newcommand{\physrep}{Phys. Rep.}
\newcommand{\aap}{A\&A}

\def\Btheo{{B_\delta^{\textrm{theo}}}}
\newcommand{\todo}[1]{{\color{red}{TODO: #1}}} 
\newcommand{\Qe}[1]{{\color{cyan}{Q: #1}}}
\newcommand{\comment}[1]{{\color{blue}{Comment: #1}}}
\newcommand{\lbar}[1]{\underline{l}_{#1}}
\newcommand{\drm}{\mathrm{d}}
\renewcommand{\d}{\mathrm{d}}
\newcommand{\gaensli}[1]{\lq #1\rq$ $}
\newcommand{\bartilde}[1]{\bar{\tilde #1}}
\newcommand{\barti}[1]{\bartilde{#1}}
\newcommand{\ti}{\tilde}
\newcommand{\oforder}[1]{\mathcal{O}(#1)}
\newcommand{\D}{\mathrm{D}}
\renewcommand{\(}{\left(}
\renewcommand{\)}{\right)}
\renewcommand{\[}{\left[}
\renewcommand{\]}{\right]}
\def\<{\left\langle}
\def\>{\right\rangle}
\newcommand{\mycaption}[1]{\caption{\footnotesize{#1}}}
\newcommand{\hattilde}[1]{\hat{\tilde #1}}
\newcommand{\mycite}[1]{[#1]}
\newcommand{\mnras}{Mon.\ Not.\ R.\ Astron.\ Soc.}
\newcommand{\apjs}{Astrophys.\ J.\ Supp.}

\def\uk{{\bf \hat{k}}}
\def\un{{\bf \hat{n}}}
\def\ur{{\bf \hat{r}}}
\def\ux{{\bf \hat{x}}}
\def\bk{{\bf k}}
\def\bn{{\bf n}}
\def\br{{\bf r}}
\def\bx{{\bf x}}
\def\bK{{\bf K}}
\def\by{{\bf y}}
\def\bl{{\bf l}}
\def\bkp{{\bf k^\pr}}
\def\brp{{\bf r^\pr}}

\newcommand{\RS}{\mathit{RS}}
\newcommand{\WX}{\mathit{WX}}
\newcommand{\YZ}{\mathit{YZ}}
\newcommand{\pprime}[2]{\mathit{#1'#2'}}
\newcommand{\PPrime}[2]{\mathit{#1'\!#2'}}
\newcommand{\WXprime}{\PPrime{W}{X}}
\newcommand{\YZprime}{\PPrime{Y}{Z}}
\newcommand{\WW}{\mathit{WW}}
\newcommand{\XX}{\mathit{XX}}
\newcommand{\UV}{\mathit{UV}}
\newcommand{\TT}{\mathit{TT}}
\newcommand{\TE}{\mathit{TE}}
\newcommand{\EB}{\mathit{EB}}
\newcommand{\BE}{\mathit{BE}}
\newcommand{\EE}{\mathit{EE}}
\newcommand{\BB}{\mathit{BB}}
\newcommand{\TTilde}[2]{\mathit{\tilde #1\!\tilde #2}}
\newcommand{\ttilde}[2]{\mathit{\tilde #1\tilde #2}}
\newcommand{\tWX}{\TTilde{W}{X}}
\newcommand{\tST}{\TTilde{S}{T}}
\newcommand{\tYZ}{\TTilde{Y}{Z}}
\newcommand{\tWW}{\TTilde{W}{W}}
\newcommand{\tXX}{\TTilde{X}{X}}
\newcommand{\tUV}{\TTilde{U}{V}}
\newcommand{\tEE}{\TTilde{E}{E}}
\newcommand{\tEB}{\TTilde{E}{B}}
\newcommand{\tBB}{\TTilde{B}{B}}

\newcommand{\fixme}[1]{{\textbf{Fixme: #1}}}
\newcommand{\detD}{{\det\!\cld}}
\newcommand{\clh}{\mathcal{H}}
\newcommand{\ud}{{\rm d}}
\renewcommand{\eprint}[1]{\href{http://arxiv.org/abs/#1}{#1}}
\newcommand{\adsurl}[1]{\href{#1}{ADS}}
\newcommand{\ISBN}[1]{\href{http://cosmologist.info/ISBN/#1}{ISBN: #1}}
\newcommand{\vort}{\varpi}
\newcommand\ba{\begin{eqnarray}}
\newcommand\ea{\end{eqnarray}}
\newcommand\be{\begin{equation}}
\newcommand\ee{\end{equation}}
\newcommand\lagrange{{\cal L}}
\newcommand\cll{{\cal L}}
\newcommand\cln{{\cal N}}
\newcommand\clx{{\cal X}}
\newcommand\clz{{\cal Z}}
\newcommand\clv{{\cal V}}
\newcommand\cld{{\cal D}}
\newcommand\clt{{\cal T}}

\newcommand\clo{{\cal O}}
\newcommand{\cla}{{\cal A}}
\newcommand{\clp}{{\cal P}}
\newcommand{\clr}{{\cal R}}
\newcommand{\uD}{{\mathrm{D}}}
\newcommand{\calE}{{\cal E}}
\newcommand{\calB}{{\cal B}}
\newcommand{\curl}{\,\mbox{curl}\,}
\newcommand\del{\nabla}
\newcommand\Tr{{\rm Tr}}
\newcommand\half{{\frac{1}{2}}}
\newcommand\fourth{{1\over 8}}
\newcommand\bibi{\bibitem}
\newcommand{\kf}{\beta}
\newcommand{\kfprod}{\mu}
\newcommand{\calS}{{\cal S}}
\renewcommand{\H}{{\cal H}}
\newcommand{\K}{{\rm K}}
\newcommand{\mK}{{\rm mK}}
\newcommand{\synch}{\text{syn}}
\newcommand{\opacity}{\tau_c^{-1}}

\newcommand{\psil}{\psi_l}
\newcommand{\bsigma}{{\bar{\sigma}}}
\newcommand{\bI}{\bar{I}}
\newcommand{\bq}{\bar{q}}
\newcommand{\bv}{\bar{v}}
\renewcommand\P{{\cal P}}
\newcommand{\numfrac}[2]{{\textstyle \frac{#1}{#2}}}

\newcommand{\la}{\langle}
\newcommand{\ra}{\rangle}
\newcommand{\lla}{\left\langle}
\newcommand{\rra}{\right\rangle}

\newcommand{\Omtot}{\Omega_{\mathrm{tot}}}
\newcommand\xx{\mbox{\boldmath $x$}}
\newcommand{\phpr} {\phi'}
\newcommand{\gam}{\gamma_{ij}}
\newcommand{\sqgam}{\sqrt{\gamma}}
\newcommand{\delk}{\Delta+3{k}}
\newcommand{\dph}{\delta\phi}
\newcommand{\om} {\Omega}
\newcommand{\dom}{\delta^{(3)}\left(\Omega\right)}
\newcommand{\rar}{\rightarrow}
\newcommand{\Rar}{\Rightarrow}
\newcommand\gsim{ \lower .75ex \hbox{$\sim$} \llap{\raise .27ex \hbox{$>$}} }
\newcommand\lsim{ \lower .75ex \hbox{$\sim$} \llap{\raise .27ex \hbox{$<$}} }
\newcommand\bigdot[1] {\stackrel{\mbox{{\huge .}}}{#1}}
\newcommand\bigddot[1] {\stackrel{\mbox{{\huge ..}}}{#1}}
\newcommand{\Mpc}{\text{Mpc}}
\newcommand{\Al}{{A_l}}
\newcommand{\Bl}{{B_l}}
\newcommand{\eAl}{e^\Al}
\newcommand{\ix}{{(i)}}
\newcommand{\ixp}{{(i+1)}}
\renewcommand{\k}{\beta}
\newcommand{\HD}{\mathrm{D}}

\newcommand{\nonflat}[1]{#1}
\newcommand{\Cgl}{C_{\text{gl}}}
\newcommand{\Cgltwo}{C_{\text{gl},2}}
\newcommand{\He}{{\rm{He}}}
\newcommand{\Mhz}{{\rm MHz}}
\newcommand{\vx}{{\mathbf{x}}}
\newcommand{\ve}{{\mathbf{e}}}
\newcommand{\vv}{{\mathbf{v}}}
\newcommand{\vk}{{\mathbf{k}}}
\newcommand{\vn}{{\mathbf{n}}}
\renewcommand{\vr}{{\mathbf{r}}}
\newcommand{\vPsi}{{\mathbf{\psi}}}

\newcommand{\vnhat}{{\hat{\mathbf{n}}}}
\newcommand{\vkhat}{{\hat{\mathbf{k}}}}
\newcommand{\taueps}{{\tau_\epsilon}}
\newcommand{\valpha}{\ensuremath{\boldsymbol\alpha}}

\newcommand{\vgrad}{{\mathbf{\nabla}}}
\newcommand{\fbarln}{\bar{f}_{,\ln\epsilon}(\epsilon)}

\newcommand{\non}{\nonumber}
\newcommand{\secref}[1]{Section \ref{sec:#1}}
\newcommand{\expt}{\mathrm{expt}}
\newcommand{\eq}[1]{(\ref{eq:#1})} 
\newcommand{\eqq}[1]{Eq.~(\ref{eq:#1})} 
\newcommand{\fig}[1]{Fig.~\ref{fig:#1}} 
\newcommand{\app}[1]{Appendix~\ref{App:#1}} 
\newcommand{\tabref}[1]{Table~\ref{tab:#1}} 
\renewcommand{\to}{\rightarrow}
\renewcommand{\(}{\left(}
\renewcommand{\)}{\right)}
\renewcommand{\[}{\left[}
\renewcommand{\]}{\right]}
\renewcommand{\vec}[1]{\mathbf{#1}}
\newcommand{\vy}{\vec{y}}
\newcommand{\vz}{\vec{z}}
\newcommand{\vq}{\vec{q}}
\newcommand{\VPsi}{\vec{\psi}}
\newcommand{\vecv}{\vec{v}}
\newcommand{\vnabla}{\vec{\nabla}}
\newcommand{\vl}{\vec{l}}
\newcommand{\VL}{\vec{L}}
\newcommand{\dl}{\d^2\vl}
\renewcommand{\L}{\mathscr{L}}
\newcommand{\hatAlens}{\hat{A}_\mathrm{lens}}
\newcommand{\Alens}{A_\mathrm{lens}}

\newcommand{\abs}[1]{\lvert #1\rvert}

\newcommand{\ul}{\underline{l}}


\newcommand{\n}{\hat{\mathbf{n}}}
\newcommand{\vecr}{\mathbf{r}}
\newcommand{\va}[1]{\ensuremath{\boldsymbol\alpha}(#1)}
\renewcommand{\k}{\mathrm{k}}
\newcommand*\Bell{\ensuremath{\boldsymbol\ell}}

\renewcommand{\(}{\left(}
\renewcommand{\)}{\right)}

\newcommand{\beq}{\begin{equation}}
\newcommand{\eeq}{\end{equation}}

\newcommand{\e}[1]{\mathrm e^{#1}}
\newcommand{\Y}[2]{\mathrm{Y}^{#1}_{\ell #2 m#2}}

\newcommand{\Fdll}[1]{\frac{\mathrm{d}^2 \Bell_{#1}}{(2 \pi)^2}}
\newcommand{\Fdkkk}[1]{\frac{\mathrm{d}^3 \mathbf{k #1}}{\twopi{3}}}
\newcommand{\Fdlll}[1]{\mathrm{\ell #1}^2\,\mathrm{d}\mathrm{\ell #1}\,}

\newcommand{\twopi}[1]{(2 \pi)^{ #1}}

\newcommand{\CMBlen}[1]{\tilde{T}\left(#1\right)}
\newcommand{\CMB}[1]{T\left(#1\right)}
\newcommand{\PS}[1]{\mathrm{C}_{\ell_#1}^{TT}}
\newcommand{\BS}[1]{\mathrm{B}_{#1}}
\newcommand{\dang}[1]{\delta\ensuremath{\boldsymbol\mu}_{#1}}

\newcommand{\Nnl}{N_L^{(3/2)}}


\thispagestyle{empty}

\title{A bias to CMB lensing measurements from the bispectrum of large-scale structure}

\author{Vanessa B\"ohm}
\affiliation{Max-Planck-Institut f\"ur Astrophysik, Karl-Schwarzschild Strasse 1, 85748 Garching, Germany}

\author{Marcel Schmittfull}
\affiliation{Berkeley Center for Cosmological Physics, Department of Physics and Lawrence Berkeley National Laboratory, University of California, Berkeley, CA 94720, USA}

\author{Blake D. Sherwin}
\affiliation{Berkeley Center for Cosmological Physics, Department of Physics and Lawrence Berkeley National Laboratory, University of California, Berkeley, CA 94720, USA}
\affiliation{Miller Institute for Basic Research in Science, University of California, Berkeley, CA, 94720, USA}

\date{\today}

\begin{abstract}
The rapidly improving precision of measurements of gravitational lensing of the Cosmic Microwave Background (CMB) also requires a corresponding increase in the precision of theoretical modeling.
A commonly made approximation is to model the CMB deflection angle or lensing potential as a Gaussian random field.  
In this paper, however, we analytically quantify the influence of the non-Gaussianity of large-scale structure lenses, arising from nonlinear structure formation, on CMB lensing measurements. In particular, evaluating the impact of the non-zero bispectrum of large-scale structure on the relevant CMB four-point correlation functions, we find that there is a bias to estimates of the CMB lensing power spectrum.
For temperature-based lensing reconstruction with CMB Stage-III and Stage-IV experiments, we find that this lensing power spectrum bias is negative and is of order one percent of the signal. This corresponds to a shift of multiple standard deviations for these upcoming experiments. We caution, however, that our numerical calculation only evaluates two of the largest bias terms and thus only provides an approximate estimate of the full bias. We conclude that further investigation into lensing biases from nonlinear structure formation is required and that these biases should be accounted for in future lensing analyses. 
\end{abstract}

\maketitle
\section{Introduction} 
The photons of the Cosmic Microwave Background (CMB) are gravitationally deflected by the large-scale matter distribution through which they pass. This effect, known as CMB lensing (see \citep{phys-repts} for a review), distorts the temperature and polarization fluctuations in the cosmic background radiation in a characteristic way, which allows reconstruction of the projected deflecting potentials.
CMB lensing probes the growth of large-scale structure over a wide range of redshifts ($0.1{<}z{<}5$). As free-streaming of massive neutrinos and the accelerated expansion of the Universe suppress the formation of structures, the  lensing signal contains valuable information about the sum of neutrino masses \citep{KaplinghatMnuCMB, lesgouMnuCMB, 2012MNRAS4251170H} and dark energy \citep{smith_hu_kaplinghat0607,2011PhRvD83b3011C, calabrese0803LensingAmplitude, 2009PhRvD80j3516C}.

First evidence of the CMB lensing effect was obtained using data from WMAP, relying on cross-correlation with other tracers of large-scale structure \citep{SmithZahnDore0705,HirataHoEtAl0801Lensing}. 
The first measurement from CMB alone (i.e.~a measurement of the lensing power spectrum) was reported by the Atacama Cosmology Telescope (ACT) collaboration \citep{ACTLensingDetectionDas1103}, followed by the South Pole Telescope (SPT) collaboration \citep{SPTLensingDetectionVanEngelen1202} and the Planck Collaboration \citep{Planck2013lensing}. 

As a consequence of Thomson scattering the CMB is polarized.
Lensing modifies the polarization pattern and in particular partly changes the parity of the modes. Extending lensing analyses to polarization data has the potential to increase the signal-to-noise of the reconstruction since the small-scale B-mode
polarization signal is expected to be solely sourced by the lensing of E-modes. First measurements of lensing power spectra based on polarization data have just recently been carried out with POLARBEAR~\citep{PolbearLensing}, SPTPol~\citep{2015SPTPollensing} and Planck~\citep{Planck2015lensing}. The ACTPol, SPT and POLARBEAR collaborations have also reported detections from cross-correlating the reconstructed polarization lensing with a measurement of the the cosmic infrared background \citep{SPTPol_lensingCIB,PB_lensingCIB,vEngelenACTPolCIBcrosspol}.  With decreasing noise levels, smaller beam sizes and larger areas observed, measurements of the CMB lensing effect have tremendously increased in precision; this rapid progress is expected to continue. Increasing precision in the measurement demands higher accuracy of reconstruction techniques and theoretical modeling of the measurements. 

CMB lensing analyses commonly rely on the assumption of Gaussianity of both the unlensed CMB temperature field as well as the lensing potential. The lensing potential is a projection of the gravitational potential, which is known to become non-Gaussian at late times due to nonlinear structure formation. However, the weighted projection which sums up the effect of all fluctuations encountered on the photon geodesic should suppress this non-Gaussianity by the central limit theorem (given a distance to the CMB of 14000 Mpc a CMB photon typically passes through $\mathcal{O}(50)$ structures of size 300 comoving Mpc, the scale at which the matter power spectrum peaks). The goal of our paper is to test this intuitive argument quantitatively by abandoning the assumption of Gaussianity of the lensing potential and investigating the consequences of a non-zero bispectrum of the lensing potential on measurements of the lensing power spectrum. The main result is a new, typically negative, reconstruction bias ($N^{(3/2)}$) that contributes to the measured lensing power spectrum and must be corrected for. Following further tests of the importance of some of the neglected terms with analytics and simulations, this bias should be subtracted from future lensing 4-point measurements.  It adds to known reconstruction power spectrum biases that arise for a Gaussian lensing potential and have been worked out in detail in \citep{kesdenCoorayN1, hanson1008, Anderes1301}.

While the effect of large-scale structure non-Gaussianity on lensing statistics has been computed in the context of galaxy weak lensing \citep{DodelsonShapiroWhite2006, shapiro0812, KrauseHirata2010}, it has not been analytically studied for CMB lensing reconstruction before. A first numerical analysis with non-Gaussian deflection fields computed from N-body simulations has just recently been carried out \citep{Antolini2014}. Higher order corrections to the lensed temperature power spectrum have been investigated analytically \citep{MerkelSchaefer2011, Hagstotz2014} and in lensing simulations \citep{Carbone2009,SPTLensingDetectionVanEngelen1202,Calabrese2015}.  Related but different 4-point CMB lensing biases were studied in \cite{VanEngelen1310,Osborne1310}, caused by unresolved radio/infrared point sources and galaxy clusters that add to CMB fluctuations.

Our paper is organized as follows. In \secref{formalism} we review the formalism of CMB lensing and lensing measurements, introduce notation and conventions used in this paper and provide an analytic expression for the lensing bispectrum. The rigorous derivation of the new reconstruction power bias is presented in \secref{effect_on_CPP} and results of its numerical evaluation are given in \secref{NumEval}.  In \secref{ValTests} we provide an overview of potential caveats in the numerical evaluation of the bias and present cross-checks that were carried out to validate the results. An extension of the bias to CMB lensing cross-correlation measurements is derived in \secref{crosscorr}. We conclude in \secref{conclusions}. In a series of appendices we provide details on the CMB lensing bispectrum and its effect on lensing reconstruction in Fourier space, large-scale and squeezed limits, the generalization of one of the contributing bias terms to polarization and a position space re-interpretation.

\subsection*{Notation and Conventions}
We will mostly work on the flat sky, denoting position space coordinates with $\vx$. We use non-unitary Fourier conventions
\begin{equation}
	\label{FourierConvention}
		\mathcal{F}[f(\vx)]=\tilde{f}(\vk)=\int \d^\mathrm{n} \vx\, f(\vx) \e{-i\vk\vx}
		\hspace{1 cm}
		\mathcal{F}^{-1}[\tilde{f}(\vk)]=f(\vx)=\int \frac{\d^\mathrm{n} \vk}{\twopi{\mathrm{n}}}\, \tilde{f}(\vk) \e{i \vk\vx}
\end{equation}
and express correlations in harmonic space in terms of power spectra, bispectra and trispectra defined in the usual manner
\begin{align}
	\label{Spectra}
		\langle A(\vk)B(\vk')\rangle&= (2\pi)^\mathrm{n}\delta^\mathrm{n}(\vk+\vk')P_{AB}(|\vk|)\\
		\langle A(\vk)B(\vk')C(\vk'')\rangle&= (2\pi)^\mathrm{n} \delta^\mathrm{n}(\vk+\vk'+\vk'')B_{ABC}(\vk,\vk',\vk'')\\
		\langle A(\vk)B(\vk')C(\vk'')D(\vk''')\rangle _c&= (2\pi)^\mathrm{n} \delta^\mathrm{n}(\vk+\vk'+\vk''+\vk''')\mathcal{T}_{ABCD}(\vk,\vk',\vk'',\vk''').
\end{align}
For compactness we also denote
\begin{equation}
  \label{eq:5}
  \int_{\vl} \equiv \int\frac{\d^2 \vl}{(2\pi)^2}
  \hspace{0.5 cm}\text{and}\hspace{0.5 cm}
  \int_{\vk} \equiv \int\frac{\d^3 \vk}{(2\pi)^3}.
\end{equation}

\section{CMB lensing and reconstruction}
\label{sec:formalism}

This section provides a short review of CMB lensing and reconstruction to set up the basic formalism needed later. 

\subsection{CMB lensing potential and its statistics}
Lensing remaps the CMB temperature field $T$ at angular position $\vx$ on the sky
\beq
\tilde{T}(\vx)=T(\vx+\va{\vx}),
\eeq
where the total deflection angle $\valpha(\vx)$ depends on the large-scale structures encountered along the line of sight. It can be expressed in terms of the  lensing potential $\phi$ through
\begin{align}
  \label{eq:alphaphi}
 \valpha(\vx)=\nabla\phi(\vx).
\end{align}
The lensed temperature $\tilde T$ can be approximated by perturbing in the lensing potential.
We restrict ourselves to this series expansion here, but note that this approximation is only accurate to about $5-10\%$ \cite{phys-repts} and could be improved by using the correlation function approach of \cite{SeljakCorrelFcn95}, which is nonperturbative in the deflection angle. 
Working under the flat-sky approximation valid on small scales and truncating at second order in $\nabla \phi$, the perturbative series can be written as
\begin{equation}
\label{eq:TaylorExpandT}
\tilde T(\vx) = T(\vx) +  \nabla T(\vx)\cdot \nabla \phi(\vx) + \frac{1}{2}\,\nabla_i \nabla_j T(\vx)\;\nabla_i \phi(\vx) \;\nabla_j \phi(\vx)   + \mathcal{O}(\phi^3).
\end{equation}
Throughout the paper,  $\nabla$ only acts on the single variable directly following it.
In harmonic space, products turn into convolutions and gradients correspond to multiplication with $-i\vl$, so that
\begin{equation}
  \label{eq:1}
  \tilde T(\vl) = T(\vl) + \delta T(\vl) + \delta^2 T(\vl) + \mathcal{O}(\phi^3)
\end{equation}
with $\mathcal{O}(\phi)$ correction
\begin{equation}
  \label{eq:deltaT}
  \delta T(\vl) = - \int_{\vl'} \vl'\cdot (\vl-\vl')
  T(\vl')\phi(\vl-\vl')
\end{equation}
and $\mathcal{O}(\phi^2)$ correction
\begin{equation}
  \label{eq:delta2T}
\delta^2 T(\vl) = \frac{1}{2} \int_{\vl'}\int_{\vl''}
\left[\vl'\cdot\vl''\right]
\left[\vl'\cdot(\vl-\vl'-\vl'')\right]
T(\vl')\phi(\vl'')
\phi(\vl-\vl'-\vl'').
\end{equation}
All perturbations are linear in the unlensed temperature $T$.

The lensing potential is a weighted projection of the gravitational potential $\psi$ along the line of sight
\beq
\phi(\n)=\int_0^{\chi_{\ast}}\,\d\chi W (\chi)\psi\left(\chi\n, \eta(\chi) \right)
\eeq
where $\chi$ denotes the comoving angular diameter distance, $W$ the lensing efficiency and $\eta$ the conformal time. In a flat Universe which is assumed here, the lensing efficiency simplifies to
\begin{equation}
W \( \chi \) = -2 \frac{\chi_\ast{-}\chi}{\chi_\ast \chi},
\end{equation}
where an asterisk is used to mark quantities at decoupling. This description of CMB lensing relies on the Born approximation, which assumes that the integration can be carried out along the unperturbed photon geodesic.

Commonly the lensing potential is modeled as a homogeneous Gaussian random field that is solely characterized by its power spectrum.
This power spectrum is well described by a Limber projection of the power spectrum of matter fluctuations, $P_\delta(k,\eta)$,
\beq
C_L^{\phi\phi}=\int_0^{\chi_{\ast}} \d\chi \frac{{W(\chi)}^2}{\chi^2} \frac{\gamma\(\chi\)^2} {(L/\chi)^4} P_{\delta}(L/\chi;\chi),
\eeq
where
\beq
\gamma(\chi)\equiv \frac{3}{2}\frac{ H_0^2\, \Omega_\mathrm{m0}}{ c^2 a(\chi)}.
\eeq
In this work we drop the assumption of Gaussianity and allow for a non-zero bispectrum of the lensing potential. 

Similarly to the lensing power spectrum, the lensing bispectrum is a projection of the bispectrum of density perturbations $B_\delta(\vk_1,\vk_2,{-}\vk_1{-}\vk_2;\chi)$:
\beq
\label{eq:bi_phi_of_B_delta}
B_{\phi}(\vl_1, \vl_2, \vl_3) =- \int_0^{\chi_{\ast}} \d\chi\,\chi^2 {W(\chi)}^3 \frac{\gamma\(\chi\)^3} {({l_1} {l_2} {l_3})^2}{B}_{\delta}(l_1/\chi,l_2/\chi,l_3/\chi;\chi).
\eeq
As summarized in \app{Bispec}, this follows by applying the Fourier space analogue of Limber's equation for bispectra (see e.g.~\citep{2000ApJ53036B,2004MNRAS.348..897T}).
On very large scales, we expect the flat-sky and the Limber approximations that we assume to break down. The lensing power spectrum is overestimated on large scales in the Limber approximation and the bispectrum could be affected similarly. We will therefore only consider multipoles $L > 100$. 

The bispectrum of matter perturbations can be modeled by standard Eulerian perturbation theory, which gives at leading order in the linear matter overdensity (see \citep{Bernardeau2002} for a review),
\beq
\label{eq:SPTbispec}
B_\delta(\vk_1,\vk_2,\vk_3;\eta)=2\,{F}_2(\vk_1,\vk_2) P_\delta(k_1,\eta) P_\delta(k_2,\eta)+ (1\leftrightarrow 3)+(2\leftrightarrow 3).
\eeq
This is quadratic in the linear matter power spectrum $P_\delta$ and involves the symmetrized kernel 
\beq
\label{eq:F2}
F_2(\vk_i,\vk_j)=\frac{5}{7}+\frac{1}{2}\left(\frac{k_i}{k_j}+\frac{k_j}{k_i}\right)\hat\vk_i\cdot\hat\vk_j+\frac{2}{7}(\hat\vk_i\cdot\hat\vk_j)^2,
\eeq
where $\hat\vk_i=\vk_i/|\vk_i|$.
The simple bispectrum model \eq{SPTbispec} is only accurate on relatively large scales that are under perturbative control (roughly $k\lesssim 0.07\,h/\mathrm{Mpc}$ at $z=0$, see e.g.~\cite{Lazanu:2015rta} for a recent study).  It can be extended to smaller, nonlinear scales by including higher-order (loop) corrections.
A simpler phenomenological modification that extends the range of validity to slightly smaller scales can be obtained by replacing $P_\delta(k)$ with a matter power spectrum with nonlinear corrections, $P^{\text{nl}}_\delta(k)$, in \eqq{SPTbispec} \cite{SPTBispecFit1}. 
On smaller scales, that cannot be modeled analytically, one needs to resort to fitting formulae calibrated against simulations.

\subsection{Lensing reconstruction}
For a fixed lensing potential, the effect of lensing is to introduce a correlation between different, in the unlensed case independent, modes of the temperature field. The resulting non-diagonal terms in the 2-point correlator of the CMB in harmonic space can be used to construct a quadratic lensing reconstruction estimator \cite{1999PhRvL82.2636S,2001ApJ...557L..79H}, which can be written on the flat sky as 
\begin{equation}
  \label{eq:phi_esti}
  \hat\phi(\VL) = A_L \int_\vl g(\vl,\VL)\tilde
  T_\expt(\vl)\tilde T_\expt(\VL-\vl),
\end{equation}
where $\tilde T_\mathrm{expt}$ are beam-deconvolved noisy temperature fluctuations. The observed temperature fluctuations are assumed to contain white noise and a Gaussian beam, so that the final power spectrum is
\begin{equation}
  \label{eq:Cltot}
  C^{\tilde T\tilde T}_{l,\expt} = C^{\tilde T\tilde T}_l + \sigma^2_N\exp\left[l(l+1)\theta^2_\mathrm{FWHM}/(8\ln 2)\right],
\end{equation}
where the instrumental noise level is specified by $\sigma_N^2$ and the beam size is given in terms of the full width at half-maximum (FWHM) $\theta_\mathrm{FWHM}$. 
The weight $g$ in \eqq{phi_esti} is chosen such that the variance of the estimator
is minimized \cite{okamotoHu0301,lewis1101,hanson1008};
\begin{equation}
  \label{eq:gweight}
  g(\vl,\VL)=\frac{(\VL-\vl)\cdot\VL C^{\tilde T\tilde T}_{|\VL-\vl|} +
    \vl\cdot\VL C_l^{\tilde T\tilde T}}{2C_{l,\expt}^{\tilde T\tilde
      T}C_{|\VL-\vl|,\expt}^{\tilde T\tilde T}}.
\end{equation}
Note that $g(\VL-\vl, \VL)=g(\vl,\VL)=g(-\vl,-\VL)$.
The normalization is given by
\begin{equation}
  \label{eq:ALdef}
  A_L^{-1} = 
2    \int_\vl g(\vl,\VL)
\vl\cdot\VL C^{ \tilde T \tilde T}_l.
\end{equation}

The power spectrum of the lensing reconstruction \eq{phi_esti} involves the lensed temperature $4$-point function,
\begin{align}
  \label{eq:reconstructionPowerExpValue}
  \la\hat{\phi}(\VL)\hat{\phi}(-\VL)\ra = 
A_L^2 \int_{\vl_1}\int_{\vl_2} g(\vl_1,\VL)\,
g(\vl_2,\VL) \,
\la\, \tilde T_\expt(\vl_1)\, \tilde T_\expt(\VL-\vl_1) 
\,\tilde T_\expt(-\vl_2)\,\tilde T_\expt(\vl_2-\VL)\,\ra.
\end{align}
This 4-point function can be split
into a disconnected part, obtained by contracting two pairs of lensed temperature fields with each other, and a connected part, given by the full 4-point function minus the disconnected part. The disconnected part leads to the $N^{(0)}$ power spectrum bias, which would be present even for Gaussian temperature fluctuations in absence of lensing.  It is called $N^{(0)}$ because it is of zeroth order in $C^{\phi\phi}$.\footnote{We follow the common power-counting practice where only explicit appearances of $C^{\phi\phi}$ are counted that are not contained in lensing contributions to  $C^{\tilde T\tilde T}$. }
Note $N_L^{(0)}=A_L$ (a consequence of optimal weighting). The connected part of the 4-point function in \eqq{reconstructionPowerExpValue} leads to the desired signal contribution $C_L^{\phi\phi}$. Additionally, it  gives rise to the $N^{(1)}$ bias which is also of order $C^{\phi\phi}$ \cite{KCK0302N1,hanson1008,Anderes1301}.   The expectation value of the measured lensing power spectrum is therefore
\begin{equation}
  \label{eq:biasesGaussian}
  \langle C^{\hat\phi\hat\phi}_L\rangle = N_L^{(0)} + C_L^{\phi\phi}+ N_L^{(1)} + \mathcal{O}[(C^{\phi\phi})^3] 
\quad\qquad \mbox{(Gaussian } \phi \mbox{)}
\end{equation}
if the lensing potential $\phi$ is assumed to be Gaussian. To obtain an unbiased estimator for the signal $C^{\phi\phi}$, the $N^{(0)}$ and $N^{(1)}$ biases are calculated (typically using simulations or simulation-data combinations) and subtracted from the measured lensing power.
\section{Effect of lensing bispectrum on measured lensing power spectrum}
\label{sec:effect_on_CPP}

\subsection{Overview}
\label{sec:N32Overview}
We now drop the assumption that the lensing potential $\phi$ is Gaussian.  In this case, $n$-point functions with an odd number of lensing potentials no longer need to vanish, and $n$-point functions no longer need be determined by the Gaussian 2-point power spectrum $C^{\phi\phi}$ alone.  We consider only a non-zero 3-point function or bispectrum, and ignore corrections from all higher-order $n$-point functions.  This approximation is motivated by the specific non-Gaussianity generated by large-scale structure modes in the mildly nonlinear regime relevant for CMB lensing.  
We also assume that the unlensed CMB is a Gaussian field. 
For simplicity, we ignore the ISW effect and its induced correlation $C^{T\phi}$, but note that accounting for it may lead to additional biases that should be investigated in the future.

Allowing a non-zero lensing potential bispectrum $B_{\phi}$, the lensed temperature 4-point function entering the expectation value for the measured lensing power spectrum \eq{reconstructionPowerExpValue} picks up additional contractions that  would vanish for a Gaussian lensing potential.  For example, using the Taylor expansion \eq{TaylorExpandT}, one new allowed contraction is of the form
\begin{align}
  \label{eq:71}
  \la\tilde T\tilde T\tilde T\tilde T\ra
\;=\;
\la\delta T\delta T\delta T T\ra +\cdots
\;=\;\la 
{
\contraction{}{T}{\!{}_{,i}\phi\!{}_{,i}}{T}
\contraction{T\!{}_{,i}\phi\!{}_{,i} T\!{}_{,j}\phi\!{}_{,j} }{T}{\!{}_{,k}\phi\!{}_{,k}}{T}
\contraction{T\!{}_{,i}\phi\!{}_{,i} T\!{}_{,j}\phi\!{}_{,j} }{T}{\!{}_{,k}\phi\!{}_{,k}}{T}
\bcontraction{T\!{}_{,i}}{\phi}{\!{}_{,i} T\!{}_{,j}}{\phi}
\bcontraction{T\!{}_{,i}}{\phi}{\!{}_{,i} T\!{}_{,j}\phi\!{}_{,j} T\!{}_{,k}}{\phi}
T\!{}_{,i}\phi\!{}_{,i} T\!{}_{,j}\phi\!{}_{,j} T\!{}_{,k}\phi\!{}_{,k} T}
\ra +\cdots,
\end{align}
where subscripts denote gradients $T_{,i}=\nabla_i T$ and $\phi_{,i}=\nabla_i\phi$.  Since the lensing change $\delta^n T$ is of order $\phi^n$ and linear in the unlensed temperature $T$, there are four qualitatively different contraction types that arise for the measured lensing power spectrum \eq{reconstructionPowerExpValue} at order $\phi^3$:
\begin{equation}
\label{eq:66}
\text{type A: }\langle \delta T\delta T\,\delta T' T' \rangle \qquad
\text{type B: }\langle\delta^2 T\delta T\,  T' T' \rangle \qquad
\text{type C: }\langle \delta^2 T  T\,\delta T'  T'\rangle \qquad
\text{type D: }\langle \delta^3 T T\,T'T' \rangle.
\end{equation}
The last two temperature fields are labeled with primes to indicate that they correspond to the second reconstruction field $\hat\phi(-\VL)$ in \eqq{reconstructionPowerExpValue}; quantities without primes correspond to the first reconstruction field $\hat\phi(\VL)$.\footnote{In position space, this corresponds to reconstructed lenses at two different positions $\vx$ and $\vx'$ on the sky, also see \app{N32Interpretations}.} 

Each type of terms allows several Wick's theorem contractions. For example, for type A there are three contractions that we label A1, A2 and A3:
\begin{alignat}{2}
  \label{eq:70}
\text{(A)} \; & \;  \langle \delta T\delta T\,\delta T' T' \rangle \;\sim\;\la &&
{
\contraction{}{T}{\!{}_{,i}\phi\!{}_{,i}}{T}
\contraction{T\!{}_{,i}\phi\!{}_{,i} T\!{}_{,j}\phi\!{}_{,j} \,}{T'}{\!{}_{,k}\phi'\!{}_{,k}}{T'}
\bcontraction{T\!{}_{,i}}{\phi}{\!{}_{,i} T\!{}_{,j}}{\phi}
\bcontraction{T\!{}_{,i}}{\phi}{\!{}_{,i} T\!{}_{,j}\phi\!{}_{,j} \,T'\!{}_{,k}}{\phi'}
T\!{}_{,i}\phi\!{}_{,i} T\!{}_{,j}\phi\!{}_{,j}\, T'\!{}_{,k}\phi'\!{}_{,k} T'}
\ra_{A1}
\;
+
\;\la
{
\contraction[0.75ex]{}{T}{\!{}_{,i}\phi\!{}_{,i} T\!{}_{,j}\phi\!{}_{,j}\, }{T'}
\contraction[1.5ex]{T\!{}_{,i}\phi\!{}_{,i}}{T}{\!{}_{,j}\phi\!{}_{,j}\, T'\!{}_{,k}\phi'\!{}_{,k}}{ T'}
\bcontraction{T\!{}_{,i}}{\phi}{\!{}_{,i} T\!{}_{,j}}{\phi}
\bcontraction{T\!{}_{,i}}{\phi}{\!{}_{,i} T\!{}_{,j}\phi\!{}_{,j} \,T'\!{}_{,k}}{\phi'}
T\!{}_{,i}\phi\!{}_{,i} T\!{}_{,j}\phi\!{}_{,j}\, T'\!{}_{,k}\phi'\!{}_{,k} T'}
\ra_{A2}
\;
+
\;\la
{
\contraction[1.5ex]{}{T}{\!{}_{,i}\phi\!{}_{,i} T\!{}_{,j}\phi\!{}_{,j}\, T'\!{}_{,k}\phi'\!{}_{,k}}{T'}
\contraction[0.75ex]{T\!{}_{,i}\phi\!{}_{,i}}{T}{\!{}_{,j}\phi\!{}_{,j}\, }{T'}
\bcontraction{T\!{}_{,i}}{\phi}{\!{}_{,i} T\!{}_{,j}}{\phi}
\bcontraction{T\!{}_{,i}}{\phi}{\!{}_{,i} T\!{}_{,j}\phi\!{}_{,j} \,T'\!{}_{,k}}{\phi'}
T\!{}_{,i}\phi\!{}_{,i} T\!{}_{,j}\phi\!{}_{,j}\, T'\!{}_{,k}\phi'\!{}_{,k} T'}
\ra_{A3}.
\end{alignat}
Similarly, the type B term has three contractions B1, B2 and B3,
\begin{alignat}{2}
  \label{eq:typeBconnected}
\text{(B)} \; & \; \langle \delta^2 T \delta T\,  T' T' \rangle \;\sim\;\la &&
{
\contraction[1ex]{}{T}{\!{}_{,ij}\phi\!{}_{,i}\phi\!{}_{,j}}{T}
\contraction[1ex]{T\!{}_{,ij}\phi\!{}_{,i}\phi\!{}_{,j}T\!{}_{,k}\phi\!{}_{,k}\, }{T'}{}{T}
\bcontraction[1ex]{T\!{}_{,ij}}{\phi}{\!{}_{,i}}{\phi}
\bcontraction[1ex]{T\!{}_{,ij}}{\phi}{\!{}_{,i}\phi\!{}_{,j}T\!{}_{,k}}{\phi}
T\!{}_{,ij}\phi\!{}_{,i}\phi\!{}_{,j}T\!{}_{,k}\phi\!{}_{,k}\, T' T'
}
\ra_{B1}
\;+\;
\la {
\contraction[0.5ex]{}{T}{\!{}_{,ij}\phi\!{}_{,i}\phi\!{}_{,j}T\!{}_{,k}\phi\!{}_{,k}\, }{T'}
\contraction[1.5ex]{T\!{}_{,ij}\phi\!{}_{,i}\phi\!{}_{,j}}{T}{\!{}_{,k}\phi\!{}_{,k}\, T'}{T'}
\bcontraction[1ex]{T\!{}_{,ij}}{\phi}{\!{}_{,i}}{\phi}
\bcontraction[1ex]{T\!{}_{,ij}}{\phi}{\!{}_{,i}\phi\!{}_{,j}T\!{}_{,k}}{\phi}
T\!{}_{,ij}\phi\!{}_{,i}\phi\!{}_{,j}T\!{}_{,k}\phi\!{}_{,k}\, T' T'
}
\ra_{B2}
\;+\;
\la {
\contraction[1.5ex]{}{T}{\!{}_{,ij}\phi\!{}_{,i}\phi\!{}_{,j}T\!{}_{,k}\phi\!{}_{,k}\,T' }{T'}
\contraction[0.5ex]{T\!{}_{,ij}\phi\!{}_{,i}\phi\!{}_{,j}}{T}{\!{}_{,k}\phi\!{}_{,k}\,}{T'}
\bcontraction[1ex]{T\!{}_{,ij}}{\phi}{\!{}_{,i}}{\phi}
\bcontraction[1ex]{T\!{}_{,ij}}{\phi}{\!{}_{,i}\phi\!{}_{,j}T\!{}_{,k}}{\phi}
T\!{}_{,ij}\phi\!{}_{,i}\phi\!{}_{,j}T\!{}_{,k}\phi\!{}_{,k}\, T' T'
}
\ra_{B3},
\quad
\end{alignat}
and the type C term has contributions C1, C2 and C3:
\begin{alignat}{2}
  \label{eq:typeCcontractions}
\text{(C)} \; & \;  \langle \delta^2 T T\, \delta T' T' \rangle \;\sim\;\la &&
{
\contraction[1ex]{}{T}{\!{}_{,ij}\phi\!{}_{,i}\phi\!{}_{,j}}{T}
\contraction[1ex]{ T\!{}_{,ij}\phi\!{}_{,i}\phi\!{}_{,j} T\,}{T'}{\!{}_{,k}\phi'\!{}_{,k}}{T'}
\bcontraction[1ex]{ T\!{}_{,ij}}{\phi}{\!{}_{,i}\phi\!{}_{,j} T\, T'\!{}_{,k}}{\phi'}
\bcontraction[1ex]{ T\!{}_{,ij}}{\phi}{\!{}_{,i}}{\phi}
   T\!{}_{,ij}\phi\!{}_{,i}\phi\!{}_{,j} T\, T'\!{}_{,k}\phi'\!{}_{,k} T'
}
\ra_{C1}
\;+\;\la
{
\contraction[0.75ex]{}{T}{\!{}_{,ij}\phi\!{}_{,i}\phi\!{}_{,j} T\,}{T'}
\contraction[1.5ex]{T\!{}_{,ij}\phi\!{}_{,i}\phi\!{}_{,j} }{T}{\, T'\!{}_{,k}\phi'\!{}_{,k} }{T'}
\bcontraction[1ex]{ T\!{}_{,ij}}{\phi}{\!{}_{,i}\phi\!{}_{,j} T\, T'\!{}_{,k}}{\phi'}
\bcontraction[1ex]{ T\!{}_{,ij}}{\phi}{\!{}_{,i}}{\phi}
   T\!{}_{,ij}\phi\!{}_{,i}\phi\!{}_{,j} T\, T'\!{}_{,k}\phi'\!{}_{,k} T'
}
\ra_{C2}
\;+\;\la
{
\contraction[1.5ex]{}{T}{\!{}_{,ij}\phi\!{}_{,i}\phi\!{}_{,j} T\, T'\!{}_{,k}\phi'\!{}_{,k}}{T'}
\contraction[0.75ex]{T\!{}_{,ij}\phi\!{}_{,i}\phi\!{}_{,j}}{T}{\,}{T'}
\bcontraction[1ex]{ T\!{}_{,ij}}{\phi}{\!{}_{,i}\phi\!{}_{,j} T\, T'\!{}_{,k}}{\phi'}
\bcontraction[1ex]{ T\!{}_{,ij}}{\phi}{\!{}_{,i}}{\phi}
   T\!{}_{,ij}\phi\!{}_{,i}\phi\!{}_{,j} T\, T'\!{}_{,k}\phi'\!{}_{,k} T'
}
\ra_{C3}.
\end{alignat}
We omit the type D terms here as these can be shown to be zero.

In our paper, we evaluate the A1 and C1 terms numerically and focus on them in the main text. We focus on these terms both because they are expected to be among the largest and because they allow for numerical evaluation on reasonable timescales. In contrast, as discussed in Appendix~\ref{App:AllContris}, the B1 term is zero, and the A2 and A3 terms are tightly coupled, which prevents evaluation (the integrals are six-dimensional), but also suggests that these terms are small. Furthermore, the C2 term should be naturally accounted for in the (realization-dependent) calculations of the $N^{(0)}$ bias which is included in modern lensing pipelines. We defer a full evaluation of the remaining B2, B3, and C3 terms to future work; we note that if they have a similar order of magnitude to $A1+C1$, our approximate calculation might underestimate the true bias.

The new contractions allowed by a non-zero lensing bispectrum lead to a new bias $N^{(3/2)}_{L,\mathrm{tot}}$ of the measured 4-point lensing power spectrum,
\begin{equation}
  \label{eq:biasesNG}
  \langle C^{\hat\phi\hat\phi}_L\rangle = N_L^{(0)} + C_L^{\phi\phi}+ N_L^{(1)}  + N_{L,\mathrm{tot}}^{(3/2)}+\mathcal{O}[(C^{\phi\phi})^{5/2}] \quad\qquad \mbox{(non-Gaussian } \phi \mbox{)}.
\end{equation}
We call the new non-Gaussian reconstruction bias  $N^{(3/2)}$ because it scales like $\phi^3\propto (C^{\phi\phi})^{3/2}$, and previously considered biases like $N^{(0)}$ and $N^{(1)}$ were labeled by the power of $C^{\phi\phi}$ they involve.
The total $N^{(3/2)}$ bias is a sum over all possible 4-point contractions listed above,
\begin{align}
  \label{eq:N32tot}
  N^{(3/2)}_\mathrm{tot} = 
\left( N^{(3/2)}_{A1}+N^{(3/2)}_{C1}\right)
+N^{(3/2)}_{A2}+N^{(3/2)}_{A3} 
+N^{(3/2)}_{B2} + N^{(3/2)}_{B3}
+N^{(3/2)}_{C2}+N^{(3/2)}_{C3}.
\end{align}
where as explained previously we focus here on the A1 and C1 terms in parentheses.

The A1 and C1 bias terms in Eqs.~\eq{70} and \eq{typeCcontractions} have a simple intuitive interpretation: They arise because the \emph{quadratic response} of the lensing reconstruction $\hat\phi(\VL)$ to the true lensing potential $\phi$ is correlated with the \emph{linear response} of the lensing reconstruction $\hat\phi(-\VL)$ to the true lensing potential $\phi'$.  This correlation involves the 3-point correlation function $\la\phi\phi\phi'\ra$ of the true lensing potential, which is nonzero in presence of nonlinear gravitational clustering. 

We proceed by discussing these A1 and C1 terms, which contribute substantially to the total bias \eq{N32tot}, in detail.  Analytical expressions for the remaining bias contributions are given in Appendix~\ref{App:AllContris}.

\subsection{A1 contribution to the \texorpdfstring{$N^{(3/2)}$}{N32} bias}

We begin by computing the lensing bias from the contraction A1 in \eqq{70}. This contraction is given by
\begin{align}
\nonumber
\langle\delta T_{\vl_1}& \delta T_{\vl_2}\delta T_{\vl_3}T_{\vl_4}\rangle_{A1}\non\\
&=-(2\pi)^2 \delta_D(\vl_1+\vl_2+\vl_3+\vl_4)
  C^{TT}_{l_4} \[(\vl_3+\vl_4)\cdot\vl_4\] \int_{\vl}  \[\vl\cdot(\vl_1-\vl)\]\[\vl\cdot(\vl_2+\vl)\] C^{TT}_l B_\phi(\vl_1-\vl,\vl_2+\vl,-\vl_1-\vl_2),
\end{align}
where we used the Fourier space expression \eq{deltaT} for the first order temperature change $\delta T$ due to lensing, and contracted temperature and lensing fields as indicated for the A1 term in \eqq{70}.\footnote{For Gaussian instrument noise that is uncorrelated with the signal, all contributions to the four point correlator $\langle \tilde T^{\mathrm{expt}}_{\vl_1} \tilde T^{\mathrm{expt}}_{\vl_2} \tilde T^{\mathrm{expt}}_{\vl_3} \tilde T^{\mathrm{expt}}_{\vl_4}\rangle$ that involve instrument noise either vanish or contribute to the Gaussian noise bias. This justifies ignoring instrument noise in the calculation of the connected four point contributions to $N^{(3/2)}$.}
Inserting this into \eqq{reconstructionPowerExpValue} yields the following A1 bias of the measured lensing power spectrum:
\begin{eqnarray}
N^{(3/2)}_{A1}(L) &=& -4 A_L^2  S_L \int_{\vl_1,\vl} g(\vl_1,\VL)
\[\vl\cdot(\vl_1{-}\vl)\]\[\vl\cdot(\VL-(\vl_1-\vl))\] 
C^{TT}_l B_\phi(\vl_1{-}\vl,\VL-(\vl_1-\vl),{-}\VL)  \\
&=&
\label{eq:N32A1}
-4 A_L^2  S_L
\int_{\vl_1,\vl} g(\vl_1,\VL)  [(\vl_1-\vl)\cdot\vl]
[(\vl_1-\vl)\cdot (\VL-\vl)]C^{TT}_{|\vl_1-\vl|}B_\phi(\vl, \VL-\vl,-\VL).
\end{eqnarray}
The prefactor $S_L$ is an integral over the filtered unlensed CMB power spectrum, 
\begin{align}
  \label{eq:Sdef}
  S_L = \int_{\vl_2} g(\vl_2,\VL) (\vl_2\cdot\VL) 
C_{l_2}^{TT},
\end{align}
satisfying $S_L\approx 1/(2A_L)$ at leading order in $C^{\phi\phi}$.
The prefactor of 4 in \eqq{N32A1} stems from the four possibilities to arrange three temperatures perturbed to first order and one unperturbed temperature in a 4-point correlator.  \eqq{N32A1} follows by changing integration variables $\vl\rightarrow \vl_1-\vl$.

\subsection{C1 contribution to the \texorpdfstring{$N^{(3/2)}$}{N32} bias}

The C1 contraction defined in \eqq{typeCcontractions} is
\begin{align}
\nonumber
  \langle \delta^2 T_{\vl_1}& T_{\vl_2}\delta T_{\vl_3}T_{\vl_4}
  \rangle_{C1}\\
\label{eq:TrispTypeCFinalMainText}
&=\;\frac{(2\pi)^2}{2}\delta_D(\vl_1+\vl_2+\vl_3+\vl_4)C_{l_2}^{TT}C_{l_4}^{TT}
\left[(\vl_3+\vl_4)\cdot\vl_4\right]
\int_{\vl} (\vl_2\cdot\vl ) \left[
\vl_2\cdot(\vl_1+\vl_2-\vl)
\right] B_\phi(\vl, \vl_1+\vl_2-\vl,-\vl_1-\vl_2).
\end{align}
Inserting this in \eqq{reconstructionPowerExpValue} gives the following C1 bias of the measured lensing power spectrum:
\begin{eqnarray}
\label{eq:N32C1}
  N^{(3/2)}_{C1}(L) &=& 4 A_L^2 
S_L
\int_{\vl_1,\vl} 
g(\vl_1,\VL)  
(\vl_1 \cdot \vl )
\left[\vl_1 \cdot(\VL-\vl) \right] 
C_{l_1}^{TT} 
B_\phi(\vl,\VL-\vl,-\VL),
\end{eqnarray}
We changed integration variables $\vl_1\rightarrow \VL-\vl_1$, and we accounted for a symmetry factor $8$ that arises because the resulting lensing bias does not change if we exchange $\vl_1\leftrightarrow \vl_2$, or $\vl_3\leftrightarrow\vl_4$, or both in \eqq{TrispTypeCFinalMainText}.

\subsection{Integral expressions for fast numerical evaluation}

The A1 and C1 biases in Eqs.~\eq{N32A1} and \eq{N32C1} involve four-dimensional integrals for every multipole  $L$, which are computationally expensive to evaluate. 
Fortunately, however, the integrands of these 4D integrals can be rewritten in a product-separable form, which allows much faster numerical evaluation by multiplying 2D integrals. In \app{fastExpressions} we demonstrate this and derive the following simply-evaluated expression for the C1 bias:
\begin{eqnarray}
\label{eq:N32decompExact}
N^{(3/2)}_{C1}(L) &=& -4 A_L^2 S_L 
\left[R_\parallel(L)\beta_\parallel(L)+R_\perp(L)\beta_\perp(L)\right],
  \end{eqnarray}
where we defined
the temperature integral $R_\parallel$ and integrated lensing bispectrum $\beta_\parallel$ as
\begin{eqnarray}
  \label{eq:Rparallel}
R_\parallel(L) &=& \int_{\vl_1}g(\vl_1,\VL)  l_1^2\cos^2(\mu_{\vl_1}) C_{l_1}^{TT}, \\
\label{eq:betaparallel}
  \beta_\parallel(L) &=& \int_{\vl}  l\cos\mu_\vl\left[ l\cos\mu_\vl - L \right] \,B_\phi(\vl,\VL-\vl,-\VL),
\end{eqnarray}
and similarly for the perpendicular component,
\begin{eqnarray}
\label{eq:Rperp}
\qquad
R_\perp(L) &=& \int_{\vl_1}g(\vl_1,\VL)  l_1^2 \sin^2(\mu_{\vl_1}) C^{TT}_{l_1},\\
  \label{eq:betaperp}
  \beta_\perp(L) &=& \int_\vl  l^2 \sin^2(\mu_\vl) \,B_\phi(\vl,\VL-\vl,-\VL),
\end{eqnarray}
where $\cos\mu_{\vl_1}=\vl_1\cdot\VL/(l_1L)$ and $\cos\mu_\vl=\vl\cdot\VL/(lL)$.
In \app{fastExpressions}, we also derive a similar fast integral expression for the A1 bias. 

\subsection{Comparison of A1 and C1 contributions to the $N^{(3/2)}$ bias}
The A1 bias of \eqq{N32A1} and the C1 bias of \eqq{N32C1} have a very similar structure. This makes sense because these biases arise from similar contractions in Eqs.~\eq{70} and \eq{typeCcontractions}.
In the limit of \eqq{N32A1} where the lensing multipole $l$ is much lower than the temperature multipole $l_1$ (i.e.~$l\ll l_1$ and $\vl_1-\vl\approx \vl_1$), the A1 and C1 biases cancel each other.
The potential cancellation in this limit demands careful numerical evaluation of the A1 and C1 contributions to the $N^{(3/2)}$ bias.
Numerically, we will find later that the range of reconstruction multipoles $L$ where this cancellation is actually relevant depends strongly on experimental specifications.
At very low reconstruction multipoles $L$, the cancellation helps to regularize the $N^{(3/2)}$ bias by cancelling individually large A1 and C1 contributions with opposite sign.

\section{Numerical evaluation}
\label{sec:NumEval}

\subsection{Implementation}
We continue by evaluating the expressions in \eq{N32A1} and \eq{N32decompExact} that follow from the type A1 and type C1 contractions. The integrals over the lensing bispectrum can be evaluated for any model of the lensing bispectrum.  We evaluate them using the leading-order standard perturbation theory expression \eq{bi_phi_of_B_delta} with $P^{\text{lin}}_\delta(k)$ replaced by the nonlinear matter power spectrum $P^{\text{nl}}_\delta(k)$, which fits simulations slightly better than the leading-order bispectrum involving $P^{\text{lin}}_\delta(k)$ (also see \secref{kNL_test} for a discussion of the validity of this bispectrum model). 

Small-scale temperature contributions to the integrals are suppressed by setting the experimental noise to an unphysically high value (irrespective of the experiment) for temperature multipoles $l\ge 3000$. This small-scale cutoff is often applied to real data to ensure the results are insensitive to astrophysical emission from dusty galaxies and the Sunyaev-Zeldovich effect which become relevant at these scales.

\begin{table}[t]
\caption{\label{tab:Experiments} Typical beam and noise specifications of current and future experiments. All resolution and noise dependent results shown are based on one of these configurations. The noise level stated in this table is for temperature measurements. For polarization we use $\sigma_N^{BB}=\sigma_N^{EE}=\sqrt{2}\sigma_N^{TT}$.}
\begin{ruledtabular}
\begin{tabular}{c|m{3cm}m{3cm}m{3cm}m{3cm}}
Representative experiment&Stage-IV\newline(CMB-S4)&Stage-III\newline(AdvancedACT-like)&Planck\\
 \hline
 $\theta_\mathrm{FWHM}$[arcmin]&1.0&1.4&7.0\\
 $\sigma_N^{TT} [\mu \text{Karcmin}]$&1.0&6.0&30.0\\
 $f_{\mathrm{sky}}$& 0.5 & 0.4 & 0.63
\end{tabular}
\end{ruledtabular}
\end{table}

To evaluate the contributions to the $N^{(3/2)}$ bias we consider different experimental setups roughly corresponding to CMB Stage-III, Stage-IV and Planck experiments. 
Beam width, noise levels and sky coverage for these representative classes of experiments are summarized in \tabref{Experiments}.

For the calculation of the fiducial power spectra of matter, CMB and lensing potential we use the publicly available CLASS code\footnote{\url{http://www.class-code.net/}} \cite{CLASS}. The computation of nonlinear corrections to the power spectrum of density fluctuations is based on the HALOFIT method \cite{halofit1,halofit2}. The underlying cosmology is a standard $\Lambda$CDM cosmology with Planck 2013 best fit parameters: $\omega_m=0.311$, $\Omega_b=0.049$, $h=0.671$, $A_s=2.215\times 10^{-9}$, $n_s=0.968$ and $T_{\text{CMB}}=2.7255\;\mathrm{K}$ \citep{Planck2013Params}.

We next discuss results for lensing measurements from the CMB temperature fluctuations and include the contribution to $N^{(3/2)}$ from the sum of the two couplings A1 and C1.  We then proceed with polarization measurements for which we only evaluate the C1 coupling because it is simpler to evaluate.

\subsection{Results for A1 and C1 contributions to the \texorpdfstring{$N^{(3/2)}$}{N32} bias for \texorpdfstring{$\TT,\TT$}{TT,TT} reconstruction}
\label{sec:results}
\renewcommand{\arraystretch}{1.4}

\ifincludefigs
\begin{figure}[tp]
\begin{center}
\scalebox{0.47}{\includegraphics{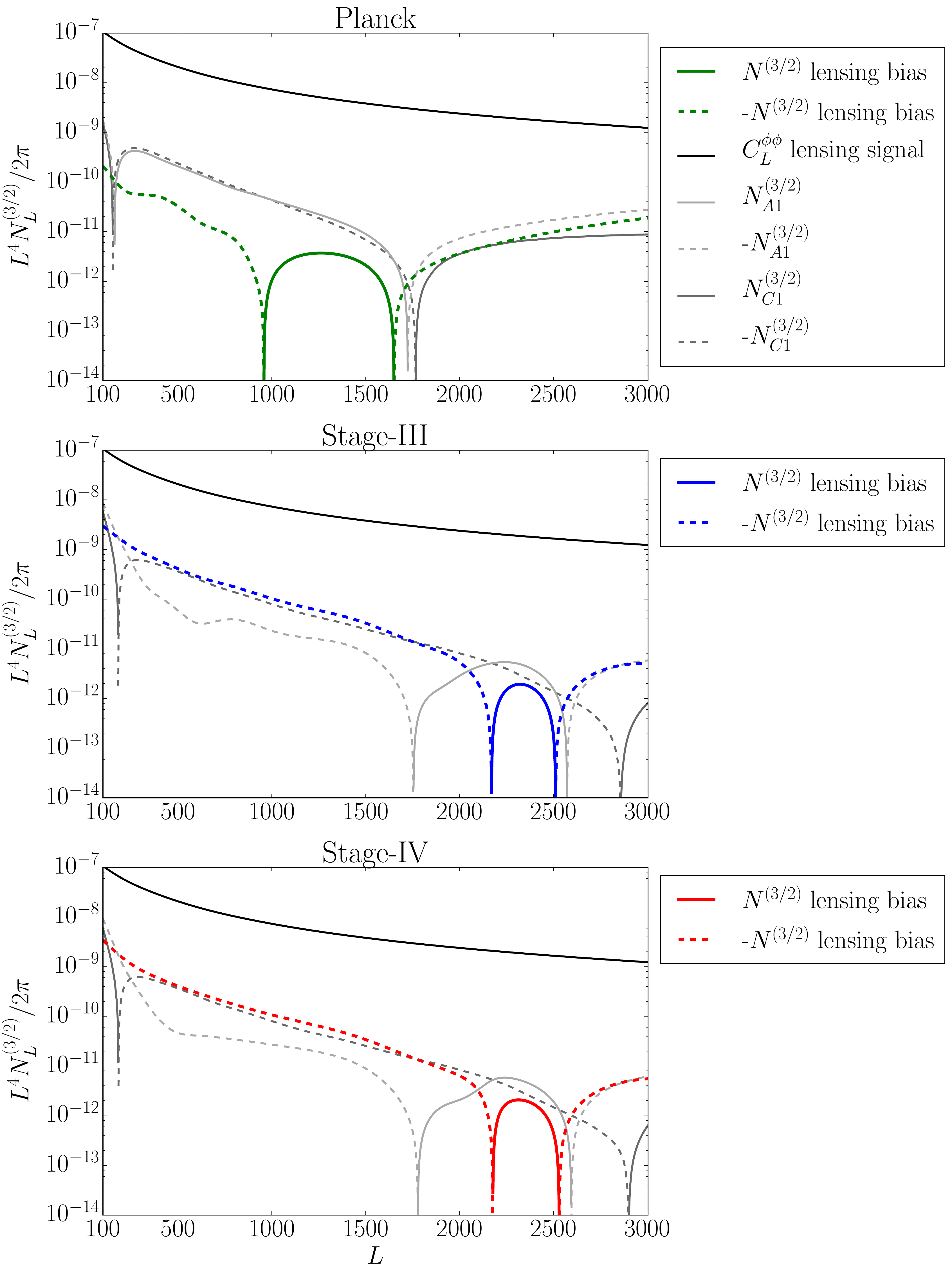}}
\caption{\label{fig:ngBias} $N^{(3/2)}$ CMB lensing bias that arises as a consequence of a non-vanishing bispectrum of large-scale structure. In this plot we show the bias on a measurement of the CMB lensing power spectrum from temperature data. The signal lensing power spectrum $C^{\phi\phi}$ is shown for comparison (black). Different panels show different experiment specifications summarized in \tabref{Experiments}. The bias appears significant for Stage-III and Stage-IV experiments. It should be noted that the bias plotted here is the sum of two out of many contributing terms to the total bias (see \secref{N32Overview}). These two terms, denoted type A1 (\eqq{N32A1}) and type C1 (\eqq{N32C1}), are likely two of the largest terms. We provide analytic expressions for the remaining terms but defer their evaluation to future work. }
\end{center}
\end{figure}
\fi

\ifincludefigs
\begin{figure}[tp]
\begin{center}
\subfloat{
\scalebox{0.43}{\includegraphics{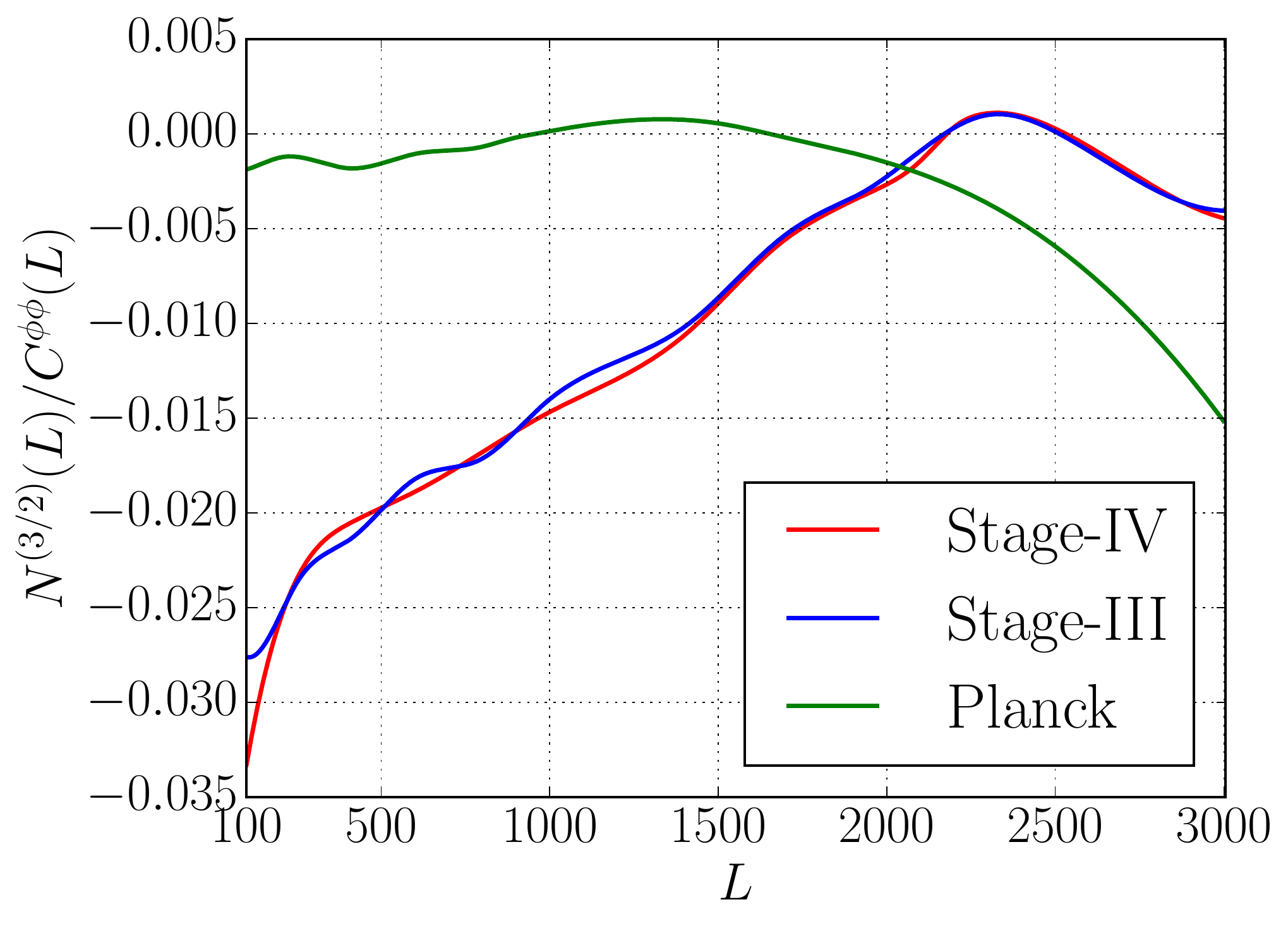}}}
\subfloat{
\scalebox{0.43}{\includegraphics{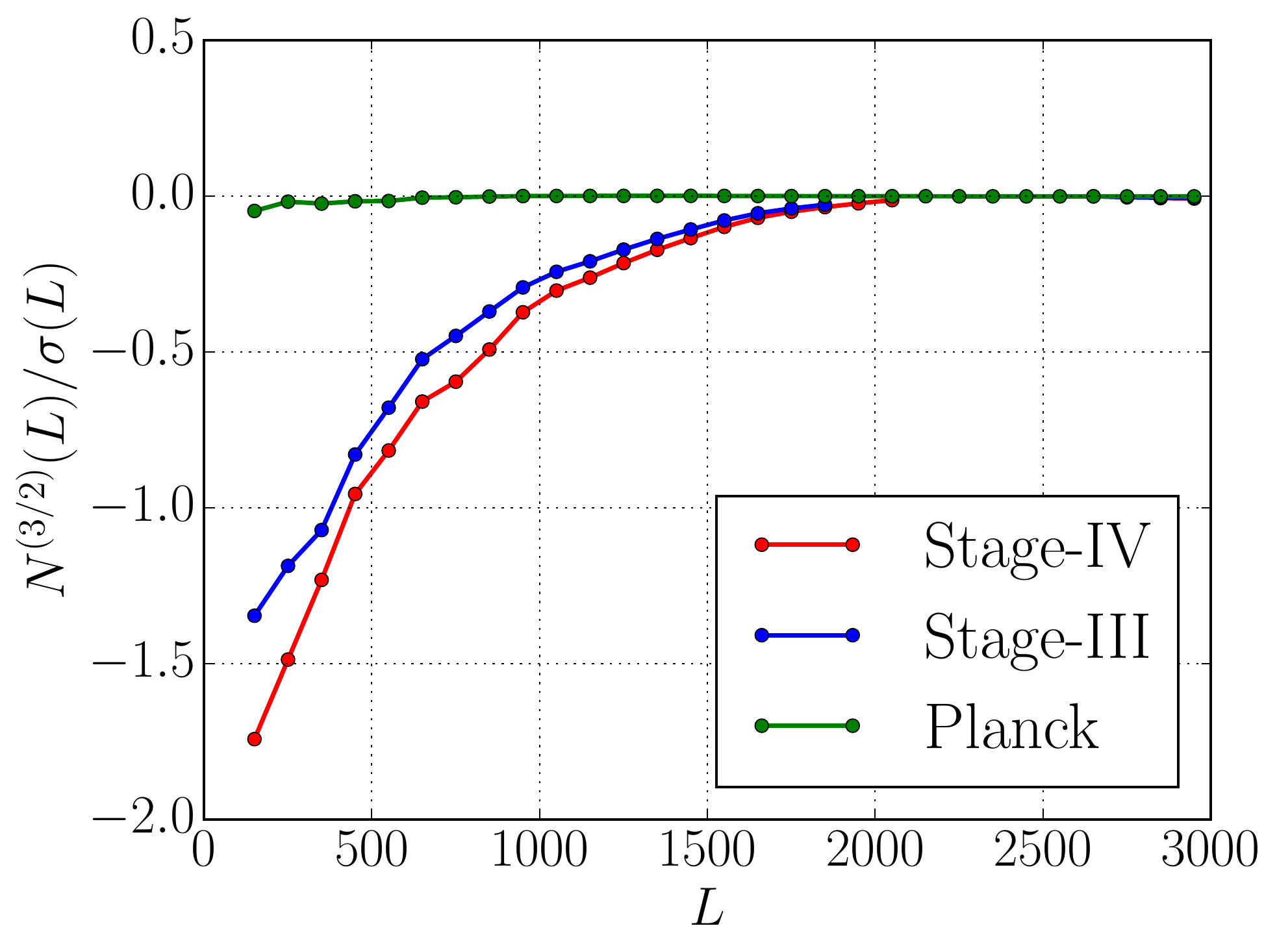}}}
\caption{\label{fig:ngBias_closeup} Left panel: Ratio of non-Gaussian $N^{(3/2)}$ bias over the signal lensing power spectrum $C^{\phi\phi}$, for $TT,TT$ reconstruction. Right panel: Ratio of $N^{(3/2)}$ bias over lensing power spectrum error $\sigma(L)$ on bandpowers of width $\Delta L= 100$. It can be seen that, while the bias appears negligible for Planck, it is a significant percent-level effect for Stage-III and Stage-IV experiments.}
\end{center}
\end{figure}
\fi

\fig{ngBias} shows the A1 and C1 contributions to the non-Gaussian reconstruction bias $N^{(3/2)}$ for the different classes of experiments summarized in \tabref{Experiments}. To assess the importance of the non-Gaussian biases, the left panel of \fig{ngBias_closeup} shows the ratio of their sum to the lensing power spectrum signal.

For the high resolution Stage-III and Stage-IV experiments the bias is of order 0.5-2.5\% of the signal, slowly decreasing towards smaller scales. The sign of the bias is negative over all relevant scales, i.e.~it reduces the measured lensing power. For Planck, the bias appears nearly an order of magnitude smaller than in the high resolution case, typically entering at negligible levels well below one percent of the signal. This is the case because for Planck the A1 and C1 contributions to the bias partially cancel each other (see also the discussion at the end of Section~\ref{sec:effect_on_CPP}). For Planck, the sign of the effect varies with angular scale.

The significance of the bias in each experiment depends on the statistical uncertainty of the measured lensing power spectrum. The Gaussian variance is given by
\beq
\label{eq:sigmaLDef}
\sigma^2(L)=
\frac{1}{f_{\mathrm{sky}}}\,
\frac{2}{(2L+1)}\(N^{(0)}_L+C_L^{\phi\phi}+N_L^{(1)}\)^2.
\eeq
The right panel of \fig{ngBias_closeup} shows the bias-to-noise ratio $N_{A1+C1}^{(3/2)}(L)/\sigma(L)$ if the measured lensing power spectrum is binned with bin width $\Delta L=100$.  
The bias is significant in low-noise, high-resolution experiments such as CMB Stage-III or Stage-IV: If the bias is ignored, the measured lensing power spectrum will be biased low by $\sim 0.5\sigma-1.5\sigma$ per bin for $L\sim 200-800$, for each bin of width $\Delta L=100$. 
The total significance of this bias is $\sim 2.5\sigma$ for Stage-III and $\sim 3\sigma$ for Stage-IV.

The total bias is thus significant and should therefore be accounted for when performing $TT,TT$ lensing reconstruction with CMB Stage-III or Stage-IV experiments.  
While Stage-IV will likely get most lensing information from polarization-based measurements so that the bias of temperature-based lensing measurements is less worrisome, a large fraction of the lensing information from Stage-III experiments will come from $TT,TT$ lensing measurements, so that accounting for the $N^{(3/2)}$ bias will be particularly important in this case.
For Planck, however, the bias appears negligible (the significance of the total bias is only $0.06\sigma$).

We emphasize that the above numbers just provide a rough estimate of the actual size of the $N^{(3/2)}$ bias because of the simplifying assumptions we made for the numerical evaluation. 
In particular, additional bias contributions from other contractions than A1 and C1 may be important for all experiments, and a more accurate model of the lensing bispectrum on small scales could change results for Stage-III and Stage-IV by an order one factor.
We will discuss these caveats in more detail in Sections~\ref{sec:ValTests} and \ref{sec:conclusions} below.

\subsection{Results for C1 contribution to the \texorpdfstring{$N^{(3/2)}$}{N32} bias for polarization}
\label{se:polarization_main_text}

We can generalize the $N^{(3/2)}$ bias to polarization-based measurements of the lensing power spectrum. In this paper we derive and evaluate the corresponding expressions for contributions from the coupling type C1 only (see \app{Pol}). The contribution from the coupling type A1 is numerically more expensive to evaluate and we defer its generalization to polarization to future work.

The left panel of \fig{ngBiasEB} shows the C1 contribution to the bias from $\EB,\EB$, $\EE,\EE$ and $\TT,\TT$ reconstruction for a Stage-IV experiment.
On most relevant scales, the $\TT,\TT$ and $\EE,\EE$ biases are similar to each other, but the $\EB,\EB$ bias is much smaller.
However, $\EB,\EB$ reconstruction is also the combination which is expected to achieve the lowest error on the lensing measurement for future polarization-sensitive experiments like CMB Stage-IV. To assess the significance of the C1 bias contribution in this case, the right panel of \fig{ngBiasEB} shows the bias divided by the reconstruction uncertainty for CMB Stage-IV and Planck (assuming \eqq{sigmaLDef} for the noise, and bin width $\Delta L=100$).  Despite the higher precision of $\EB,\EB$ reconstruction, the bias still appears rather small, $0.1-0.3\sigma$ per $L$-bin of width $\Delta L=100$.

We emphasize again that the bias is expected to change if contributions from the A1 and other contractions for polarization are included (like in the temperature-only case where the A1 contribution is rather important), and additional changes may arise from more accurate models for the matter bispectrum on small scales. Note also that the use of an iterative EB estimator could enhance the relative importance of the bias with respect to the lensing measurement error by roughly a factor of 3 for CMB Stage-IV, 
although the form of the bias may also be different for such a lensing estimator.

\ifincludefigs
\begin{figure}[tp]
\begin{center}
\subfloat{
\scalebox{0.43}{\includegraphics{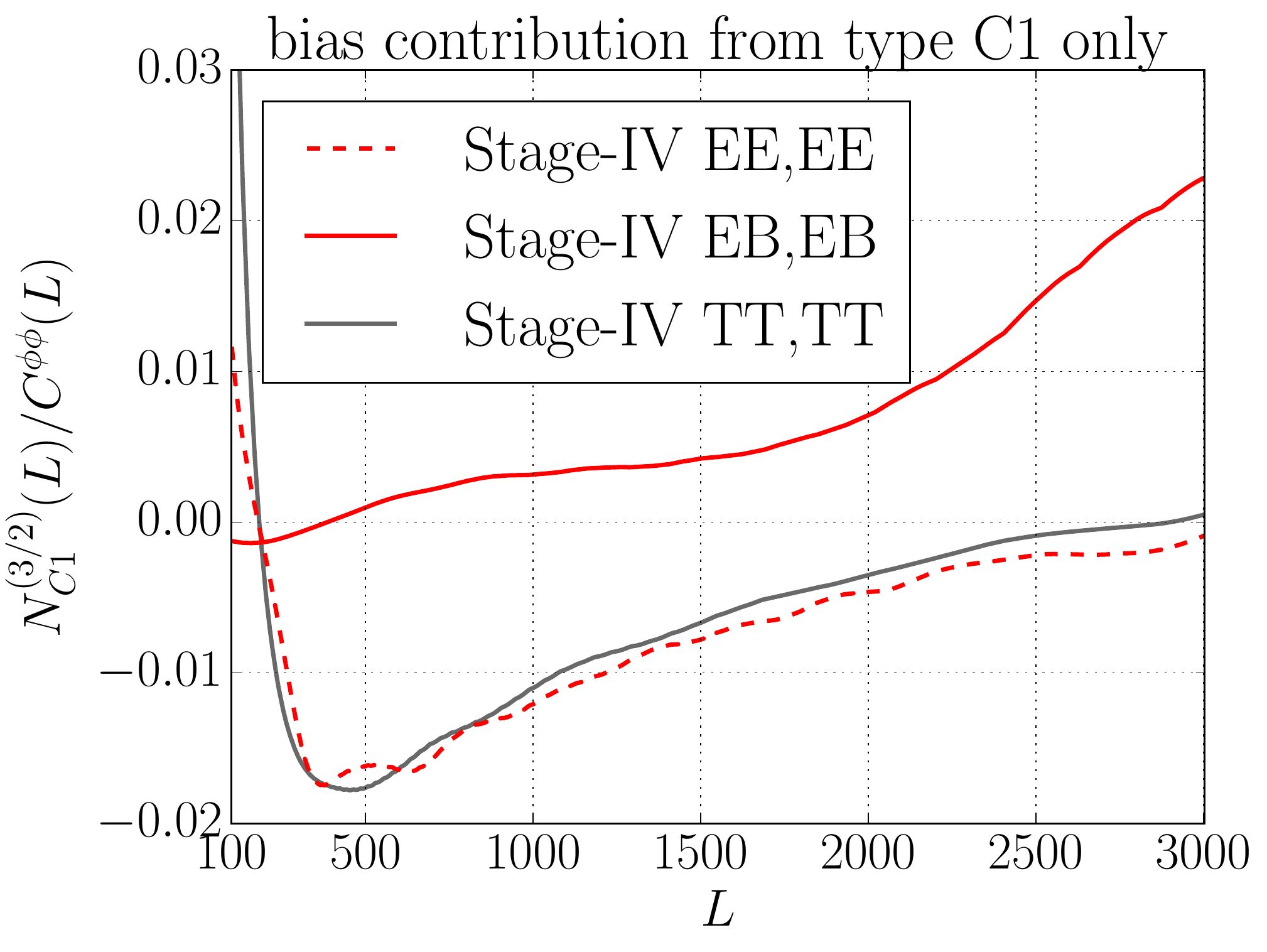}}}
\subfloat{
\scalebox{0.43}{\includegraphics{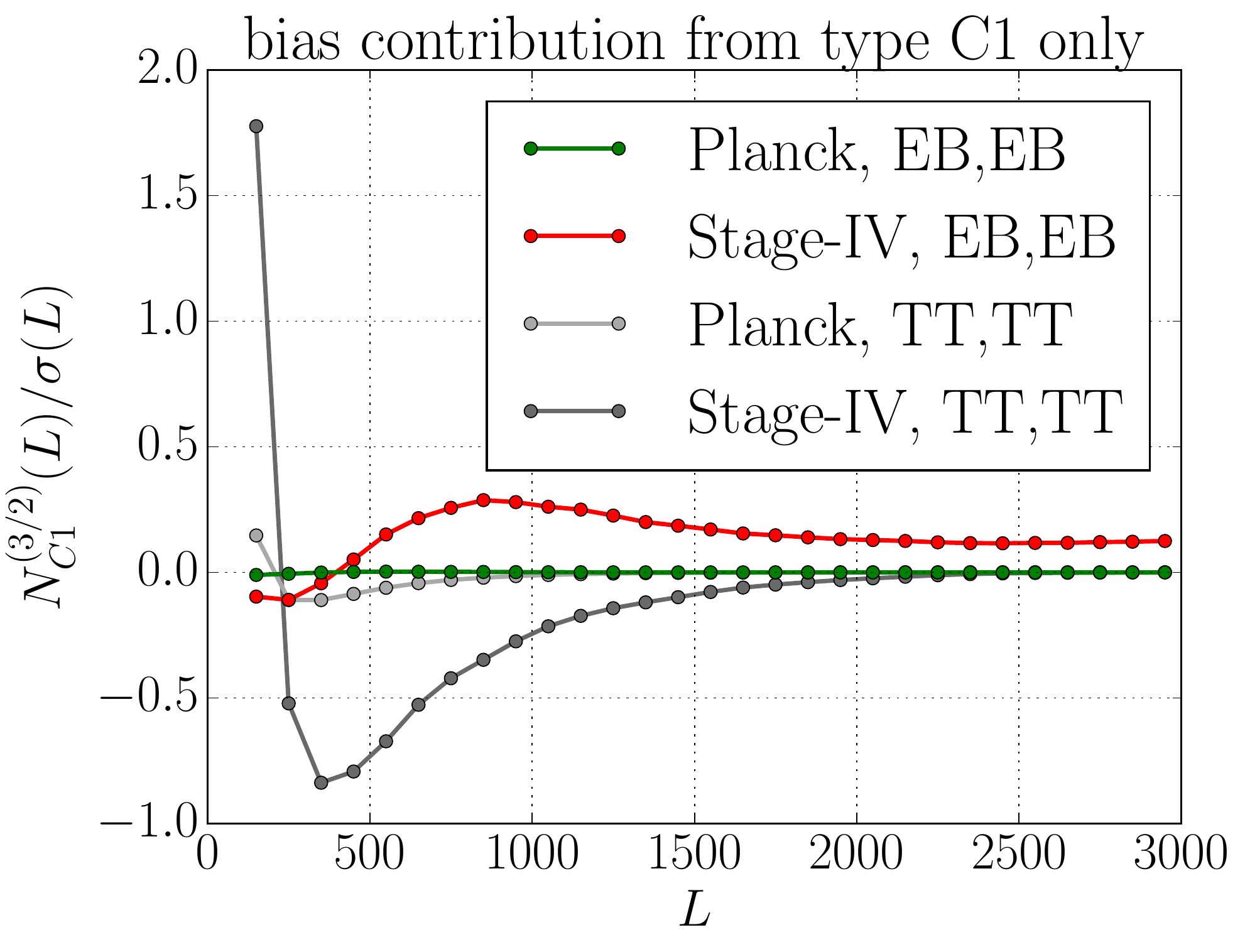}}}
\caption{\label{fig:ngBiasEB} Generalization of one of the bias contributions, from coupling type C1, to polarization-based measurements of the CMB lensing power spectrum (see \app{Pol} for details). All curves ignore the similar type A1 bias contribution. This provides some qualitative idea of how the bias changes for polarization, but should not be confused with the full expected lensing bias. A thorough quantification, which involves the evaluation of the remaining non-negligible term(s), is deferred to future work.
In the left panel, we show the C1 bias contribution for a Stage-IV experiment divided by the signal power spectrum for different estimators. In the right panel, we divide the C1 bias contribution by the error on the lensing power spectrum measurement. In both panels the type C1 bias contribution for a temperature-based measurement is plotted for comparison. It can be seen that the C1 bias is less important for polarization lensing measurements using the $EB$ estimator than it is for lensing measurements from temperature. Note that the use of an iterative $EB$ estimator for polarization could enhance the relative importance of the bias with respect to the error by roughly a factor of 3 for CMB Stage-IV at high $L$.}
\end{center}
\end{figure}
\fi

\section{Discussion and validation of the calculations}
\label{sec:ValTests}

In the following sections we discuss potential caveats in the evaluation of $N^{(3/2)}$ like its strong dependence on $\sigma_8$, cross-checks of our numerical implementation and assumptions made in the derivation and evaluation of $N^{(3/2)}$. Further, we explain how we have tested the influence of nonlinear modes on the results and we comment on the sensitivity of the bias on the large-scale structure bispectrum model that is used. Some of our results have been derived for the C1 term only, but we expect them to apply similarly to the other relevant terms. We begin by discussing the scaling with $\sigma_8$.

\subsection{Dependence of \texorpdfstring{$N^{(3/2)}$}{N32} bias on \texorpdfstring{$\sigma_8$}{sigma8}}
Since the bispectrum of the lensing potential $B_\phi$ is quadratic in the power spectrum of the matter density, we expect the $N^{(3/2)}$ bias to scale with the fourth power of the normalization of matter fluctuations, $\sigma_8$. Thus relatively small changes of $\sigma_8$ lead to large changes of the $N^{(3/2)}$ bias. 
Computing the $N^{(3/2)}$ bias for a fiducial cosmology with slightly wrong $\sigma_8$ may therefore leave a significant residual bias. This may raise concerns because $\sigma_8$ is not very well known in practice. 
However, we found that the effect of $\sigma_8$ on the $N^{(3/2)}_{C1}$ bias is relatively well approximated by rescaling $N^{(3/2)}_{C1}$ with a scale-independent factor $(\sigma_8/\sigma_8^\mathrm{fiducial})^4$. Therefore, the $\sigma_8$ dependence of the bias could easily be included when fitting cosmological parameters to data. 
This can in principle increase the precision of $\sigma_8$, because it includes information from the non-Gaussianity of the lensing potential $\phi$ (though more optimal methods for extracting this non-Gaussian information could be used instead).

\subsection{Cross-check with large-lens and squeezed bispectrum limits}
The numerical evaluation of the contributions to the $N^{(3/2)}$ bias involves several steps and relies on numerical approximations such as discretization schemes. Therefore, any of the computed results should be validated. Apart from code internal tests we have derived analytic large-lens and squeezed bispectrum limits for the various numerical integrals involved in evaluating $N^{(3/2)}$, evaluated them independently and compared them to full code results. These limits do not only provide a cross-check of the implementation, but are also useful to qualitatively understand the behavior and dependencies of the contributing terms. We have found excellent agreement between the analytic limits and our numerical calculations of the C1 contribution to the bias; for a detailed description of these tests we refer the reader to \app{lllimits}.

\subsection{Higher-order corrections to the matter bispectrum}
\label{sec:kNL_test}

As discussed in Section~\ref{sec:formalism}, the simple model of \eqq{SPTbispec} for the dark matter bispectrum from Eulerian standard perturbation theory at leading order is only valid for large-scale LSS modes. It breaks down for small-scale LSS modes that can have large overdensities $\delta\gg 1$ due to gravitational collapse. 
We use the simple leading-order model of \eqq{SPTbispec} to get an approximate, conservative estimate of the expected size of the $N^{(3/2)}$ bias.
In reality, higher-order (and ultimately non-perturbative) gravitational collapse on small scales generates a larger bispectrum that may lead to a larger $N^{(3/2)}$ lensing bias, especially for small-scale lenses (high $L$). 
For actual data analyses of experiments where the $N^{(3/2)}$ bias is relevant, fitting formulae for the matter bispectrum calibrated against $N$-body simulations should be used for more accurate predictions of the $N^{(3/2)}$ bias from small-scale LSS modes.
Our expressions for the lensing bias take an arbitrary matter bispectrum model as input so that it is straightforward to include more realistic bispectrum models.
To get a rough estimate for the importance of small-scale LSS modes on the $N^{(3/2)}$ bias, we compute the bias with the bispectrum set to zero if any of the contributing LSS modes is larger than a nonlinear cutoff scale $k_\mathrm{NL}(z)$ defined by
\beq
\label{eq:kNL_def}
\frac{k_\mathrm{NL}^3(z) P(k_\mathrm{NL},z)}{2 \pi^2}>1,
\eeq
and compare it to the full result.
This test reveals that the contribution from these scales to the type C1 bias makes up $\sim 30\%$ of the signal at $L=3000$ for CMB-S4 (and less for Stage-III and Planck experiments). Up to $L\sim 1000$ it lies below $10\%$ for all experiments. At least up to this multipole range, the leading-order bispectrum \eq{SPTbispec} seems an acceptable approximation for the coupling of type C1.
For the second coupling that we consider, type A1, we find somewhat different results. For a Planck-like experiment the contribution of small-scale LSS modes at $L\sim 1000$ is of $\mathcal{O}(10\%)$ and thus similar to the type C1 term. For a Stage-IV experiment, however, these small modes contribute significantly even at lower multipoles. In particular, we find that the type A1 bias at $L=1000$ has a different sign if modes smaller than the cutoff scale are excluded.

\ifincludefigs
\begin{figure}[tpb]
\begin{center}
\scalebox{0.46}{\includegraphics{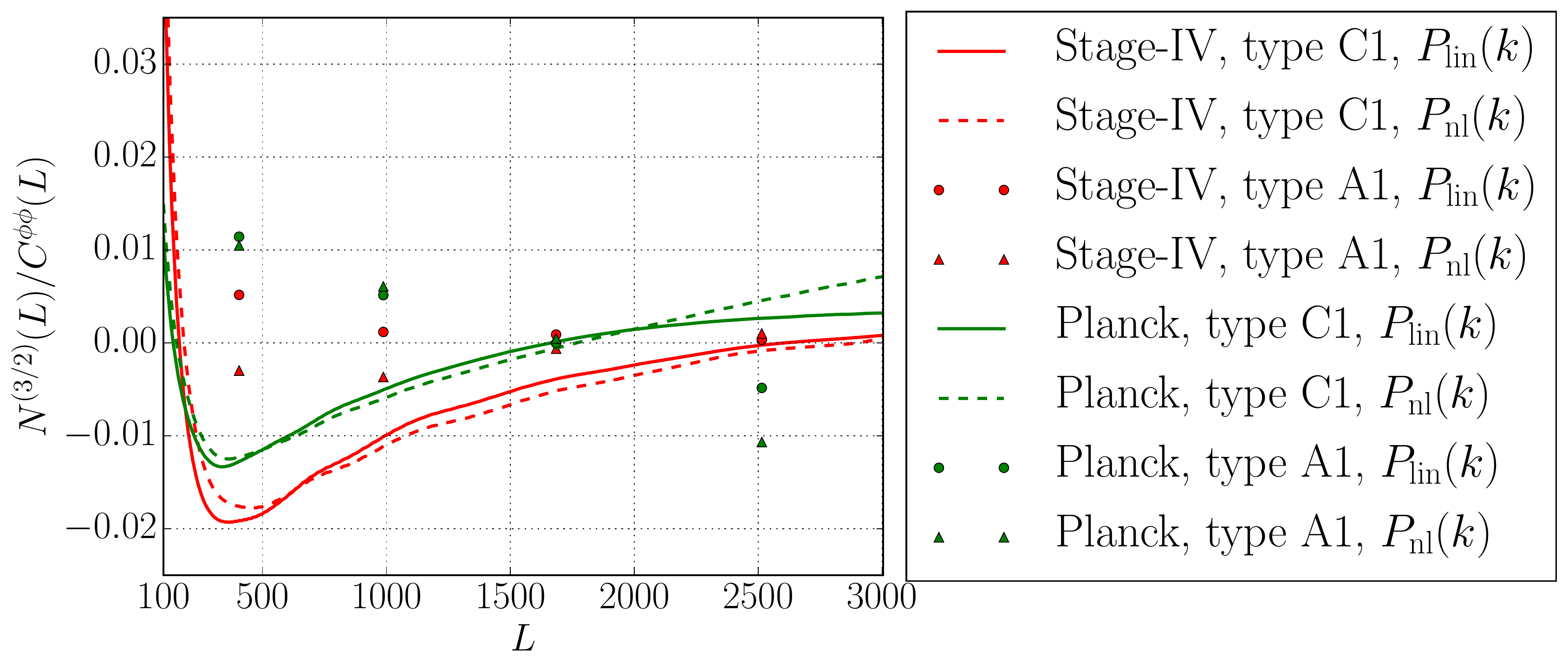}}
\caption{\label{fig:IkNLratio} The fractional contribution of the nonlinear bias from coupling type C1 (lines) and type A1 (symbols) (similar to the left panel in \fig{ngBias_closeup}) computed from the standard Eulerian perturbation theory bispectrum at leading order and from a modified form where the linear matter power spectrum was replaced by a nonlinear one (dashed lines). For the C1 contribution to the $N^{(3/2)}$ bias the standard leading-order bispectrum and the modified bispectrum (with enhanced nonlinearity) give similar results at multipoles $L<2000$.
For the A1 contribution, which is computationally much more expensive to evaluate, the results are similar for Planck-like experiments, but for CMB Stage-IV higher-order corrections to the bispectrum seem to be important even at intermediate $L$.}
\end{center}
\end{figure}
\fi

Another simple test of the impact of small-scale LSS modes is obtained by comparing the $N^{(3/2)}$ bias evaluated with the standard perturbative bispectrum formula \eq{SPTbispec} and the bias computed from a modified bispectrum model where the linear matter power spectrum is replaced by the nonlinear one, $P_{\text{lin}}\rightarrow P_{\text{nl}}$. 
This comparison is shown in \fig{IkNLratio}. Lines indicate results for the type C1 contribution to the bias, symbols indicate the A1 contribution. 
The C1 bias contribution changes by $\mathcal{O}(10\%)$ or less for lensing multipoles $L<2000$ if the nonlinear instead of linear matter power spectrum is used.
At higher multipoles $L>2000$ the change can be larger. 

For the similar A1 contribution to the bias we restrict this test to a few points (indicated by markers), because  evaluation is much more computationally expensive. In this case, we find that the importance of nonlinear corrections strongly depends on the experimental specifications. For a Planck-like experiment, the corrections seem similarly small as for the type C1 term. For a CMB Stage-IV experiment, however, the modification of the matter power spectrum leads to a significant change of the bias even at intermediate $L$. 

We conclude that for high resolution experiments the leading-order perturbation theory bispectrum model may not be sufficient for obtaining an exact estimate of the size of the $N^{(3/2)}$ bias, but instead can only provide an approximate estimate. For Planck, however, the leading-order model appears to be accurate. A thorough quantification of the bias for Stage-III and Stage-IV experiments requires a more accurate modeling of the LSS bispectrum for small LSS modes. This could be achieved by using fitting formulae for the matter bispectrum calibrated by numerical N-body simulations (e.g.~\citep{SPTBispecFit1,SPTBispecFit2,Lazanu:2015rta,2015arXiv151102022L}).

\subsection{Prospects for comparison with results from numerical simulations}
\label{se:ComparisonWithSims}
The derived form of the nonlinear bias relies on the validity of certain assumptions, including e.g.~the validity of the bispectrum approximation, the domination of the two contributions of type A1 and C1 to the bias over all other contributions, and the negligibility of nonlinear corrections that are higher than third order in the lensing potential. An independent test of their correctness could be obtained by a comparison with N-body simulations that provide a full nonlinear lensing potential and do not rely on a perturbative approach. 
We defer an analysis of the nonlinear bias in CMB lensing simulations based on N-body simulations to future work.

\ifcross
\section{Cross-correlation of CMB lensing with an external LSS tracer}
	\label{sec:crosscorr}

While our paper focuses on the auto-power spectrum of the quadratic lensing reconstruction, $\la\hat\phi\hat\phi\ra$, it is also worthwhile to cross-correlate the lensing reconstruction $\hat\phi$ with other external LSS tracers $\phi_\mathrm{ext}$ like the cosmic infrared background, galaxy weak lensing, galaxy or quasar catalogs, or Lyman-alpha observations; see e.g.~\cite{SmithZahnDore0705,HirataHoEtAl0801Lensing,sherwin1207,2015PhRvD..91f2001H,2015PhRvD..92f3517L,SPTPol_lensingCIB,PB_lensingCIB,vEngelenACTPolCIBcrosspol}.
The cross-correlation $\la\hat\phi\phi_\mathrm{ext}\ra$ between the quadratic CMB lensing reconstruction $\hat\phi$ and the external LSS tracer $\phi_\mathrm{ext}$ then picks up a similar bias arising from the large-scale structure bispectrum generated by nonlinear structure formation.  In this section we compute this cross-spectrum bias similarly to the calculations above, under the assumption that the observed external LSS tracer is uncorrelated with the unlensed CMB.

The bias of the cross-spectrum induced by a nonzero LSS bispectrum is caused by the correlation of the external LSS tracer with the second order response of the reconstructed lensing potential to the true lensing potential.  Similarly to the A1 and C1 contributions to the auto-spectrum bias in  Eqs.~\eq{70} and \eq{typeCcontractions}, this bias to the cross-spectrum follows schematically from two contractions `A1cross' and `C1cross':
\begin{eqnarray}
  \label{eq:17}
  \la\tilde T\tilde T \phi_\mathrm{ext}\ra_{\mathcal{O}[(C^{\phi\phi})^{3/2}]} &=&
\la \delta T\delta T\phi_\mathrm{ext}\ra + 2\la\delta^2 TT\phi_\mathrm{ext}\ra\\
&=&
{
\contraction{\la}{T}{\!{}_{,i}\phi\!{}_{,i}}{T}
\bcontraction{\la T\!{}_{,i}}{\phi}{\!{}_{,i} T\!{}_{,j}}{\phi}
\bcontraction{\la T\!{}_{,i}}{\phi}{\!{}_{,i} T\!{}_{,j}\phi\!{}_{,j}\,}{\phi}
\la T\!{}_{,i}\phi\!{}_{,i} T\!{}_{,j}\phi\!{}_{,j}\,\phi_\mathrm{ext}\ra_{A1\mathrm{cross}}
}
+2
{
\contraction{\la}{T}{\!{}_{,ij}\phi\!{}_{,i}\phi\!{}_{,j}}{T}
\bcontraction{\la T\!{}_{,ij}}{\phi}{\!{}_{,i}}{\phi}
\bcontraction{\la T\!{}_{,ij}}{\phi}{\!{}_{,i}\phi\!{}_{,j}T\,}{\phi}
\la T\!{}_{,ij}\phi\!{}_{,i}\phi\!{}_{,j}T\,\phi_\mathrm{ext}\ra_{C1\mathrm{cross}}
}.
\end{eqnarray}
These are all contractions allowed for the cross-spectrum, so that the full expectation value of the cross-spectrum up to fifth order in LSS perturbations is\footnote{An additional $N^{(2)}$ bias of order $(C^{\phi\phi})^2$ also arises, but we avoid it by using lensed CMB power spectra in the normalization $A_L$ \eqq{ALdef} and in the numerator of the weight $g$ in \eqq{gweight} \cite{hanson1008,lewis1101}.}
\begin{align}
  \label{eq:29}
  \la C^{\hat\phi\phi_\mathrm{ext}}_L\ra = C^{\phi\phi_\mathrm{ext}}_L + N^{(3/2)}_{A1\mathrm{cross}}(L)
+ N^{(3/2)}_{C1\mathrm{cross}}(L) + \mathcal{O}(\phi^5),
\end{align}
where the new bispectrum-induced biases are
\begin{eqnarray}
  \label{eq:N32A1cross}
  N^{(3/2)}_{A1\mathrm{cross}}(L) &=&
- A_L \int_{\vl,\vl_1} g(\vl_1,\VL) [(\vl_1-\vl)\cdot\vl] [(\vl_1-\vl)\cdot(\VL-\vl)] C_{|\vl_1-\vl|}^{TT}B_{\phi\phi\phi_\mathrm{ext}}(\vl,\VL-\vl,-\VL)
\end{eqnarray}
and
\begin{eqnarray}
  \label{eq:N32C1cross}
  N^{(3/2)}_{C1\mathrm{cross}}(L) &=&
A_L\int_{\vl,\vl_1} g(\vl_1,\VL) (\vl_1\cdot\vl) [\vl_1\cdot(\VL-\vl)] C^{TT}_{l_1}
B_{\phi\phi\phi_\mathrm{ext}}(\vl,\VL-\vl,-\VL).
\end{eqnarray}
Here, $B_{\phi\phi\phi_\mathrm{ext}}$ is the mixed bispectrum between two CMB lensing modes and one external LSS tracer.

The cross-spectrum biases \eq{N32A1cross} and \eq{N32C1cross} are similar to the A1 and C1 auto-spectrum biases in Eqs.~\eq{N32A1} and \eq{N32C1}.  Indeed, if the external tracer were equal to the true lensing potential modulo uncorrelated noise, $\phi_\mathrm{ext}=\phi+n$, the cross biases would be half the auto-spectrum biases at leading order in the lensing potential power:
\begin{align}
  \label{eq:44}
  N^{(3/2)}_{A1\mathrm{cross}}(L) \approx \frac{1}{2}N^{(3/2)}_{A_1}(L)
\qquad\text{and}\qquad
  N^{(3/2)}_{C1\mathrm{cross}}(L) \approx \frac{1}{2}N^{(3/2)}_{C_1}(L).
\end{align}
In practice, the external LSS tracer is typically different from the lensing potential, e.g.~because of different redshift kernels, so that the cross-bias should be evaluated with the full Eqs.~\eq{N32A1cross} and \eq{N32C1cross}.
Fast-to-evaluate expressions for these biases take the same form as those for the A1 and C1 auto-spectrum biases if the lensing bispectrum is replaced by the mixed lensing-lensing-tracer bispectrum $B_{\phi\phi\phi_\mathrm{ext}}$.
 We note that for lower-redshift tracers, the non-linearity is enhanced, so that cross-correlation biases may be larger than the biases for CMB lensing alone.
\fi

\section{Conclusions}
\label{sec:conclusions}

This paper investigates the effect of large-scale structure non-Gaussianity on CMB lensing reconstruction.  The bispectrum of the CMB lensing potential generated by nonlinear structure formation leads to a bias of the measured CMB lensing power spectrum that has been neglected so far. We call the bias $N^{(3/2)}$ because it involves $\phi^3\sim (C^{\phi\phi})^{3/2}$.  For an unbiased measurement, this bias must be calculated and subtracted from measured lensing power spectra. 
We derive an analytical expression for this lensing bias, which splits into several contributions that involve the CMB power spectrum and the dark matter bispectrum.

The magnitude of the $N^{(3/2)}$ bias depends on experiment specifications and field combinations used for the lensing reconstruction. For CMB Stage-III and Stage-IV experiments, we find that the lensing power spectrum measurements are biased low by 0.5-2.5\% (for Planck, the bias is at a negligible sub-percent level) if temperature data is used. For future experiments, this negative bias will shift measurements of the lensing power spectrum by multiple standard deviations and must thus be accounted for.
For Stage-III a large fraction of the lensing signal-to-noise is expected from the temperature-based reconstruction, so accounting for the bias is particularly important in this case.
We focus on temperature-only lensing reconstructions, but we demonstrate for one of the bias contributions how it can be straightforwardly generalized to polarization-based reconstructions.

Our first results on this non-Gaussian bias, including the expected size of the bias, rely on a number of simplifying assumptions that should be tested in future work:

\begin{enumerate}
\item Some contributions to the non-Gaussian lensing bias involve high-dimensional integrals that are computationally challenging to evaluate.  
Therefore, for numerical evaluations, we consider only two bias contributions that can be evaluated in reasonable timescales.
They arise from particular contractions denoted type A1 (\eqq{N32A1}) which contributes to $\<\delta T \delta T \delta T T\>$, and type C1 (\eqq{N32C1}) which contributes to $\<\delta^2 T T \delta T T\>$.
Intuitively, we suspect that these two contributions to $N^{(3/2)}$ are among the largest contributions, because they have relatively simple, separable forms in Fourier space.
For all other bias contributions we present analytical expressions but do not evaluate them numerically in the present work. 
Future work should check if these additional bias contributions are relevant, e.g.~by performing the required numerical integrations or by comparing against estimates of the same non-Gaussian lensing bias from ray-traced N-body simulations.
 
\item While our analytical expressions can take arbitrary matter bispectrum models as their input, our numerical evaluations assume a simple matter bispectrum model that follows from leading-order Eulerian standard perturbation theory. 
While this is valid in the regime where only large-scale lensing modes contribute, more accurate results for the non-Gaussian lensing bias can be obtained by using more accurate matter bispectrum models on small scales.
Our tests indicate that such corrections are likely small for Planck but significant for future CMB Stage-III or Stage-IV experiments.

\item Our analytical expressions follow by perturbing lensed CMB fluctuations in the lensing deflection angle.  This perturbative expansion does not converge well on all scales, although corrections from nonperturbative approaches are typically less than $10\%$. 
Again, the accuracy of this approximation should be checked in the future.
\item Our calculation for the lensing potential bispectrum $B_\phi$  induced by nonlinear structure formation assumes the flat sky approximation and Limber's projection. 
This is valid on intermediate and small scales, but breaks down on very large scales. 
We therefore restrict the discussion of the bias to multipoles $L\ge 100$. Although it would be interesting to extend our result to the full sky, we note that CMB experiments have most lensing information at multipoles $L\ge 100$.

\end{enumerate}

Apart from testing each of the above assumptions in more detail, there are various other directions to extend and generalize our work in the future.
For example, while we regard the non-Gaussianity of the lensing potential and the induced lensing power bias as a \emph{nuisance}, it could equally well be regarded as a new \emph{signal}.  
Pushing this further, one could envision more optimal estimators to extract information from the non-Gaussianity of the lensing potential, e.g.~by measuring the skewness or bispectrum of the reconstructed lensing potential, as investigated very recently by Namikawa \cite{2016arXiv160408578N}.  We leave such exciting extensions to future studies.  
We also note that we have assumed the standard quadratic lensing estimator when deriving the $N^{(3/2)}$ bias. However, future polarization-sensitive experiments like CMB Stage-IV will benefit significantly from likelihood-based lensing estimators \cite{2003PhRvD..67d3001H}; the impact of large-scale structure non-Gaussianity on these estimators should be considered.

More generally, accounting for the bispectrum and nonlinearity of large-scale structure is just one of many possible extensions to refine theoretical modeling of CMB lensing. 
While leading-order modeling of CMB lensing is often rather accurate, the highly increased sensitivity of upcoming CMB Stage-III and Stage-IV experiments may require additional modeling corrections that should be investigated in the future.

$ $

While finalizing our draft, Namikawa \cite{2016arXiv160408578N} pointed out that the CMB lensing bispectrum can also be regarded as a potential future signal from the CMB 6-point function rather than a bias of lensing 4-point measurements, which is the focus of our paper. While our papers are complementary in most parts, they both demonstrate the future importance of the non-Gaussianity of the CMB lensing potential. We checked that our theoretical CMB lensing bispectrum from leading-order standard perturbation theory agrees with \cite{2016arXiv160408578N}.

\begin{acknowledgements}
We thank Anthony Challinor, Colin Hill, Antony Lewis and Uro\v{s} Seljak for numerous crucial discussions and for comments on an earlier version of our paper.  We also thank Torsten En{\ss}lin, Jia Liu, Bj\"orn Malte Sch\"afer, Alex van Engelen, Martin White and Oliver Zahn for useful discussions. 
VB thanks the organizers of the Moriond conference for the opportunity to present this work.
\end{acknowledgements}

\appendix

\section{CMB lensing bispectrum}
	\label{App:Bispec}
The nonlinear bias is the consequence of a non-vanishing bispectrum of the lensing potential. In this appendix we provide the full-sky expression for the Limber-projected CMB lensing bispectrum (see e.g.~\citep{2000ApJ53036B,2004MNRAS.348..897T}). In the flat-sky limit this reduces to the expression in \eqq{bi_phi_of_B_delta}. We evaluate this expression with a matter bispectrum at leading order in standard Eulerian perturbation theory and show the cumulative contributions from different redshifts and wavenumbers for equilateral configurations.

We start out with the three-point correlation function of the lensing potential in angular coordinates
\begin{align}
\nonumber
\langle \phi(\n) \phi(\n')\phi(\n'')\rangle&=\int_0^{\chi_{\ast}}\!\d\chi \int_0^{\chi_{\ast}}\!\d\chi'\int_0^{\chi_{\ast}}\!\d\chi''
W(\chi)W(\chi')W(\chi'')
\langle \psi\left(\chi\n, \eta_0{-}\chi \right) \psi\left(\chi'\n', \eta_0{-}\chi' \right)\psi\left(\chi''\n'', \eta_0{-}\chi'' \right)\rangle\\
&=
\int_0^{\chi_{\ast}}\!\d\chi \int_0^{\chi_{\ast}}\!\d\chi'\int_0^{\chi_{\ast}}\!\d\chi''
W(\chi)W(\chi')W(\chi'')
\int_{\vk,\vk',\vk''}(2\pi)^{3} \delta_D(\vk+\vk'+\vk'') \non\\
&\quad \times B_\psi\(k,k',k'';\eta,\eta', \eta''\)
\e{i \vk\cdot\chi\n}\e{i \vk'\cdot\chi'\n'}\e{i \vk''\cdot \chi''\n''},
\end{align}
where we introduced the bispectrum of the Newtonian potential,
\beq
\langle \psi\(\vk, \eta\) \psi\(\vk', \eta' \)\psi\(\vk'', \eta'' \)\rangle=\(2\pi\)^{3}\delta_D\(\vk+\vk'+\vk''\) B_\psi\(\vk,\vk',\vk'';\eta,\eta', \eta''\).
\eeq
Here, $\eta$, $\eta'$ and $\eta''$ denote the conformal times at which the photon encounters the potentials of wavevectors $\vec{k}$, $\vec{k}'$ and $\vec{k}''$, respectively.

Expanding the lensing potential $\phi$ in spherical harmonics yields
\beq
\label{eq:FullSkybispec}
\langle\phi_{\ell m} \phi_{\ell' m'} \phi_{\ell''m''}\rangle = \mathcal{G}^{m m' m''}_{\ell \ell' \ell''}  \int_0^{\chi_{\ast}}\frac{\d\chi}{\chi^4}W(\chi)^3B_\psi\(\frac{\ell}{\chi},\frac{\ell'}{\chi},\frac{\ell''}{\chi};\eta\).
\eeq
This is obtained by expanding plane waves and Dirac delta's in spherical harmonics $Y_{\ell m}$ and spherical Bessel functions $j_\ell$, 
 performing all angular integrals, using the closure relation for spherical Bessel functions (which enforces $\eta=\eta'=\eta''$), and applying Limber's approximation by replacing $k$ by $l/\chi$ (see e.g.~\cite{huHarmonicCMBLensing0001303} for similar calculations).  We also used the Gaunt integral
\begin{eqnarray}
\nonumber
\mathcal{G}^{m m' m''}_{\ell \ell' \ell''}
&=&\int \d \Omega\ 
Y_{lm}(\n)Y_{l'm'}(\n)Y_{l''m''}(\n)
\\
\nonumber
&=&\sqrt{\frac{(2\ell+1)(2\ell'+1)(2\ell''+1)}{4\pi}}
\begin{pmatrix}
  \ell & \ell' & \ell''\\
  0 & 0 & 0
\end{pmatrix}
\begin{pmatrix}
  \ell & \ell' & \ell''\\
   m & m' & m''
\end{pmatrix},
\end{eqnarray}
imposing $\ell + \ell' + \ell''$ = even.
The flat-sky expression corresponding to \eqq{FullSkybispec} is
\beq
\langle\phi(\vl)\phi(\vl')\phi(\vl'')\rangle=(2 \pi)^2\delta_D(\vl+\vl'+\vl'') \int_0^{\chi_{\ast}}\frac{\d\chi
}{\chi^4}W(\chi)^3
B_\psi\(\frac{l}{\chi},\frac{l'}{\chi},\frac{l''}{\chi};\eta\).
\eeq

The bispectrum of the potential $\psi(\vk,\chi)$ due to nonlinear gravitational clustering is obtained by noting that the potential is sourced by the fractional overdensity $\delta(\vk,\chi)$ through the Poisson equation,
\beq
\psi(\vk,\chi)=-\frac{3}{2}\frac{H_0^2\, \Omega_\mathrm{m0}}{c^2 {k}^2}\frac{\delta(\vk,\chi)}{a(\chi)} \equiv - \frac{\gamma(\chi)}{{k}^2}\delta(\vk,\chi),
\eeq
so that
\beq
{B}_\psi(\vk_1,\vk_2,\vk_3;\chi)=-\frac{\gamma\(\chi\)^3}{{k}_1^2 {k}_2^2 {k}_3^2}{B}_{\delta}(\vk_1,\vk_2,\vk_3;\chi).
\eeq
The lensing potential bispectrum $B_\phi$ is then given by the following line-of-sight integral over the matter bispectrum:
\beq
\label{eq:bi_phi_of_B_delta2}
B_{\phi}(\vl_1, \vl_2, \vl_3) =- \int_0^{\chi_{\ast}} \d\chi\,\chi^2 W(\chi)^3 \frac{\gamma\(\chi\)^3} {({l_1} {l_2} {l_3})^2}{B}_{\delta}(l_1/\chi,l_2/\chi,l_3/\chi;\chi).
\eeq

In this paper we evaluate this formula by inserting a slightly modified version of the standard perturbation theory result for the LSS bispectrum at leading order (\eqq{SPTbispec}), where the linear matter power spectrum $P_{\mathrm{lin}}$ is replaced by a power spectrum with nonlinear corrections $P_{\mathrm{nl}}$. This modification extends the validity of the model to slightly smaller scales of the large-scale structure.

\begin{figure}[tpb]
\begin{center}
\subfloat{
\scalebox{0.43}{\includegraphics{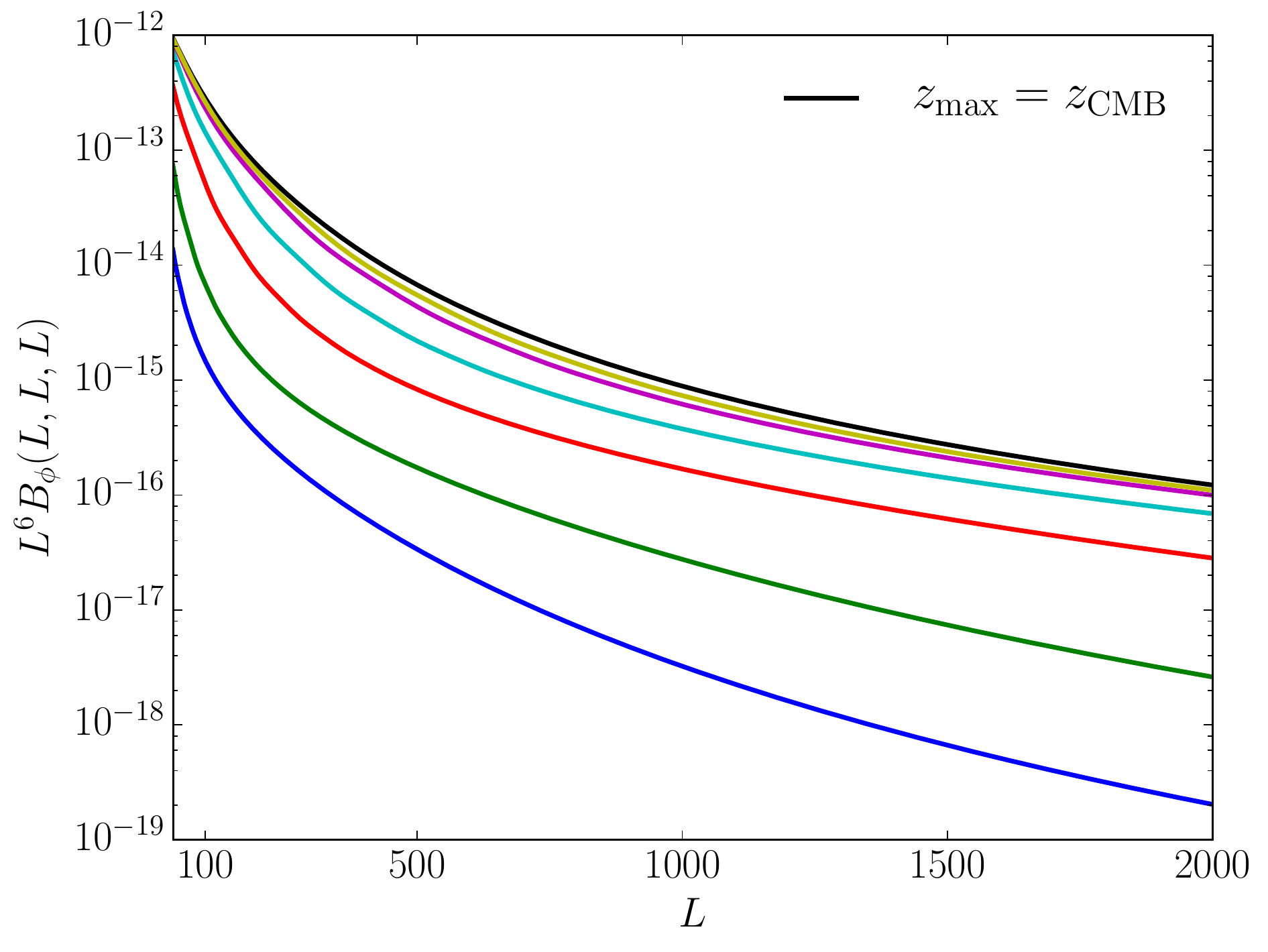}}}
\subfloat{
\scalebox{0.43}{\includegraphics{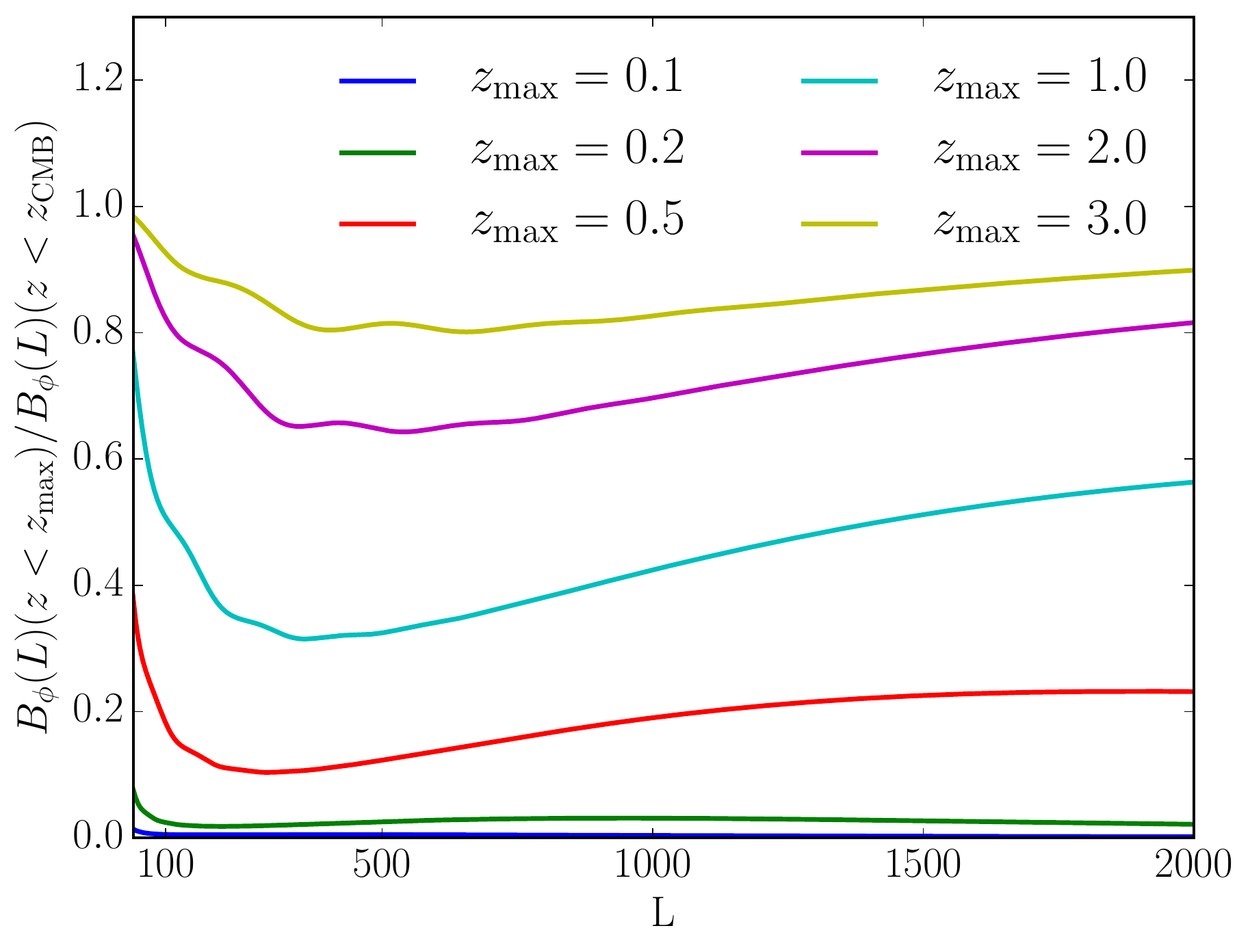}}}
\end{center}
\caption{\label{fig:BiphiCum} Cumulative contribution to the equilateral lensing bispectrum $B_\phi(L,L,L)$ from different redshifts. In the left panel we plot lensing bispectra obtained by integrating to different redshifts. In the right panel we show their relative contribution to the full lensing bispectrum. On large lensing scales, $L \sim 100$, the bispectrum is dominated by contributions from structures at low redshifts, $z\lesssim 1$. Their importance decreases with increasing $L$. On smaller lensing scales, $L\gtrsim $ few hundred, this trend is reversed and their contributions regain relevance. The resulting dip in the right plot is a consequence of replacing $P_{\text{lin}}$ by $P_{\text{nl}}$ in the LSS bispectrum model.}
\end{figure}

We plot $B_\phi$ for equilateral triangle configurations and its cumulative contribution from different redshifts in \fig{BiphiCum}.
The individual contributions can best be analyzed in the right panel of \fig{BiphiCum}, where we plot the lensing bispectrum integrated to different redshifts divided by the full lensing bispectrum. 
On large lensing scales (low $L$) we find that the bispectrum is mainly sourced by nearby structures at low redshifts ($z\lesssim 1$). Going to smaller lensing scales (higher $L$), it gets more and more contributions from structures at higher redshifts.
This trend continues until nonlinear corrections from the nonlinear matter power spectrum used in the numerical evaluation of $B_\phi$ become relevant. 
They enhance contributions from lower redshifts to smaller lensing scales. This enhancement leads to the turn-around at a scale of $L \sim 500$, which would be absent if $P_{\text{lin}}$ was used in the LSS bispectrum model. 

\begin{figure}[tpb]
\begin{center}
\scalebox{0.43}{\includegraphics{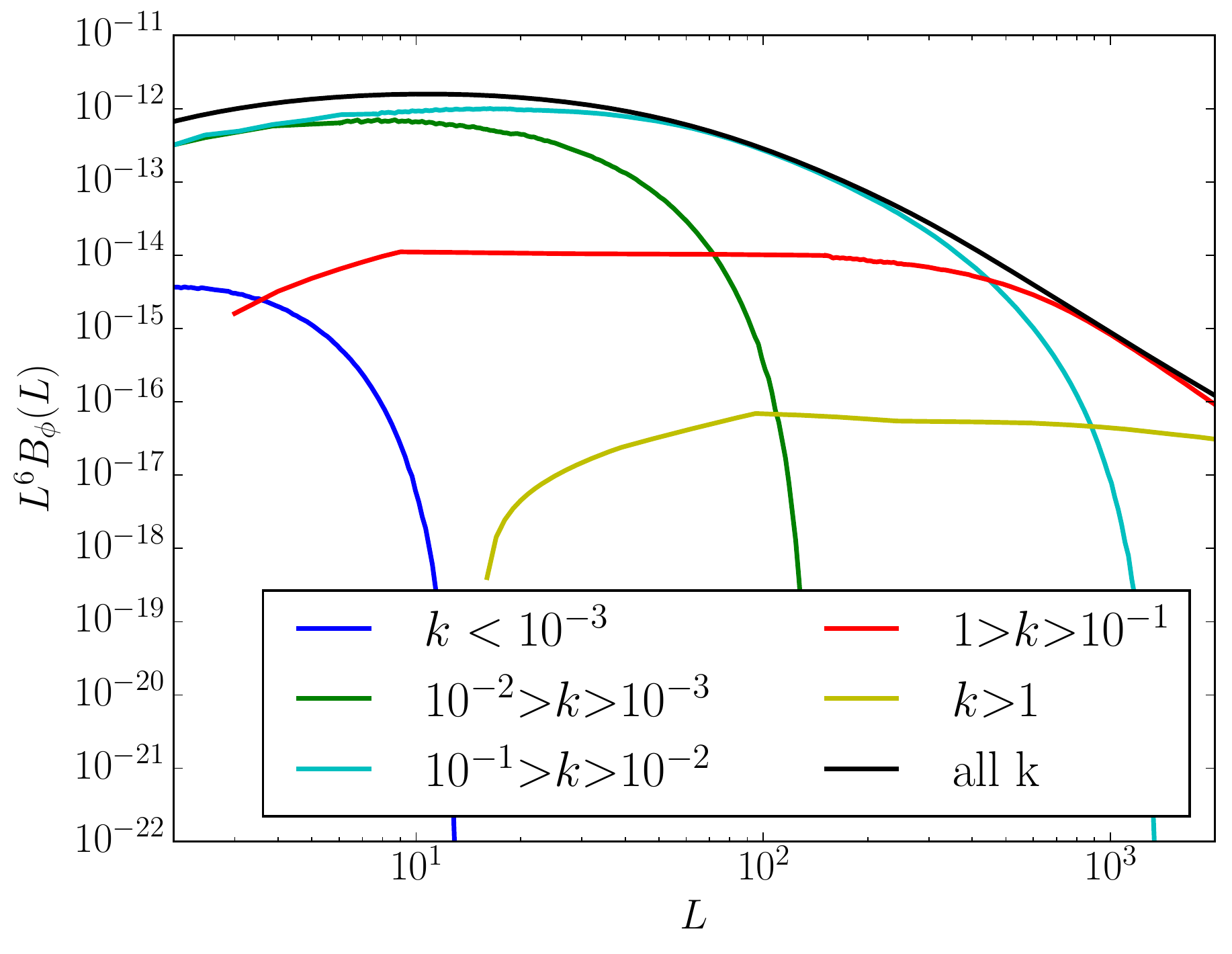}}
\end{center}
\caption{\label{fig:BiphiCum_k} Contribution to the equilateral lensing bispectrum $B_\phi(L,L,L)$ from different wavenumbers of LSS modes (in $\mathrm{Mpc}^{-1}$). At multipoles of $L\gtrsim 500$ the lensing bispectrum starts to become dominated by contributions from LSS modes with wavenumbers $k>0.1\,\mathrm{Mpc}^{-1}$. On these scales and at low redshifts the perturbation theory bispectrum model at leading order underestimates the matter bispectrum in N-body simulations. A lensing potential bispectrum based on this model is likely similarly underestimated at higher multipoles.}
\end{figure}

Contributions to the lensing potential bispectrum from different wavenumbers of LSS modes are shown in \fig{BiphiCum_k}. Up to intermediate lensing multipoles ($L\sim 500$) the lensing potential bispectrum is sourced by LSS modes with $k<0.1\, \mathrm{Mpc}^{-1}$. The LSS bispectrum on these scales is sufficiently described by the standard perturbation theory bispectrum model at leading order. Using this model in the evaluation of $B_\phi$ should therefore provide accurate results up to at least intermediate $L$. On smaller lensing scales (higher $L$), we find significant contributions from LSS modes with $k>0.1\, \mathrm{Mpc}^{-1}$. The leading-order perturbation theory model for the LSS bispectrum fails to accurately describe the LSS bispectrum in N-body simulations at low redshifts on these scales. An improved estimate of the lensing potential bispectrum on small scales would therefore require a more accurate model for the matter bispectrum for small LSS modes.

\section{Bias integral expressions for faster numerical evaluation}
\label{App:fastExpressions}

The A1 and C1 biases in Eqs.~\eq{N32A1} and \eq{N32C1} involve four-dimensional integrals for every multipole  $L$, which are computationally expensive to evaluate. 
Fortunately, however, the integrands of these 4D integrals can be rewritten in a product-separable form, which allows much faster numerical evaluation by multiplying 2D integrals.
The next two subsections will show this explicitly for the C1 and A1 contributions to the bias, with the final results given by Eqs.~\eq{N32decompExact} and \eq{N32A1Fast}, which have a simple form.

\subsection{Fast expression for C1 bias by separation of integrals}

We start with the C1 contribution to the $N^{(3/2)}$ bias because it is somewhat simpler to speed up than the A1 contribution.
The C1 contribution given by \eqq{N32C1} involves a 4D integral over $\vl_1$ and $\vl$ for every value of $L$, which is computationally expensive.  
To separate the integrand, we rewrite scalar products between wavevectors using the angle addition theorem for the cosine: If we define $\cos\mu_{\vl_1}=\vl_1\cdot\VL/(l_1L)$ and $\cos \mu_{\vl}=\vl\cdot \VL/(lL)$, then the angle between $\vl_1$ and $\vl$ is $\mu_{\vl_1}{-}\mu_{\vl}$, so that
\begin{align}
  \label{eq:tradeAngle}
  \vl_1\cdot\vl = l_1l \cos(\mu_{\vl_1}-\mu_{\vl}) 
= l_1l\left[\cos(\mu_{\vl_1})\cos(\mu_{\vl}) + \sin(\mu_{\vl_1})\sin(\mu_{\vl})\right].
\end{align}
Then, using basic trigonometric identities we obtain for the expression in the integrand of \eqq{N32C1}
\begin{align}
  \left[\vl_1 \cdot(\vl-\VL) \right]\left[\vl_1 \cdot \vl \right]
= 
l_1^2l\Big\{&
\cos^2(\mu_{\vl_1})\cos\mu_\vl\left[l\cos\mu_\vl-L\right]\non\\
&+
\cos\mu_{\vl_1}\sin\mu_{\vl_1}\sin\mu_\vl\left[2l\cos\mu_\vl-L\right]
\non\\
  \label{eq:DoubleGeoProductInIntegrand}
&+ \sin^2(\mu_{\vl_1})\, l\sin^2(\mu_\vl)
\Big\},
\end{align}
which is a sum of terms that are separable in $\mu_{\vl_1}$ and $\mu_\vl$ as desired. 
The first term on the right hand side of \eqq{DoubleGeoProductInIntegrand} involves $l_{1,\parallel}^2=(\vl_1\cdot\hat\VL)^2=l_1^2\cos^2(\mu_{\vl_1})$ which measures the component of the temperature multipole $\vl_1$ along the reconstruction multipole $\VL$.  The third term involves $l_{1,\perp}^2=l_1^2\sin^2(\mu_{\vl_1})$ which measures the component of the temperature multipole $\vl_1$ perpendicular to the reconstruction multipole $\VL$.  The second term in \eqq{DoubleGeoProductInIntegrand} is a cross term involving a product of these two components, $l_{1,+}^2= l_{1,\parallel}l_{1,\perp}$.  Using \eqq{DoubleGeoProductInIntegrand}, the C1 contribution \eq{N32C1} to the $N^{(3/2)}$ bias therefore turns into the following simple form of \eqq{N32decompExact}:
\begin{eqnarray}
\label{eq:N32decompExact_b}
N^{(3/2)}_{C1}(L) &=& -4 A_L^2 S_L 
\left[R_\parallel(L)\beta_\parallel(L)+R_\perp(L)\beta_\perp(L)\right],
  \end{eqnarray}
where we defined
the temperature integral $R_\parallel$ and integrated lensing bispectrum $\beta_\parallel$ as
\begin{eqnarray}
  \label{eq:Rparallel_b}
R_\parallel(L) &=& \int_{\vl_1}g(\vl_1,\VL)  l_1^2\cos^2(\mu_{\vl_1}) C_{l_1}^{TT}, \\
\label{eq:betaparallel_b}
  \beta_\parallel(L) &=& \int_{\vl}  l\cos\mu_\vl\left[ l\cos\mu_\vl - L \right] \,B_\phi(\vl,\VL-\vl,-\VL),
\end{eqnarray}
and similarly for the perpendicular component,
\begin{eqnarray}
\label{eq:Rperp_b}
\qquad
R_\perp(L) &=& \int_{\vl_1}g(\vl_1,\VL)  l_1^2 \sin^2(\mu_{\vl_1}) C^{TT}_{l_1},\\
  \label{eq:betaperp_b}
  \beta_\perp(L) &=& \int_\vl  l^2 \sin^2(\mu_\vl) \,B_\phi(\vl,\VL-\vl,-\VL).
\end{eqnarray}
The cross term from the second line of \eqq{DoubleGeoProductInIntegrand} yields $R_+\beta_+=0$; see \app{betaplusRplus0}.  
We will use \eqq{N32decompExact} for numerically evaluating the C1 contribution to the $N^{(3/2)}$ bias, because it only involves 2D integrals that are much faster to evaluate than the 4D integral in \eqq{N32C1}.
  
A slightly simpler approximate expression follows by noting
that $S_L\approx 1/(2A_L)$ at leading order in $C^{\phi\phi}$:
\begin{align}
\label{eq:N32decomp_b}
 N^{(3/2)}_{C1}(L) 
\approx -2 A_L 
\left[R_\parallel(L)\beta_\parallel(L)+R_\perp(L)\beta_\perp(L)\right].
\end{align}
The separation into parallel and perpendicular components with respect to the reconstruction multipole $\VL$ also follows by working in position space (see Appendix~\ref{App:N32Interpretations}, which also interprets this result using a scalar-tensor decomposition of correlation functions between derivatives of temperature or lensing fields).

\subsection{Fast expression for A1 bias by evaluating Fourier-space convolution as position-space product}

Numerical evaluation of the 4D integral appearing in the A1 contribution of \eqq{N32A1} to the $N^{(3/2)}$ bias can also be accelerated by suitably rewriting the integral.  The idea is that, for fixed $\vl$, the integral over $\vl_1$ in \eqq{N32A1} is a convolution in Fourier space, which can be evaluated efficiently as a product in position space (similarly to References~\cite{MarcelZvonimirPat1603} and \cite{2016arXiv160304826M} which used the same idea to accelerate large-scale structure perturbation theory integrals). 
This gives the following fast expression for the A1 bias of \eqq{N32A1}:\footnote{This follows by introducing $\vl=\vl_1-\vl'$ in \eqq{N32A1} with a Dirac delta,
\begin{align}
\int_{\vl_1,\vl'} g(\vl_1,\VL) C^{TT}_{|\vl_1-\vl'|} &[(\vl_1-\vl')\cdot\vl']
[(\vl_1-\vl')\cdot (\VL-\vl')]B_\phi(\vl', \VL-\vl',-\VL)\non\\
&= \int_{\vl_1,\vl,\vl'}
 (2\pi)^2 \delta_D(\vl-\vl_1+\vl') g(\vl_1,\VL)
C^{TT}_{l}
 (\vl\cdot\vl')[\vl\cdot(\VL-\vl')]
B_\phi(\vl',\VL-\vl',-\VL),
\end{align}
expanding the Dirac delta in plane waves, and separating the scalar products using \eqq{DoubleGeoProductInIntegrand}.}
\begin{align}
  \label{eq:N32A1Fast}
N^{(3/2)}_{A1}(L)
  = 4 A_L^2 
S_L
 \int\d^2 \vr\,
\xi_g(\vr,\VL)
\left[
\bar{\beta}_\parallel(\vr,\VL)\bar{\xi}^{TT}_\parallel(\vr)
+\bar{\beta}_\perp(\vr,\VL)\bar{\xi}^{TT}_\perp(\vr)
\right].
\end{align}
The structure of this is very similar to the fast expression for the C1 term given by \eqq{N32decompExact}, but it involves a 2D $\vr$-integral over the following 2D Fourier transforms:
\begin{align}
  \label{eq:xi_g_def}
  \xi_g(\vr,\VL)=\int_{\vl_1}e^{i\vl_1\cdot\vr}\,g(\vl_1,\VL),
\end{align}
and
\begin{eqnarray}
  \label{eq:barbetaparallel}
  \bar{\beta}_\parallel(\vr,\VL) &=& \int_{\vl}e^{-i\vl\cdot\vr}\,l \cos\mu_{\vl}(l\cos\mu_{\vl}-L)B_\phi(\vl,\VL-\vl,-\VL)\\
  \label{eq:barbetaperp}
  \bar{\beta}_\perp(\vr,\VL) &=& \int_{\vl}e^{-i\vl\cdot\vr}\,(l)^2 \sin^2(\mu_{\vl})B_\phi(\vl,\VL-\vl,-\VL),
\end{eqnarray}
which satisfies $\bar\beta_{n}(0,\VL)=\beta_n(\VL)$. 
We also defined temperature correlation functions
\begin{eqnarray}
  \label{eq:barxiparallel}
  \bar{\xi}^{TT}_\parallel(\vr) &=& \int_{\vl}e^{-i\vl\cdot\vr}\,l^2\cos^2(\mu_\vl) C^{TT}_l
= \frac{1}{2\pi}\left[
\frac{1}{r}\int_0^\infty \d l\,l^2J_1(lr) C_l^{TT}
- \cos^2(\nu_\VL)\int_0^\infty \d l\,l^3J_2(lr)C_l^{TT}\right],
 \\
\label{eq:barxiperp}
  \bar{\xi}^{TT}_\perp(\vr) &=& \int_{\vl}e^{-i\vl\cdot\vr}\,l^2\sin^2(\mu_\vl) C^{TT}_l
= \frac{1}{2\pi}\int_0^\infty\d l\,l^3 J_0(lr)C_l^{TT}
-  \bar{\xi}^{TT}_\parallel(\vr,\VL).
\end{eqnarray}
On the right hand sides, 2D Fourier transforms reduce to 1D Hankel transforms by using the cosine angle addition theorem to express $\cos\mu_{\vl}=\hat\vl\cdot\hat\VL$ in terms of $\cos\nu_{\VL}=\hat\VL\cdot\hat\vr$ and $\cos\nu_\vl=\hat\vl\cdot\hat\vr$ (similarly to \eqq{tradeAngle}).  The angular integrals then lead to Bessel functions of the first kind, $J_n$.  
The 1D Hankel transforms can be evaluated efficiently with 1D FFTs using e.g.~FFTLog \cite{hamiltonfftlog}.
A somewhat slower but still feasible approach is to evaluate the 2D Fourier transforms on a grid using 2D FFTs.

\section{Effect of lensing bispectrum on measured lensing power spectrum: Remaining contractions}
\label{App:AllContris}

Having discussed the contractions A1 and C1 contributing to the $N^{(3/2)}$ lensing bias in detail in the main text and in the previous section, this section derives analytical expressions for the non-Gaussian lensing bias from the remaining contractions A2, A3, B, C2, C3 and D as outlined in \secref{N32Overview}.

For easier reference of the contractions, we categorize them by their temperature pairings: \emph{Intra}-temperature contractions $\la TT\ra$ and $\la T'T'\ra$, which appear in A1, B1 and C1 terms, involve two temperature fields that belong to the same lensing reconstruction $\hat\phi$. \emph{Inter}-temperature contractions $\la TT'\ra$, which appear in A2, A3, B2, B3, C2 and C3 terms, involve one temperature field belonging to $\hat\phi(\VL)$ and another temperature field belonging to $\hat\phi(-\VL)$.\footnote{Roughly speaking, intra-temperature correlations are zero-lag terms of \emph{filtered} temperature fields, and inter-temperature correlations are correlations of two filtered temperature fields with non-zero separation. However, the filtering of observed temperature maps is non-local in position space, so that strictly speaking intra-temperature correlations are not zero-lag in the \emph{observed} temperature.}

\subsection{Type C biases from inter-temperature contractions C2 and C3 of \texorpdfstring{$\langle\delta^2 TT\delta T'T'\rangle$}{<d2T T dT' T'>}}
	\label{App:N32}

We start with the inter-temperature C2 and C3 contributions to the $N^{(3/2)}$ bias following from the contractions of $\la\delta^2 T T \delta T' T'\ra$ defined in \eqq{typeCcontractions}. 

\subsubsection{C2 contraction in Fourier space}
The C2 contraction in  \eqq{typeCcontractions} is given by
\begin{align}
  \label{eq:4}
  \la\delta^2 T_{\vl_1}T_{\vl_2}\delta T_{\vl_3}T_{\vl_4}\ra_{C2}
= 
\la\delta^2 T_{\vl_1}\delta T_{\vl_3}\ra
\la T_{\vl_2}T_{\vl_4}\ra.
\end{align}
This involves the correction $\la\delta^2 T\delta T\ra$ of the lensed temperature power spectrum generated by a non-zero lensing bispectrum (another correction would be $\la\delta^3 T T\ra)$.  Based on analytical \cite{MerkelSchaefer2011} and numerical investigations \citep{Carbone2009,SPTLensingDetectionVanEngelen1202,Calabrese2015} this correction is expected to be small. 
Further, it should be automatically accounted for when using realization-dependent subtraction of the Gaussian $N^{(0)}$ bias which is common in modern lensing pipelines. 
We do not investigate this term here further.

\subsubsection{C3 contraction in Fourier space}
\label{App:CorrTerm}
The C3 contraction is given by
\begin{align}
  \nonumber
  \langle & \delta^2 T_{\vl_1} T_{\vl_2}\delta T_{\vl_3}T_{\vl_4}
  \rangle_{C3}\\
  \label{eq:trispecTermC}
  &=  -\frac{(2\pi)^2 }{2} \delta_D(\vl_1+\vl_2+\vl_3+\vl_4) \int_{\vl_2'}
  \left[\vl_4\cdot(\vl_1+\vl_4-\vl'_2) \right]\left[\vl_4 \cdot \vl'_2 \right] \left[ \vl_2 \cdot (\vl_1+\vl_4)\right]
C_{l_2}^{TT}C_{l_4}^{TT}
   B_\phi(\vl'_2,\vl_1+\vl_4-\vl'_2,-(\vl_1+\vl_4)) 
\end{align}
The induced bias of the measured lensing power spectrum is
\begin{align}
\nonumber
N^{(3/2)}_{C3}(L)
=\,&- \frac{8}{2} A_L^2 \int_{\vl_1,\vl_2}g(\vl_1,\VL) g(
\vl_2,\VL) \left[ (\vl_1-\VL) \cdot (\vl_1+\vl_2-\VL)\right]
C^{TT}_{|\VL- \vl_1|}C^{TT}_{|\vl_2-\VL|} \\
&\times \int_{\vl}
  \left[(\vl_2-\VL)\cdot(\vl-\vl_1-\vl_2+\VL) \right]\left[(\vl_2-\VL) \cdot \vl \right] B_\phi(\vl,-\vl_2+\VL-\vl_1,\vl_1+\vl_2-\VL-\vl),
\end{align}
where we have accounted for all possibilities to place the (perturbed) temperatures in the four-point correlator by including a symmetry factor of 8.
Changing integration variables $\vl_1\rightarrow \VL-\vl_1$ and $\vl_2\rightarrow \VL-\vl_2$, we obtain
\begin{align}
\nonumber
N^{(3/2)}_{C3}(L)
&= 4 A_L^2 \int_{\vl_1,\vl_2} g(\vl_1,\VL) g(
\vl_2,\VL) \left[ \vl_1 \cdot (\VL-\vl_1-\vl_2)\right] 
C^{TT}_{l_1}C^{TT}_{l_2}\\
\label{eq:SubDomTypeC_step1}
&\quad \times\int_{\vl}
  \left[\vl_2\cdot(\vl-(\VL-\vl_1-\vl_2)) \right](\vl_2 \cdot \vl)
B_\phi(\vl, (\VL-\vl_1-\vl_2)-\vl, -(\VL-\vl_1-\vl_2)),
\end{align}
where we used $g(\vl,\VL)=g(\VL-\vl,\VL)$.

\subsubsection{Fast expression for C3 contraction}

The C3 contribution to the $N^{(3/2)}$ bias given by \eqq{SubDomTypeC_step1} involves a 6D integral over $\vl_1$, $\vl_2$ and $\vl$ for every value of $L$, which can be regarded as a 3-loop integral. Evaluating this numerically is prohibitively computationally expensive. Fortunately, however, the integral can be rearranged to allow much faster evaluation. This follows by noting that the integral over the lensing bispectrum reduces to $\beta_n(|\VL-\vl_1-\vl_2|)$ defined in Eqs.~\eq{betaparallel} and \eq{betaperp}.  The total integral over $\vl_1$ and $\vl_2$ is then an integral over functions of $\vl_1$, $\vl_2$ and $\VL-\vl_1-\vl_2$, which is a double convolution.  This can be evaluated efficiently by rewriting it as a product in position space, similarly to the fast expressions of Eqs.~\eq{N32decompExact} and \eq{N32A1Fast} for the C1 and A1 contractions discussed in \app{fastExpressions} (also see \cite{MarcelZvonimirPat1603} and \cite{2016arXiv160304826M}).

To see this explicitly, we introduce a Dirac delta enforcing $\vl_3=\VL-\vl_1-\vl_2$, 
\begin{align}
  \label{eq:48}
  N^{(3/2)}_{C3}(L) &= 4A_L^2 \int_{\vl_1,\vl_2,\vl_3} (2 \pi)^2 \delta_D(\vl_3-\VL+\vl_1+\vl_2) g(\vl_1,\VL) g(\vl_2,\VL)( \vl_1 \cdot \vl_3)
C^{TT}_{l_1}C^{TT}_{l_2}\\
\label{eq:SubDomTypeC_step2}
&\quad \times\int_{\vl}
  \left[\vl_2\cdot(\vl-\vl_3) \right](\vl_2 \cdot \vl)
B_\phi(\vl, \vl_3-\vl, -\vl_3).
\end{align}
Parameterizing orientations of $\vl_2$ and $\vl$ in terms of $\cos\vartheta_{\vl_2}=\vl_2\cdot\vl_3/(l_2l_3)$ and $\cos\vartheta_\vl=\vl\cdot\vl_3/(ll_3)$, the last integrand can be cast separable using \eqq{DoubleGeoProductInIntegrand} (replacing $\vl_1$ by $\vl_2$ and $\VL$ by $\vl_3$ there). The Dirac delta also becomes separable by expressing it in terms of plane waves,
\begin{align}
  \label{eq:DiracDeltaInPlaneWaves}
  (2\pi)^2 \delta_D(\vl_3-\VL+\vl_1+\vl_2) = \int\d^2\vr\, e^{-i(\vl_3-\VL+\vl_1+\vl_2)\cdot \vr}.
\end{align}
We thus get
\begin{align}
\label{eq:SubDomTypeC_step3}
    N^{(3/2)}_{C3}(L) &= 4A_L^2
\int\d^2\vr\, e^{i\VL\cdot\vr}\int_{\vl_3}\,e^{-i\vl_3\cdot\vr} 
\left[\int_{\vl_1}\,e^{-i\vl_1\cdot\vr}\,g(\vl_1,\VL)\,\vl_1\cdot\vl_3 \,C_{l_1}^{TT}\right]\;
 \sum_{n=\parallel,\perp}\beta_n(l_3)\xi^{TT}_n(\vr, \VL, \hat\vl_3),
\end{align}
where we used $\beta_+=0$ from \app{betaplusRplus0} and defined weighted temperature correlation functions
\begin{align}
\label{eq:xiparallel}
\xi^{TT}_\parallel(\vr, \VL, \hat\vl_3) 
 &= \int_{\vl_2}\,e^{-i\vl_2\cdot\vr}\, g(\vl_2,\VL)  l_2^2\cos^2(\vartheta_{\vl_2}) C_{l_2}^{TT}\\
\label{eq:xiperp}
\xi^{TT}_\perp(\vr, \VL, \hat\vl_3) &= \int_{\vl_2}\,e^{-i\vl_2\cdot\vr}\,\,g(\vl_2,\VL)  l_2^2 \sin^2(\vartheta_{\vl_2}) C^{TT}_{l_2}.
\end{align}
\eqq{SubDomTypeC_step3} can be simplified for easier numerical evaluation  and interpretation. 
Similarly to \eqq{tradeAngle}, angles $\vartheta$ with respect to $\vl_3$ can be expressed in terms of cosines $\cos\mu_{\vl_2}=\hat\vl_2\cdot\hat\VL$ and $\cos\mu_{\vl_3}=\hat\vl_3\cdot\hat\VL$ with respect to $\VL$, so that
\begin{align}
  \label{eq:51}
  \cos^2(\vartheta_{\vl_2}) &= (\hat\vl_2\cdot\hat\vl_3)^2 = 
(\cos\mu_{\vl_2}\cos\mu_{\vl_3} + \sin\mu_{\vl_2}\sin\mu_{\vl_3})^2\\
&=\sum_{m=0}^2 {2 \choose m} \cos^{2-m}(\mu_{\vl_2})\cos^{2-m}(\mu_{\vl_3}) \sin^{m}(\mu_{\vl_2})\sin^{m}(\mu_{\vl_3}).
\end{align}
A similar expression follows for $\sin^2(\vartheta_{\vl_2})=1-\cos^2(\vartheta_{\vl_2})$. The temperature correlation functions thus become
\begin{eqnarray}
  \label{eq:27}
  \xi^{TT}_{\parallel}(\vr,\VL,\hat\vl_3) &=& 
\sum_{m=0}^2 {2 \choose m} \cos^{2-m}(\mu_{\vl_3})\sin^{m}(\mu_{\vl_3})
\,\check{\xi}^{TT}_{2,m}(\vr,\VL),\\
  \xi^{TT}_{\perp}(\vr,\VL,\hat\vl_3) &=& \check{\xi}^{TT}_{00}(\vr,\VL) - \xi^{TT}_{\parallel}(\vr,\VL,\hat\vl_3),
\end{eqnarray}
where we defined
\begin{align}
  \label{eq:xi_jm}
\check\xi^{TT}_{jm}(\vr,\VL) \equiv
  \int_{\vl}\,e^{-i\vl\cdot\vr}\,g(\vl,\VL)\,l^j
\cos^{j-m}(\mu_{\vl})\sin^{m}(\mu_{\vl})\,C^{TT}_{l}
\end{align}
where $\cos\mu_\vl=\hat\vl\cdot\hat\VL$.  For fixed $\VL$, \eqq{xi_jm} can be computed as a 2D Fourier transform of the integrand (regarded as a function of $\vl$ on a 2D grid). Ignoring the weight $g$, $\check \xi^{TT}_{jm}$ is the correlation function of the temperature $T(\vx)$ with a second  derivative of the temperature (i.e.~it is related to $\langle T(\vx)\partial_0^{j-m}\partial_1^m T(\vx')\rangle$ if  $\VL$ is aligned with the 0-axis).  Note that we recover some of the zero-lag/intra-temperature correlation $R_L$ integrals for $\vr=0$.
The square brackets in \eqq{SubDomTypeC_step3} similarly reduce to
\begin{align}
  \label{eq:52}
\int_{\vl_1}\,e^{-i\vl_1\cdot\vr}\,g(\vl_1,\VL)\,\vl_1\cdot\vl_3 \,C_{l_1}^{TT}
=
\sum_{m'=0}^1
 l_3\cos^{1-m'}(\mu_{\vl_3})\sin^{m'}(\mu_{\vl_3}) \,\xi^{TT}_{1,m'}(\vr,\VL),
\end{align}
which is related to the correlation function between the temperature and a first derivative of the temperature ($\langle T(\vx)\partial_{m'} T(\vx')\rangle$ if $\VL$ is aligned with the 0-axis).

The C3 $N^{(3/2)}$ bias of \eqq{SubDomTypeC_step3} thus turns into the following fast-to-evaluate expression:
  \begin{align}
\non
    N^{(3/2)}_{C3}(L) 
= 4A_L^2
\int\d^2\vr\,e^{i\VL\cdot\vr}
\Bigg\{&
\sum_{m'=0}^1 
\check\xi^{TT}_{00}(\vr,\VL)\check\xi^{TT}_{1,m'}(\vr,\VL)
 \check \beta_{m'}(\vr,\hat\VL)\\
\label{eq:SubDomTypeC_step4}
& + 
\sum_{m=0}^2
\sum_{m'=0}^{1}\,
 {2 \choose m}\,
\check\xi^{TT}_{2,m}(\vr,\VL)\check\xi^{TT}_{1,m'}(\vr,\VL)
\check \beta_{mm'}(\vr,\hat\VL)
\Bigg\}
\end{align}
where we defined
\begin{align}
  \label{eq:54}
  \check \beta_{m'}(\vr,\hat\VL) &\equiv \int_{\vl_3}e^{-i\vl_3\cdot\vr}\,l_3
\beta_{\perp}(l_3)
\cos^{1-m'}(\mu_{\vl_3})\sin^{m'}(\mu_{\vl_3})\\
 \check{\beta}_{mm'}(\vr, \hat\VL) &\equiv
\int_{\vl_3}e^{-i\vl_3\cdot\vr}\,l_3
[\beta_{\parallel}(l_3)-\beta_\perp(l_3)]
\cos^{3-m-m'}(\mu_{\vl_3})\sin^{m+m'}(\mu_{\vl_3}).
\end{align}
These are functions of $\vr$ that can be computed with a 2D Fourier transform.
Alternatively, the angular integrals can be done analytically so that the integrals become 1D integrals involving Bessel functions.

For fixed $\VL$, all  factors in the integrand of \eqq{SubDomTypeC_step4} can be evaluated with 2D Fourier transforms. The final integration over $\vr$ can be evaluated as a 2D integral for fixed $\VL$. Alternatively, it can be obtained by fixing $\VL$,  regarding the curly brackets as a function of $\vr$, computing its 2D Fourier transform $\mathcal{F}(\VL')$, and picking the entry $\mathcal{F}(\VL'=\VL)$. 
Without loss of generality we choose $\VL$ to be aligned with the 0-axis of the 2D grid.  Once $\beta_n(l)$ are computed, the bias at a given $L$ can be computed with $\mathcal{O}(10 \,l_\mathrm{max}^2 \log l_\mathrm{max})$ operations because it involves only 2D FFTs. 
A very preliminary implementation of the C3 contribution to the $N^{(3/2)}$ bias gave results that were much smaller than the A1 and C1 contributions, but future work should evaluate this term more carefully to check its importance, also in combination with consistency checks against simulations.

\subsection{Type A biases from inter-temperature contractions A2 and A3 of \texorpdfstring{$\langle\delta T \delta T \delta T'T'\rangle$}{<dT dT dT' T'>}}

\label{App:SubdomTypeAB}

The coupling of type A, $\langle\delta T_{\vl_1} \delta T_{\vl_2}\delta T_{\vl_3}T_{\vl_4}\rangle$, measures the correlation between three temperatures perturbed to first order in the lensing potential and one unlensed temperature. 
It has three contractions A1, A2 and A3 defined in \eqq{70}.  Having discussed the A1 term in detail in the main text, we consider the remaining A2 and A3 term here. 

The A2 contraction in \eqq{70} is given by
\begin{align}
  \langle\delta T_{\vl_1}& \delta T_{\vl_2}\delta T_{\vl_3}T_{\vl_4}\rangle_{A2}\non\\
&=
-(2\pi)^2 \delta_D(\vl_1+\vl_2+\vl_3+\vl_4)
C^{TT}_{l_4} \[\vl_4\cdot(\vl_2+\vl_4)\] 
\int_{\vl} C^{TT}_l \[\vl\cdot(\vl_1-\vl)\]\[\vl\cdot(\vl+\vl_3)\] B_\phi(\vl_1-\vl,\vl+\vl_3,\vl_2+\vl_4).
\end{align}
The resulting contribution to the $N^{(3/2)}$ bias is
\begin{eqnarray}
  \label{eq:13}
  N^{(3/2)}_{A2}(L) &=& 
-4A_L^2\int_{\vl_1\vl_2}
g(\vl_1,\VL)g(\vl_2,\VL)
[(\vl_2-\vl_1)\cdot(\vl_2-\VL)]
C^{TT}_{|\VL-\vl_2|}
\int_{\vl}[\vl\cdot(\vl_1-\vl)][\vl\cdot(\vl-\vl_2)]
C^{TT}_l
B_\phi(\vl_1-\vl, \vl_2-\vl_1, \vl-\vl_2)\non\\
&=& 
-4A_L^2\int_{\vl_1\vl_2}
g(\vl_1,\VL)g(\vl_2,\VL)
[(\vl_2-\vl_1)\cdot(\vl_2-\VL)]
C^{TT}_{|\VL-\vl_2|}\non\\
&&\quad \times
\int_{\vl}[(\vl_1-\vl)\cdot\vl][(\vl_1-\vl)\cdot(\vl_1-\vl_2-\vl)]
C^{TT}_{|\vl_1-\vl|}
B_\phi(\vl, (\vl_1-\vl_2)-\vl, -(\vl_1-\vl_2))
\end{eqnarray}
where we changed integration variables $\vl\rightarrow \vl_1-\vl$ to simplify the bispectrum arguments and included a symmetry factor of 4. 
The integrand involves functions with different arguments in all three integration variables $\vl_1$, $\vl_2$ and $\vl$, leading to a tightly coupled 6D integral for every $L$, which is computationally prohibitively expensive. Since the integral does not seem to have an obvious convolution-like structure, we do not investigate further if it can be accelerated, and leave numerical evaluation and discussion of its importance for future work (also noting that it may be more efficient to first check if the other simpler bias contributions can already explain simulation results).  
The $N^{(3/2)}$ bias from the A3 contraction equals that of the A2 contraction because the bias is invariant under exchanging $\vl_1\leftrightarrow \VL-\vl_1$ in \eqq{reconstructionPowerExpValue}.

\subsection{Type B biases from \texorpdfstring{$\langle\delta^2 T \delta T T'T'\rangle$}{<d2T dT T' T'>}}

Terms of type B are of the form $\<\delta^2 T \delta T T' T'\>$ with both perturbed temperatures coupling to the same estimator $\hat\phi$ (both perturbed temperatures are on the same side of the correlator). There are 4 possibilities to form such a term, resulting in a symmetry factor of 4.
Three different contractions of CMB fields contribute to type B:
\begin{alignat}{2}
  \label{eq:typeBconnectedAppendix}
\langle \delta^2 T \delta T\,  T' T' \rangle \;\sim\;\la &&
{
\contraction[1ex]{}{T}{\!{}_{,ij}\phi\!{}_{,i}\phi\!{}_{,j}}{T}
\contraction[1ex]{T\!{}_{,ij}\phi\!{}_{,i}\phi\!{}_{,j}T\!{}_{,k}\phi\!{}_{,k}\, }{T'}{}{T}
\bcontraction[1ex]{T\!{}_{,ij}}{\phi}{\!{}_{,i}}{\phi}
\bcontraction[1ex]{T\!{}_{,ij}}{\phi}{\!{}_{,i}\phi\!{}_{,j}T\!{}_{,k}}{\phi}
T\!{}_{,ij}\phi\!{}_{,i}\phi\!{}_{,j}T\!{}_{,k}\phi\!{}_{,k}\, T' T'
}
\ra_{B1}
\;+\;
\la {
\contraction[0.5ex]{}{T}{\!{}_{,ij}\phi\!{}_{,i}\phi\!{}_{,j}T\!{}_{,k}\phi\!{}_{,k}\, }{T'}
\contraction[1.5ex]{T\!{}_{,ij}\phi\!{}_{,i}\phi\!{}_{,j}}{T}{\!{}_{,k}\phi\!{}_{,k}\, T'}{T'}
\bcontraction[1ex]{T\!{}_{,ij}}{\phi}{\!{}_{,i}}{\phi}
\bcontraction[1ex]{T\!{}_{,ij}}{\phi}{\!{}_{,i}\phi\!{}_{,j}T\!{}_{,k}}{\phi}
T\!{}_{,ij}\phi\!{}_{,i}\phi\!{}_{,j}T\!{}_{,k}\phi\!{}_{,k}\, T' T'
}
\ra_{B2}
\;+\;
\la {
\contraction[1.5ex]{}{T}{\!{}_{,ij}\phi\!{}_{,i}\phi\!{}_{,j}T\!{}_{,k}\phi\!{}_{,k}\,T' }{T'}
\contraction[0.5ex]{T\!{}_{,ij}\phi\!{}_{,i}\phi\!{}_{,j}}{T}{\!{}_{,k}\phi\!{}_{,k}\,}{T'}
\bcontraction[1ex]{T\!{}_{,ij}}{\phi}{\!{}_{,i}}{\phi}
\bcontraction[1ex]{T\!{}_{,ij}}{\phi}{\!{}_{,i}\phi\!{}_{,j}T\!{}_{,k}}{\phi}
T\!{}_{,ij}\phi\!{}_{,i}\phi\!{}_{,j}T\!{}_{,k}\phi\!{}_{,k}\, T' T'
}
\ra_{B3}.
\end{alignat}
The lensing power bias \eq{reconstructionPowerExpValue} resulting from the B1 contraction vanishes for $L>0$,
\begin{align}
  \label{eq:15}
  N^{(3/2)}_{B1}(L) = 0.
\end{align}
The B2 contraction is
\begin{align}
\nonumber
\langle\delta^2 T_{\vl_1} \delta T_{\vl_2} T_{\vl_3}T_{\vl_4}\rangle_{B2}=\,&-\frac{(2\pi)^2}{2} \delta_D(\vl_1+\vl_2+\vl_3+\vl_4) C^{TT}_{l_3} C^{TT}_{l_4}\\
\label{eq:TTTT_TermB}
&\times \int_\vl \[\vl_3\cdot(\vl{-}\vl_1{-}\vl_3)\]\[\vl_3\cdot\vl\][(\vl_1+\vl_2+\vl_3)\cdot(\vl_3+\vl_1)] B_\phi(\vl,\vl_1+\vl_3{-}\vl,{-}\vl_1{-}\vl_3).
\end{align}
The lensing bias resulting from this and the similar B3 contraction is
\begin{align}
\nonumber
N^{(3/2)}_{B2}(L)+N^{(3/2)}_{B3}(L)=\;& - 2 A_L^2  \int_{\vl_1,\vl_2}g(\vl_1,\VL)g(\vl_2,\VL)
[(\VL-\vl_2)\cdot(\vl_1-\vl_2)]
C^{TT}_{l_2} C^{TT}_{|\VL-\vl_2|}
\\
&\times \int_\vl \[\vl_2\cdot(\vl-(\vl_1{-}\vl_2))\](\vl_2\cdot\vl)\,
 B_\phi(\vl,(\vl_1{-}\vl_2)-\vl,-(\vl_1{-}\vl_2))
+(\vl_2 \leftrightarrow \VL{-}\vl_2)\vphantom{\int_{\vl}}.
\end{align}
The bispectrum integral over $\vl$ is a convolution similar to the bispectrum integrals arising e.g.~for the C1 or C3 contributions in Eqs.~\eq{N32C1} and \eq{SubDomTypeC_step1}, so it can likely be rewritten in a fast way similarly to Eqs.~\eq{N32decompExact} or \eq{SubDomTypeC_step4}. 
These biases of type B2 and B3 should be investigated further in future work.

\subsection{Type D bias from \texorpdfstring{$\langle\delta^3 T T T'T'\rangle$}{<d3T T T' T'>}}
	\label{App:SubdomTypeD}

The last type of coupling, type D, involves the lensed temperature perturbed to third order in $\phi$. It picks up the three-point function of the components of the lensing deflection at the same location $\langle \alpha_i(\vx) \alpha_j(\vx) \alpha_k(\vx) \rangle$. This correlation must vanish by statistical isotropy \citep{MerkelSchaefer2011}. 
This can also be seen analytically. The coupling can be written as
\begin{align}
  \nonumber
  \langle &\delta^3 T_{\vl_1} T_{\vl_2}T_{\vl_3}T_{\vl_4}
  \rangle\\
  \nonumber
  & =-\frac{1}{6}  \int_{\vl_1'}\int_{\vl_1''}\int_{\vl'''_1}
  \left[\vl'_1\cdot(\vl_1{-}\vl'_1{-}\vl''_1{-}\vl'''_1) \right]\left[\vl'_1\cdot \vl''_1 \right] \left[\vl'_1\cdot \vl'''_1 \right]\langle \phi(\vl_1{-}\vl'_1{-}\vl''_1{-}\vl'''_1)\phi(\vl''_1)\phi(\vl'''_1)\rangle_{(\phi)} \langle T_{\vl'_1}  T_{\vl_2}   T_{\vl_3}  T_{\vl_4}\rangle_{(T)}\\
  &=-\frac{1}{6}\zeta(\vl_1)
\langle T_{\vl_1}  T_{\vl_2}   T_{\vl_3}  T_{\vl_4}\rangle_{(T)},
\end{align}
where we defined
\begin{align}
  \label{eq:67}
  \zeta(\vl_1) \equiv \int_\VL
(\vl_1\cdot\VL)
\int_{\vl} 
(\vl_1\cdot\vl)[\vl_1\cdot(\vl-\VL)]B_\phi(\vl,\VL-\vl,-\VL).
\end{align}
The integral over $\vl$ is the same as that already encountered in \eqq{N32C1}. Using the same trick of \eqq{DoubleGeoProductInIntegrand} to make the integral separable leads to
 \begin{align}
   \label{eq:68}
   \zeta(\vl_1) = l_1^3 \int_\VL \left[
 \cos^3(\mu_\VL)\beta_\parallel(L)
 +\cos\mu_\VL\sin^2(\mu_\VL)\beta_\perp(L)\right] = 0,
 \end{align}
 where $\cos\mu_\VL=\cos\mu_{\vl_1}=\hat\vl_1\cdot\hat\VL$. This vanishes after performing the angular integration over $\mu_\VL$.

\subsection{Vanishing cross integrals}
	\label{App:betaplusRplus0}
The fast expression \eq{N32decompExact} for the type C1 bias has an additional contribution $\beta_+R_+$, where
\begin{align}
\label{eq:Rcross}
R_+(\VL) &= \int_{\vl_1}g(\vl_1,\VL)  l_1^2\cos\mu_{\vl_1}\sin\mu_{\vl_1} C^{TT}_{l_1}=0\\
\label{eq:betacross}
\beta_+(\VL) &= \int_\vl l \sin\mu_\vl  \left[ 2l\cos\mu_\vl - L \right] \,B_\phi(\vl,\VL-\vl,-\VL)=0.
\end{align} 
Here we show that both integrals $R_+$  and $\beta_+$  vanish. We start by writing out the weight in \eqq{Rcross},
\begin{align}
  \label{eq:63}
R_+(\VL) &= \int_{\vl}
\frac{\vl\cdot\VL C^{\tilde T\tilde T}_{l}+(\VL-\vl)\cdot\VL C_{|\VL-\vl|}^{\tilde T\tilde T}}{2C^{\tilde T\tilde T}_{l,\expt}C^{\tilde T\tilde T}_{|\VL-\vl|,\expt}}
  l^2\cos\mu_{\vl}\sin\mu_{\vl} C^{TT}_{l}.
\end{align}
Choosing a coordinate system where the $x$-axis is aligned with $\VL$ gives
$\vl\cdot\VL=l_xL$, 
$l\cos\mu_\vl=l_x$ and $l\sin\mu_\vl=l_y$, so that
\begin{align}
  \label{eq:Rplustmp}
  R_+(L) = \frac{1}{(2\pi)^2}\int_{-\infty}^\infty\d l_x\int_{-\infty}^\infty\d l_y\; \frac{l_xLC^{\tilde T\tilde T}_{(l_x^2+l_y^2)^{1/2}} + (L^2-l_xL)C^{\tilde T\tilde T}_{(L^2+l_x^2+l_y^2-2Ll_x)^{1/2}}}{2C^{\tilde T\tilde T}_{(l_x^2+l_y^2)^{1/2},\expt} C^{\tilde T\tilde T}_{(L^2+l_x^2+l_y^2-2Ll_x)^{1/2},\expt} }\,
l_x\, l_y \, C^{TT}_{(l_x^2+l_y^2)^{1/2}}
\;=\;0.
\end{align}
The integrand changes sign under $l_y\rightarrow -l_y$ so that the integral over $l_y$ vanishes and thus $R_+=0$.  Although we chose a particular coordinate system aligned with $\VL$ to show this, the fact that $R_+=0$ is coordinate-independent (in coordinate-independent terms, the 2D integral can be split into two 1D integrals parallel and perpendicular to $\VL$; the latter integral vanishes). Note that $R_\parallel$ and $R_\perp$ do not vanish because they involve even powers of $l_x$ and $l_y$ in the integrand. Following the same line of argument the very similar integral of type $R^{BE}_+$ can be shown to be zero.

To show $\beta_+=0$ we proceed similarly. Choosing a coordinate system with $x$-axis aligned with $\VL$ and writing \eqq{betacross} in components,
\begin{align}
  \label{eq:60}
  \beta_+(L) = \frac{1}{(2\pi)^2}\int_{-\infty}^\infty\d l_x\int_{-\infty}^\infty\d l_y\; l_y\,(2l_x-L) \, 
B_\phi\left((l_x^2+l_y^2)^{1/2}, (L^2+l_x^2+l_y^2-2Ll_x)^{1/2}, L\right)
\;=\;0.
\end{align}
The integral over $l_y$ vanishes again because the integrand changes sign under $l_y\rightarrow -l_y$.  

We also confirmed numerically that $R_+$ and $\beta_+$ vanish by evaluating the integrals in Eqs.~\eq{Rcross} and \eq{betacross} directly.

\section{Low-$L$, large-scale lens and squeezed limits}
\label{App:lllimits}
In this section we consider certain limits where the $N^{(3/2)}$ bias simplifies, e.g.~the limit of reconstructing large-scale lenses from small-scale temperature fluctuations, or the squeezed limit of the lensing bispectrum.  This is useful to understand the qualitative behavior of the bias and check the robustness of numerical evaluations. 
We first discuss the C1 term in \eqq{typeCcontractions}, and then the A1 term in \eqq{70}.

\subsection{Limit of C1 contribution to $N^{(3/2)}$ bias}

We first consider the limit of reconstructing large-scale lensing modes $\phi(\VL)$ from temperature fluctuations $T(\vl)$ on much smaller scales, i.e.~$L\ll l$. 
Taylor expanding $|\vl-\VL|$ around $l$ yields for the lensing reconstruction weight
\begin{equation}
  \label{eq:largeLensWeight}
\lim_{L\ll l}  g(\vl,\VL) = \frac{L^2}{2} \frac{C^{\tilde
    T\tilde T}_{l}}{(C^{\tilde T\tilde T}_{l,\expt})^2}\Bigg\{
\Bigg[1+\frac{\d\ln C^{\tilde T\tilde T}_l}{\d\ln l}\cos^2\mu_\vl\Bigg]
+ \frac{L}{l}
\Bigg[
\frac{\d\ln C^{\tilde T\tilde T}_l}{\d\ln l}\frac{\d\ln C^{\tilde T\tilde T}_{l,\expt}}{\d\ln l}
\cos^3\mu_\vl
 -\frac{\d\ln\big(
 \frac{C^{\tilde T\tilde T}_{l}}{C^{\tilde T\tilde T}_{l,\expt}}
 \big)}{\d\ln l}\cos\mu_\vl
\Bigg]
\Bigg\},
\end{equation}
where $\cos \mu_\vl = \vl\cdot\VL/(lL)$.
The terms in the first square brackets of \eqq{largeLensWeight} are of
order $(L/l)^0$ and involve only even powers of $\cos\mu_\vl$, while the
second square bracket is of order $(L/l)\ll 1$ and involves only odd powers
of $\cos\mu_\vl$.

Using this, we can compute the large-lens limit of $R_\parallel$ defined in \eqq{Rparallel}. The terms in the second square brackets vanish upon angular integration so that
\begin{eqnarray}
	\nonumber
  \lim_{L\rightarrow 0} R_\parallel(L) &=& \frac{L^2}{8 \pi} \int \d l_1 l_1^3 \frac{C^{TT}_{l_1}C^{\tilde
    T\tilde T}_{l_1}}{(C^{\tilde T\tilde T}_{l_1,\expt})^2}
\Bigg[1+\frac{3}{4}\frac{\d\ln C^{\tilde T\tilde T}_{l_1}}{\d\ln l_1}\Bigg]+\mathcal{O}\left((L/l_1)^2\right)\\
&\approx & \frac{3 L^2}{32 \pi} \int \d l_1 l_1^3 \left(\frac{C^{\tilde
    T\tilde T}_{l_1}}{C^{\tilde T\tilde T}_{l_1,\expt}}\right)^2
\Bigg[\frac{\d\ln l_1^{4/3}C^{\tilde T\tilde T}_{l_1}}{\d\ln l_1}\Bigg]+\mathcal{O}\left((L/l_1)^2\right).
\label{eq:RparalowL}
\end{eqnarray}
The power spectrum ratio in the integrand is unity on scales where the temperature power spectrum is signal-dominated and gets exponentially suppressed when it becomes noise-dominated. The $l_1^3$ weight upweights high $l_1$ in the signal-dominated regime but cannot compete against the exponential fall-off in the noise-dominated regime. These two factors are thus maximal at the highest $l_1$ that are still signal-dominated. In this regime, typically $l_1\sim$ few thousand, the derivative is mostly negative, so that the overall large-lens limit of $R_\parallel$ is negative.  Its amplitude is determined by the multipole at which the temperature power becomes noise dominated, i.e.~it is very sensitive to the noise and beam specifications of the experiment under consideration.

The large-lens limit of $R_\perp$ reads
\begin{eqnarray}
  \nonumber
  \lim_{L\rightarrow 0} R_\perp(L) &=& \frac{L^2}{8 \pi} \int\d l_1\,
l_1^3 
\frac{C^{ T T}_{l_1}C^{\tilde T\tilde T}_{l_1}}{\left(C^{\tilde T\tilde T}_{l_1,\expt}\right)^2}
\left[
1+\frac{1}{4}\,\frac{\d\ln C^{\tilde T\tilde T}_{l_1}}{\d\ln {l_1}}
\right]
+\mathcal{O}\left((L/l_1)^2\right).
\\
\label{eq:RperplowL}
&\approx & \frac{L^2}{32 \pi} \int\d l_1\,
l_1^3 
\left(
\frac{C^{\tilde T\tilde T}_{l_1}}{C^{\tilde T\tilde T}_{l_1,\expt}}
\right)^2
\frac{\d\ln( l_1^4 C^{\tilde T\tilde T}_{l_1})}{\d\ln l_1}
+\mathcal{O}\left((L/l_1)^2\right).
\end{eqnarray}
which has the same structure as the expression that was derived for $R_\parallel$, the only difference being a suppression of the derivative term by a factor of 3. It is thus similarly sensitive to beam and noise specifications as $R_\parallel$, but smaller, since the dominant contribution stems from the derivative. They are plotted as grey lines in \fig{RandBeta_perp_para}.

\newcommand{\kparam}{\frac{k_1}{k_2}}
\newcommand{\invparam}{\frac{k_2}{k_1}}
\newcommand{\cosasqr}{\cos^2{\mu_{\vl}}}
\newcommand{\param}{\frac{L}{l}}
\newcommand{\cosa}{\cos{\mu_{\vl}}}

Additionally to the $R_n$ integrals the $N^{(3/2)}_{C1}$ bias involves $\beta_n$ integrals over the lensing bispectrum defined in Eqs.~\eq{betaparallel} and \eq{betaperp}.
One limit where these simplify is the squeezed limit of the lensing bispectrum, where the reconstructed lensing mode is on much larger scales than the other two internal lensing modes, i.e.~$L\ll l\approx |\vl-\VL|$ and
\begin{equation}
\label{eq:Biphi_lim}
\lim_{L\ll l} B_{\phi}(L,l,-\cosa) =\int_0^{\chi_*} \d \chi \frac{W(\chi)^3}{\chi^4}\frac{\gamma^3(\chi)}{(L/ \chi)^2(l/\chi)^4} \left(1+2\param \cosa\right)\lim_{L\ll l}B_\delta(L/\chi,l/\chi,{-}\cosa;\chi),
\end{equation}
where we insert the squeezed limit of the matter bispectrum (e.g.~\citep{2014JCAPChiang})
\begin{eqnarray}
\nonumber
\lim_{k_1\ll k_2} B_{\delta}(k_1,k_2,\cosa)&=&\left[\frac{13}{7}+\cosasqr \left(\frac{8}{7}-\frac{\partial \ln P_{k_2}(\chi)}{\partial \ln k_2}\right)\right.\\
\nonumber
& & \left. +\kparam \left(\frac{8}{7}-\frac{8}{7}\cosasqr +\left(\frac{3}{7}\frac{\partial \ln P_{k_2}(\chi)}{\partial \ln k_2}-1\right) \cosa +\left(\frac{4}{7}\frac{\partial \ln P_{k_2}(\chi)}{\partial \ln k_2}+1\right) \cos^3{\mu_{\vl}} \right) \right]\\
\label{Bdeltalim}
& & \times P_{k_1}(\chi)P_{k_2}(\chi),
\end{eqnarray}
assuming that limits in $l$ translate to limits in $k$ in the Limber approximation. In this approximation the angle $\mu_\vl$ between $\VL$ and $\vl$ is the same as between the 3D modes $\vec{k}_1$ and $\vec{k}_2$.\footnote{Note that the integrals over the lensing bispectrum (Eqs.\ref{eq:betaperp} and \ref{eq:betaparallel}) integrate over $B_\phi(\vec{l},\vec{L}-\vec{l},-\vec{L})$. The minus sign in front of the bispectrum's third argument, $\vec{L}$, induces a minus sign in front of $\cos{\mu_{\vec{l}}}$ when inserting the squeezed limit of the matter bispectrum into these integrals.}

Using \eqq{Biphi_lim} in the expression for $\beta_\parallel$ we find for the contribution from squeezed bispectrum configurations
\begin{eqnarray}
\nonumber
\lim_{L\ll l} \beta_\parallel(L)&=&\int_{\vl} l\(l \cos^2(\mu_\vl)-L \cos(\mu_\vl)\) \lim_{L\ll l} B_{\phi}(L,l,-\cosa)\\
&=&\int_{\vl}  l^2\( \cos^2(\mu_\vl)-\param \cos(\mu_\vl)\) \int_0^{\chi*} \d \chi W(\chi)^3 \chi^2 \frac{\gamma^3(\chi)}{L^2 l^4} \left(1+2\param \cosa\right)\lim_{L\ll l}B_\delta(L/\chi,l/\chi,-\cosa;\chi),\qquad
\end{eqnarray}
Upon angular integration the squared cosine picks up all terms that are even in the cosine. This includes all contributions of order $\mathcal{O}((\param)^0)$,
\begin{eqnarray}
\label{eq:beta_para_limit}
\lim_{L\ll l} \beta_\parallel(L)&=& \frac{1}{L^2}\int_0^{\chi_*} \d \chi W(\chi)^3 \chi^2 {\gamma^3(\chi)} \int \frac{\d \ln l}{4\pi} \left[ \frac{13}{7}+\frac{3}{4}\(\frac{8}{7}-n(l,\chi)\)+\frac{2}{7}\param\right]P_{L/\chi}(\chi)P_{l/\chi}(\chi),
\end{eqnarray}
where we defined the spectral index of the matter power spectrum
\begin{align}
  \label{eq:spectralIndex}
n(l,\chi)=\frac{\partial \ln P_{l/\chi}(\chi)}{\partial \ln (l/\chi)}.  
\end{align}

The squeezed limit of $\beta_\perp$ is
\begin{eqnarray}
\nonumber
\lim_{L\ll l} \beta_\perp(L)&=&\int_{\vl} l^2 \sin^2(\mu_\vl) \lim_{L\ll l} B_{\phi}(L,l,-\cosa)\\
\label{eq:beta_perp_limit}
&=&\frac{1}{L^2}\int_0^{\chi_*} \d \chi W(\chi)^3 \chi^2 {\gamma^3(\chi)} \int\frac{\d \ln l}{4\pi} \left[\frac{13}{7}+\frac{1}{4}\left(\frac{8}{7}-n(l,\chi)\right)+\frac{6}{7}\param\right]P_{L/\chi}(\chi)P_{l/\chi}(\chi),
\end{eqnarray}
which is similar to the limit of $\beta_\parallel$ but smaller since the zeroth order term in round brackets gets suppressed by a factor of 3 (the first order term is enhanced by the same factor). The limits of both integrals, $\beta_\perp$ and $\beta_\parallel$, are positive for any realistic value of the spectral index $n$. This agrees with the results obtained by numerical integration over the full bispectrum.

\begin{figure}[tpb]
\begin{center}
\subfloat{
\scalebox{0.33}{\includegraphics{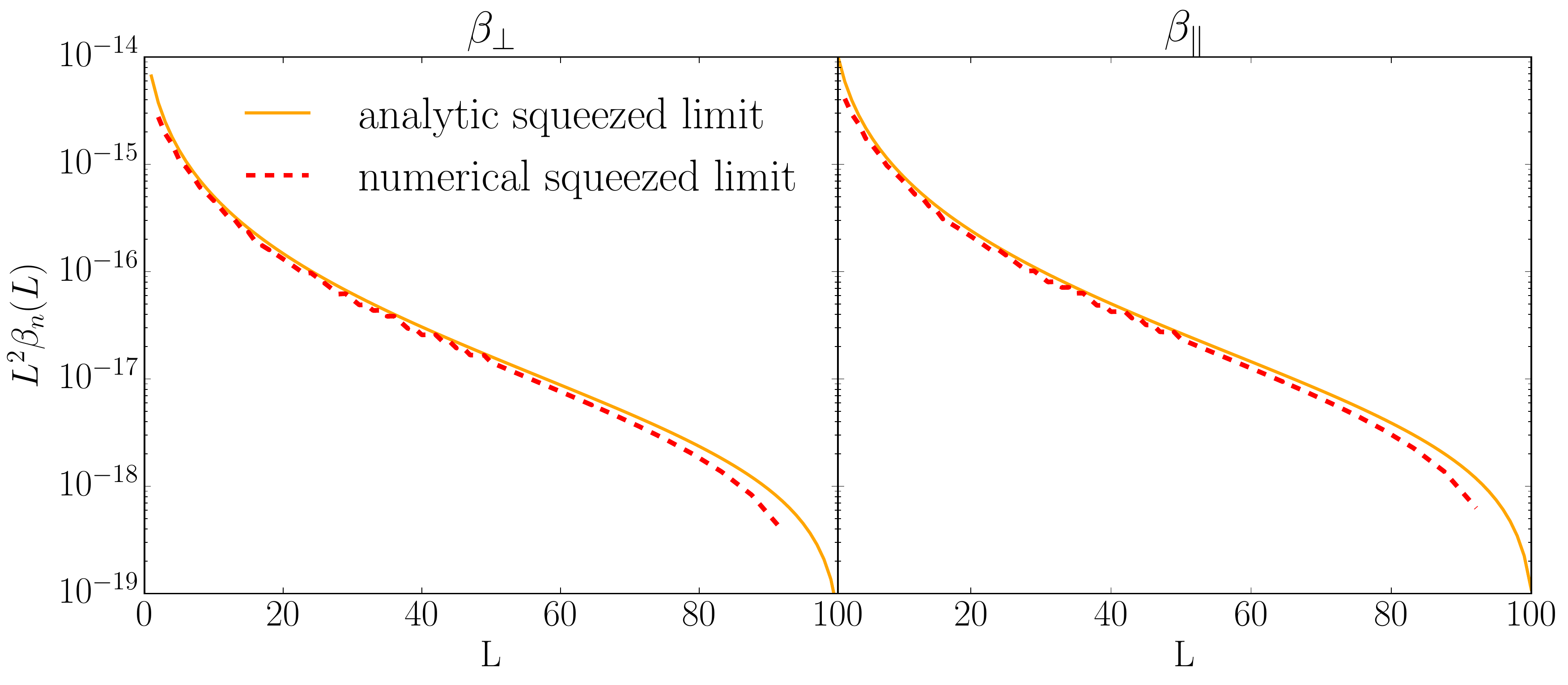}}}
\subfloat{
\scalebox{0.33}{\includegraphics{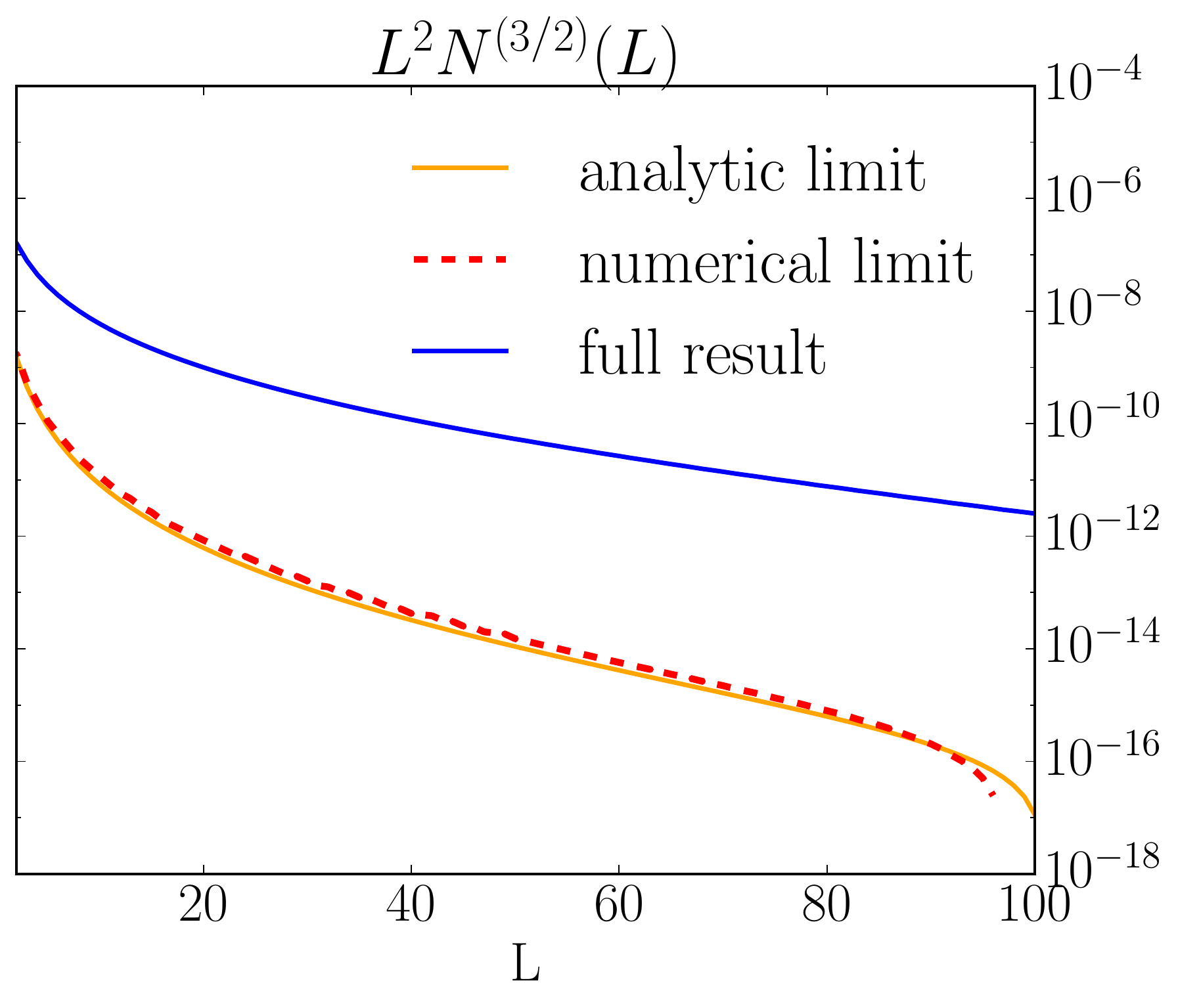}}}
\caption{\label{fig:IandNlimits} Left figures: Integrals of type $\beta_\parallel$ and $\beta_\perp$ calculated numerically with a restriction to squeezed triangle configurations ($l/L{>}100$) and their analytic squeezed limits (\eqq{beta_para_limit}, \eq{beta_perp_limit}). Right figure: The large-lens and squeezed limit of $\lim_{L\rightarrow 0} -2{A_L}\beta_\parallel(L)R_\parallel(L)$ \eqq{N32limit}, the dominant contribution to $N_{C1}^{(3/2)}$ for small $L$. Since we take the limit where only squeezed triangle configurations contribute to the lensing or matter bispectrum,  we also have to restrict the numerical evaluation to squeezed triangle configurations to find agreement.}
\end{center}
\end{figure}

In the two left panels of \fig{IandNlimits} we plot $\beta_\parallel$ and $\beta_\perp$ and their squeezed limits. Since the squeezed configuration excludes triangle configurations with small and comparable side lengths, the squeezed limits do not coincide with the full integrals at low $L$. For a valid comparison, the numerical result has to be restricted to squeezed configurations. After this modification they agree with the analytically derived limits.

To obtain the large-lens and squeezed limit of the $N^{(3/2)}$ bias we also need the large-scale limit of $A_L$  which is \cite{hanson1008}
\begin{equation}
  \label{eq:AL_lowL}
  \lim_{L\rightarrow 0} A_L = \frac{8\pi}{L^4}\left[
\sum_{l}(2l+1)\left(
\frac{C^{\tilde T\tilde T}_l}{C^{\tilde T\tilde T}_{l,\expt}}
\right)^2 D_l
\right]^{-1},
\end{equation}
where 
\begin{equation}
  \label{eq:14}
  D_l = 1+ \frac{\d\ln C_l^{TT}}{\d\ln l} + \frac{3}{8}\left(
\frac{\d\ln C_l^{TT}}{\d\ln l}
\right)^2.
\end{equation}
Putting all these results together we obtain for the large-lens and squeezed limit
of $N^{(3/2)}$ (which is dominated by the $\beta_\parallel R_\parallel$ term)
\begin{align}
\nonumber
  \lim_{L\rightarrow 0}& -2{A_L}
  R_\parallel(L)[\beta_\parallel(L)]_\mathrm{squeezed}\\
\label{eq:N32limit}
&\;\; = 
-2\left[\sum_{l}(2l+1)
\left(
\frac{C^{\tilde T\tilde T}_{l}}{C^{\tilde T\tilde T}_{l,\expt}}
\right)^2
D_{l}\right]^{-1}\left[
\sum_{l_1}\frac{(l_1)^3}{(L+1)^2}
\left(\frac{C^{\tilde T\tilde T}_{l_1}}{C^{\tilde T\tilde T}_{l_1,\expt}}\right)^2
\frac{\d\ln( l^2 C^{\tilde T\tilde T}_{l_1})}{\d\ln l_1}
\right]
\times \lim_{L\rightarrow 0} [\beta_\parallel(L)]_\mathrm{squeezed}.
\end{align}
The comparison with the full result is shown in \fig{IandNlimits}. For $\lim_{L\rightarrow 0} [\beta_\parallel(L)]_\mathrm{squeezed}$ we use the squeezed limit of $\beta_\parallel$ which we obtain in two ways: (1) by evaluating the analytic limit given in \eqq{beta_para_limit}, and (2) by restricting the full numerical result to squeezed triangle configurations of the bispectrum.

\subsection{Limit of A1 contribution to $N^{(3/2)}$ bias}
\label{App:A1_limit}

The type A1 contribution to the $N^{(3/2)}$ bias can be rearranged as in \eqq{N32A1Fast}.  Similarly to the last section, we consider the squeezed limit for bispectrum integrals and the low-$L$ limit for other integrals to obtain simplified expressions that are useful for checking numerical implementations.

For the bispectrum integrals $\bar\beta$ in Eqs.~\eq{barbetaparallel} and \eq{barbetaperp}, we consider the limit where only squeezed bispectrum configurations contribute, i.e.~$L\ll l$.
Using the squeezed limit bispectrum \eq{Biphi_lim}, Taylor expanding in $L/l\ll 1$, and using the cosine angle addition theorem to express $\cos\mu_{\vl}=\hat\vl\cdot\hat\VL$ in terms of $\cos\nu_{\VL}=\hat\VL\cdot\hat\vr$ and $\cos\nu_\vl=\hat\vl\cdot\hat\vr$ (similarly to \eqq{tradeAngle}), we obtain for example the squeezed limit of the $\bar\beta_\parallel$ integral defined in \eqq{barbetaparallel}:
\begin{align}
  \lim\limits_{L\ll l}\bar\beta_\parallel(\vr,\VL) = \frac{1}{2\pi\, L^2}&
\int_0^{\chi_*}\d \chi\,W(\chi)^3\chi^2\gamma^3(\chi)P_{L/\chi}(\chi)
\non\\
\times\Bigg\{&-\frac{13}{7}\cos^2(\nu_\VL)\int_0^\infty\d l\,J_2(lr) \frac{P_{l/\chi}(\chi)}{l}
\;+\; \frac{13}{7r}\int_0^\infty\d l \,J_1(lr)\frac{P_{l/\chi}(\chi)}{l^2}\non\\
&+\frac{3}{r^2}\left[\cos^4(\nu_\VL)-6\cos^2(\nu_\VL)\sin^2(\nu_\VL)+\sin^4(\nu_\VL)\right]
 \int_0^\infty\d l\, J_2(lr)\frac{P_{l/\chi}(\chi)}{l^3}\left[\frac{8}{7}-n(l,\chi)\right]\non\\
&-\cos^4(\nu_\VL)\int_0^\infty\d l\, J_2(lr)\frac{P_{l/\chi}(\chi)}{l}\left[\frac{8}{7}-n(l,\chi)\right]\non\\
\label{eq:barbetaparallellowL}
&+ \frac{6}{r}\cos^2(\nu_\VL)\sin^2(\nu_\VL)\int_0^\infty\d l\,J_1(lr)\frac{P_{l/\chi}(\chi)}{l^2}\left[\frac{8}{7}-n(l,\chi)\right]
\Bigg\}.
\end{align}
Here, $J_n$ are Bessel functions of the first kind that follow from performing the angular integrations over $\nu_\vl$, $P$ is the matter power spectrum, and $n$ is its spectral index defined in \eqq{spectralIndex}. 
 The integrals over $l$ are 1D Hankel transforms, which can be evaluated efficiently with 1D FFTs using e.g.~FFTLog \cite{hamiltonfftlog}. 

Similarly, using \eqq{largeLensWeight}, the low-$L$ limit of the Fourier transform of the lensing weight $\xi_g$ defined in \eqq{xi_g_def} becomes
\begin{align}
  \label{eq:xig_lowL}
  \lim\limits_{L \rightarrow 0}\xi_g(\vr,\VL) =\,& \frac{L^2}{4\pi}\Bigg[
\int_0^\infty \d l\,l\, J_0(lr) 
\frac{C^{\tilde T\tilde T}_l}{(C^{\tilde T\tilde T}_{l,\expt})^2}
+
\frac{1}{r}\int_0^\infty\d l\,J_1(lr)
\frac{C^{\tilde T\tilde T}_l}{(C^{\tilde T\tilde T}_{l,\expt})^2}
\frac{\d\ln C^{\tilde T\tilde T}_l}{\d\ln l}\non\\
&\qquad -\cos^2(\nu_\VL)\int_0^\infty\d l\,l\,J_2(lr)
\frac{C^{\tilde T\tilde T}_l}{(C^{\tilde T\tilde T}_{l,\expt})^2}
\frac{\d\ln C^{\tilde T\tilde T}_l}{\d\ln l}
\Bigg].
\end{align}

The low-$L$ and squeezed limit of the type A1 contribution to the $N^{(3/2)}$ bias of \eqq{N32A1Fast} follows by combining Eqs.~\eq{barxiparallel}, \eq{barxiperp}, \eq{AL_lowL}, \eq{barbetaparallellowL} and \eq{xig_lowL}.  The angular integration over $\hat\vr$ in \eqq{N32A1Fast} can then be performed analytically, leaving 1D integrals over $r$ over the Hankel transforms appearing in Eqs.~\eq{barxiparallel}, \eq{barxiperp}, \eq{barbetaparallellowL} and \eq{xig_lowL}.  This provides a useful consistency check of \eqq{N32A1Fast} at low $L$ if only squeezed bispectrum configurations are taken into account.

\section{Position space interpretation and scalar-tensor decomposition}

\label{App:N32Interpretations}

It is not immediately straightforward to deduce an intuitive physical interpretation of the bias from the pure Fourier space calculation provided in the main text. 
In this appendix we therefore provide a heuristic position space picture that helps to interpret the origin of the C1 contribution to the $N^{(3/2)}$ bias.

\subsection{Position space interpretation of type C1 \texorpdfstring{$N^{(3/2)}$}{N32} coupling}
\label{App:ConfigSpaceN32}

Considering a toy model where normalization $A_L$ and weight $g$ are ignored, the bias from the lensing bispectrum on the lensing reconstruction power spectrum due to the intra-temperature correlation type C1 term in \eqq{typeCcontractions} is given by
\begin{alignat}{2}
\label{eq:N32toyNew}
  N^{(3/2)}_\mathrm{C1, toy}(L)
\supseteq && \frac{1}{2}
\int_{\vx,\vx'}e^{-i\VL\cdot(\vx-\vx')}\left\la {
\contraction[1ex]{}{T}{(\vx)\,\nabla_i\nabla_j }{T}
\contraction[1ex]{T(\vx)\,\nabla_i\nabla_j T(\vx)\phi_{,i}(\vx)\phi_{,j}(\vx)\,}{T}{(\vx')\,\nabla_k }{T}
\bcontraction[1ex]{T(\vx)\,\nabla_i\nabla_j T(\vx)}{\phi}{_{,i}(\vx)}{\phi}
\bcontraction[1ex]{T(\vx)\,\nabla_i\nabla_j T(\vx)}{\phi}{_{,i}(\vx)\phi_{,j}(\vx)\,T(\vx')\,\nabla_k T(\vx')}{\phi}
T(\vx)\,\nabla_i\nabla_j T(\vx)\phi_{,i}(\vx)\phi_{,j}(\vx)\,T(\vx')\,\nabla_k T(\vx')\phi_{,k}(\vx') }\right\ra,
\end{alignat}
where $\phi_{,i}=\nabla_i\phi=\alpha_i$ is the $i$-component of the deflection angle. 
The bias \eq{N32toyNew} involves the average of the product of temperature and temperature derivatives at the same location. These intra-temperature correlation terms are given by integrals over the unlensed temperature power spectrum:
\begin{eqnarray}
  \label{eq:gTTschematic}
  \mathcal{F}\[\la g\, T(\vx)\nabla_k T(\vx)  \ra\] &=& i \int_\vl\, g(\vl,\VL) l_k C_l^{TT}
\equiv i S_k(\VL),\\
  \label{eq:gTTschematic2}
  \mathcal{F}\[\la g\, T(\vx)\nabla_i\nabla_j T(\vx)  \ra\] &=& - \int_\vl\,g(\vl,\VL) l_il_j C_l^{TT}\equiv -R_{ij}(\VL),
\end{eqnarray}
where we schematically included the weight $g$ to account for the lensing-optimized filtering of the observed temperature  (the first integral would vanish otherwise).
Additionally, the bias \eq{N32toyNew} depends on the correlation between the quadratic deflection tensor $\phi_{,i}(\vx)\phi_{,j}(\vx)$ at location $\vx$ and the deflection $\phi_{,k}(\vx')$ at another location $\vx'$,
\begin{align}
  \label{eq:zetaijkDef}
\zeta_{ijk}(\vx-\vx')\equiv  \la\phi_{,i}(\vx)\phi_{,j}(\vx)\phi_{,k}(\vx')\ra  = - i \int_{\VL'} e^{i\VL'\cdot(\vx-\vx')} L'_k \beta_{ij}(\VL').
\end{align}
On the right hand side we introduced the tensor $\beta_{ij}$ which is the Fourier-space cross-spectrum between the quadratic deflection tensor $\phi_{,i}(\vx)\phi_{,j}(\vx)$ and the lensing potential $\phi$:
\begin{align}
  \label{eq:betaijDef}
   \beta_{ij}(\VL') 
\equiv \la [\phi_{,i}(\vx)\phi_{,j}(\vx)](\VL')\;\phi(-\VL') \ra
= -\int_\vl \,l_i\,(\VL'-\vl)_j \, B_\phi(\vl,\VL'-\vl,-\VL').
\end{align}
This is an integral over the bispectrum of the lensing potential. The Fourier transform of $\beta_{ij}(\VL)$ is the 2-point correlation function $\tilde\beta_{ij}(\vr)$ between deflection tensor $\phi_{,i}(\vx)\phi_{,j}(\vx)$ and lensing potential $\phi(\vx+\vr)$ as a function of their separation $\vr$.
With Eqs.~\eq{gTTschematic}, \eq{gTTschematic2} and \eq{zetaijkDef}, the simplified reconstruction bias \eq{N32toyNew}  becomes after integration over $\vx$ and $\vx'$
\begin{align}
  \label{eq:N32toyContraction}
  N^{(3/2)}_\mathrm{C1,toy}(L) = -\frac{1}{4}\frac{A_\mathrm{sky}}{A_L}
\,  R_{ij}(\VL) \beta_{ij}(\VL)
\end{align}
where  $A_\mathrm{sky}=\int_\vx$ is the sky area  and we used $2S_k(\VL)\sim  A_L^{-1}$, the inverse lensing normalization. This shows that 
the expected $N^{(3/2)}$ bias is given by the mean product $R_{ij}$ of temperature and temperature Hessian times the integrated lensing potential bispectrum $\beta_{ij}$, 
corresponding to the cross-spectrum between the deflection tensor 
$\phi_{,i}(\vx)\phi_{,j}(\vx)$ and the lensing potential.

Since all indices in \eqq{N32toyContraction} are contracted, the sum that gives the total bias is 
independent of the orientation of the coordinate system with respect to which 
the component indices of $\beta_{ij}$ and $R_{ij}$ are defined. We are 
therefore free to choose the orientiation of the basis vectors. For example, we can choose the 
first axis to be aligned with $\VL$, and the second one orthogonal to that in 
the flat sky 2D plane.
In this coordinate system, we find the correspondences
\begin{align}
  \label{eq:64}
 R_\parallel=R_{00}, \qquad R_+ =R_{01}=R_{10},
\qquad \mbox{and} \qquad R_\perp = R_{11}
\end{align}
 by comparing Eqs.~\eq{Rparallel} and \eq{Rperp} with \eqq{gTTschematic2}.  The reconstruction bias obtained in the position space picture  
\eqq{N32toyContraction} is then indeed equivalent to \eqq{N32decompExact} derived in the
Fourier space picture (up to normalization and symmetry prefactors which we ignored in the heuristic position space calculation).

\subsection{Scalar-tensor decomposition }
\label{App:ScalTensDecomp}

Instead of choosing a coordinate system as in the last section, we can derive an equivalent expression for the type C1 lensing bias by employing a scalar-tensor decomposition as follows: The 
2-tensor $\beta_{ij}$ in \eqq{betaijDef} can be decomposed into a scalar trace part $\beta_s$, which is invariant under 
rotations of the coordinate system, and a trace-free tensor part $\beta_t$
\begin{align}
  \label{eq:3}
  \beta_{ij}(\VL) = \frac{1}{2}\beta_s(L)\delta_{ij}  + 2\beta_t(L) \Big[\hat\VL_i\hat\VL_j-\frac{1}{2}\delta_{ij}\Big] ,
\end{align}
where $i,j\in\{0,1\}$.  The trace is given by 
\begin{align}
  \label{eq:betas1}
  \beta_s(L) = \delta_{ij}\beta_{ij}(\VL) = -\int_{\vl}\vl\cdot(\VL-\vl)B_\phi(\vl,\VL-\vl,-\VL),
\end{align}
which follows from $\delta_{ij}\delta_{ij}=2$, $\hat{\VL}_i\hat\VL_i=1$ and \eqq{betaijDef}.
The trace $\beta_s(L)$ is thus the Fourier transform of the correlation function
\begin{align}
  \label{eq:11}
  \tilde \beta_s(r) = \la\alpha^2(\vx)\,\phi(\vx+\vr)\ra
\end{align}
between the squared deflection magnitude $\alpha^2=\phi_{,i}\phi_{,i}$ and the lensing potential $\phi$. The trace-free tensor part of $\beta_{ij}$ is $\beta_t(L)=[\hat\VL_i\hat\VL_j-\frac{1}{2}\delta_{ij}]\beta_{ij}(\VL)$, which evaluates to
\begin{eqnarray}
  \label{eq:10}
  \beta_t(L) &=& -\int_\vl \Big\{(\vl\cdot\hat\VL)[(\VL-\vl)\cdot\hat\VL]-\frac{1}{2}\vl\cdot(\VL-\vl)\Big\}B_\phi(\vl,\VL-\vl,-\VL).
\end{eqnarray}
This is the Fourier transform of the correlation function
\begin{align}
  \label{eq:8}
\tilde \beta_t(r) = \left\la \alpha_i(\vx)\alpha_j(\vx)\,\left[\frac{\nabla_i\nabla_j}{\nabla^2}-\frac{1}{2}\delta_{ij}\right]\phi(\vx+\vr) \right\ra  
\end{align}
between the deflection tensor $\alpha_i\alpha_j$ and the tidal tensor constructed from the lensing potential.

Similarly, the 2-tensor $R_{ij}$ can also be decomposed into scalar 
part $R_s$ and tensor part  $R_t$:
\begin{align}
  \label{eq:32}
  R_{ij}(\VL) = \frac{1}{2}R_s(L)\delta_{ij} + 2R_{t}(L) \Big[\hat\VL_i\hat\VL_j-\frac{1}{2}\delta_{ij}\Big].
\end{align}
The scalar part is
\begin{align}
  \label{eq:34}
  R_s(L) = -\la g T(\vx)\nabla^2T(\vx)\ra = \int_\vl g(\vl,\VL)l^2 C_l^{TT} 
\end{align}
and the tensor part is
\begin{eqnarray}
  \label{eq:36}
  R_t(L) &=& \int_\vl g(\vl,\VL) \left[(\vl\cdot\hat\VL)(\vl\cdot\hat\VL) - \frac{l^2}{2}\right] C_l^{TT}.\\
\end{eqnarray}
Note that the tensor part would be zero in absence of the reconstruction weight $g$.
With these scalar-tensor decompositions the reconstruction bias becomes
\begin{align}
  \label{eq:37}
N^{(3/2)}_{C1}(L)\sim \left[ \frac{1}{2}R_{s}(L)\beta_s(L)
+ 2 R_t(L)\beta_t(L)
\right].
\end{align}
The first term in the brackets is the trace of the mean product $R_{ij}$ of (weighted) temperature and temperature Hessian, coupled to the trace of the lensing 3-point statistic $\beta_{ij}$.  The second term couples the trace-free tensor parts of $R_{ij}$ and $\beta_{ij}$.

The scalar and tensor parts are connected to the tensor components by
\begin{align}
  \label{eq:38}
  \beta_s = \beta_{00}+\beta_{11}\qquad \text{and} \qquad \beta_t = \frac{1}{2}(\beta_{00}-\beta_{11})
\end{align}
and the same relations hold for $R_{ij}$. Then we get
\begin{align}
  \label{eq:43}
  \frac{1}{2}R_s\beta_s  + 2 R_t\beta_t = R_{00}\beta_{00}+R_{11}\beta_{11} = R_{\parallel}\beta_{\parallel}+R_{\perp}\beta_{\perp},
\end{align}
where we aligned the x-axis of the coordinate system with the lensing wavevector $\VL$ in the last step.

Integrals of type $R$ and $\beta$ in their decomposition in scalar and tensor 
contributions are shown \fig{RandBeta_scal_tens}. 
Note that the plotted quantities already include the prefactors with which the components enter into the bias. Similar plots for the decomposition into parallel and perpendicular components are shown in \fig{RandBeta_perp_para}.
\ifincludefigs
\begin{figure}[tpb]
\begin{center}
\scalebox{0.46}{\includegraphics{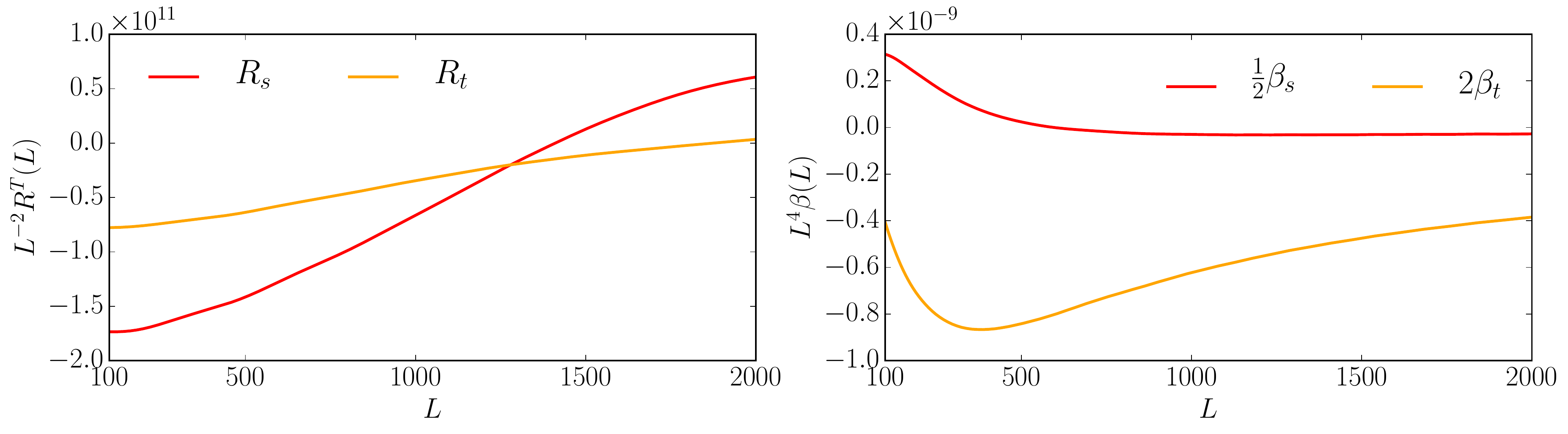}}
\caption{\label{fig:RandBeta_scal_tens} Left panel: Scalar- and tensor-like contributions to the weighted mean product of temperature and temperature Hessian $R_{ij}$. Noise and beam specifications correspond to a Planck-like experiment. Right panel: The lensing potential-deflection tensor correlation $\beta_{ij}$ decomposed in the same manner. We have included prefactors such that contributing terms to the type C1 bias can be constructed by multiplying lines of the same color (compare \eqq{37}).}
\end{center}
\end{figure}
\fi
\ifincludefigs
\begin{figure}[tpb]
\begin{center}
\scalebox{0.46}{\includegraphics{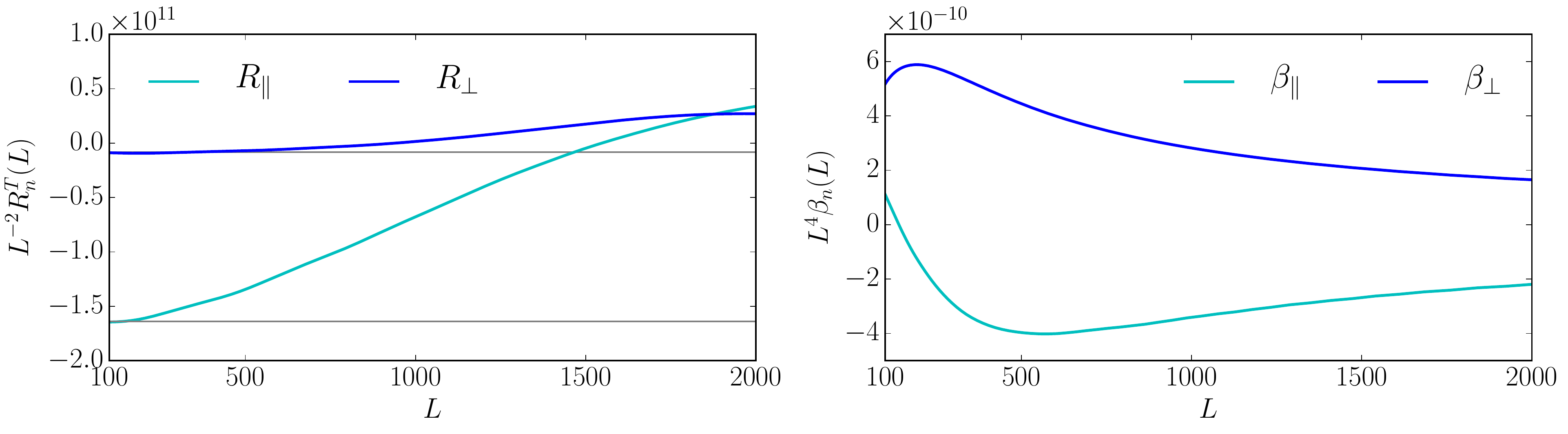}}
\caption{\label{fig:RandBeta_perp_para}  Left panel: Parallel and perpendicular contributions to $R_{ij}$. Noise and beam specifications correspond to a Planck-like experiment. In grey we plot the large-lens limits derived in \app{lllimits}. Right panel: The lensing potential-deflection tensor correlation $\beta_{ij}$ decomposed in the same manner. The bias contribution from the type C1 term can be constructed by multiplying lines of the same color (compare \eqq{37}).}
\end{center}
\end{figure}
\fi

\section{\texorpdfstring{$N^{(3/2)}$}{N32} type C1 bias for polarization-based lensing reconstructions}
\label{App:Pol}
The $N^{(3/2)}$ bias also exists for polarization-based lensing reconstructions. In this paper, we show how to generalize the non-Gaussian bias from the coupling type C1 (\eqq{typeCcontractions} to reconstructions from arbitrary field combinations but leave a generalization of other contributing terms to future work. The results for the single coupling do not provide a proper quantitative estimate of the general bias, but give some idea of the qualitative changes and other terms can be derived in a similar fashion.

\subsection{General polarization-based lensing reconstruction}
We first consider the most general case where the lensing potential is reconstructed from two placeholder fields $W$ and $X$ that can each be $T$, $E$ or $B$, and from two potentially different fields $Y$ and $Z$ that can again each be $T$, $E$, or $B$, and then the cross-spectrum of these reconstructions is used to estimate the lensing power, i.e.~we consider $\la\hat{\phi}^\WX\hat{\phi}^\YZ\ra$ where $W,X,Y,Z\in\{T,E,B\}$ throughout this section.

Lensing changes the CMB fields according to
\begin{equation}
  \label{eq:20}
\tilde X(\vl) = X(\vl) + \delta X(\vl)+\delta^2 X(\vl),
\end{equation}
where in absence of primordial gravitational waves \cite{huHarmonicCMBLensing0001303,Hu0111606}
\begin{eqnarray}
  \label{eq:21}
  \delta X(\vl) &=& -\int_{\vl_1}\bar{X}(\vl_1)\phi(\vl-\vl_1)h_X(\vl_1,\vl)(\vl-\vl_1)\cdot \vl_1 \\
  \delta^2 X(\vl) &=& -\frac{1}{2}\int_{\vl_1,\vl_2}\bar{X}(\vl_1)\phi(\vl_2)\phi(\vl-\vl_1-\vl_2)h_X(\vl_1,\vl)(\vl_1\cdot\vl_2)\left[(\vl_1+\vl_2-\vl)\cdot\vl_1\right].
\end{eqnarray}
To simplify the notation we also defined
\begin{equation}
  \label{eq:22}
  \bar{T}\equiv T,\qquad\bar{E}\equiv E,\qquad \bar{B} \equiv E,
\end{equation}
and
\begin{equation}
  \label{eq:hXDef}
  h_X(\vl_1,\vl) \equiv 
  \begin{cases}
    1 & \mbox{if } X=T, \\
    \cos(2(\varphi_{\vl_1}-\varphi_\vl)) &  \mbox{if } X=E, \\
    \sin(2(\varphi_{\vl_1}-\varphi_\vl)) & \mbox{if } X=B,
  \end{cases}
\end{equation}
which satisfies $h_X(\vl_1,\vl)=h_X(-\vl_1,-\vl)$ and $h_X(\vl_1,\vl)=h_X(\vl_1,-\vl)=h_X(-\vl_1,\vl)$.

The general lensing reconstruction estimator is \cite{Hu0111606}
  \begin{equation}
    \label{eq:23}
    \hat\phi^\WX(\VL) = A_L^\WX\int_\vl g_\WX(\vl,\VL)\tilde W_\expt(\vl)\tilde{X}^*_\expt(\vl-\VL),
  \end{equation}
with normalization
\begin{equation}
  \label{eq:24}
  A_L^\WX = \left[\int_\vl f_\WX(\vl,\VL-\vl)g_\WX(\vl,\VL)\right]^{-1}
\end{equation}
and weight
\begin{equation}
  \label{eq:25}
  g_\WX(\vl,\VL) = \frac{C_{l,\expt}^\tXX C_{|\VL-\vl|,\expt}^\tWW f_\WX(\vl,\VL-\vl) - C_{l,\expt}^\tWX C^\tWX_{|\VL-\vl|,\expt}f_\WX(\VL-\vl,\vl)}{C_{l,\expt}^\tWW C^\tXX_{|\VL-\vl|,\expt}C^\tXX_{l,\expt}C^\tWW_{|\VL-\vl|,\expt} - (C^\tWX_{l,\expt}C^\tWX_{|\VL-\vl|,\expt})^2},
\end{equation}
where $f_\WX$ is defined by $\la \tilde W(\vl)\tilde X(\VL-\vl)\ra_\mathrm{CMB}=f_\WX(\vl,\VL-\vl)\phi(\VL)$ and can be found in \cite{Hu0111606}. We assume a slightly modified form where unlensed spectra are replaced by lensed ones \cite{hanson1008} to avoid the $N^{(2)}$ bias.  Note that $f_\WX(\vl,\VL-\vl)=f_\mathit{XW}(\VL-\vl,\vl)$ and thus 
\begin{align}
  \label{eq:gSymmetry}
g_\WX(\vl,\VL)=g_\mathit{XW}(\VL-\vl,\VL).   
\end{align}

\subsection{C1 bias contribution for general polarization-based reconstruction}
\label{App:C1pol}
The type C contribution to the $N^{(3/2)}$ bias of the general reconstruction power $\la\hat\phi^\WX\hat\phi^\YZ\ra$ is
\begin{equation}
  \label{eq:N32polStep1}
  N^{(3/2),\mathrm{typeC}}_{\WX,\YZ}(L) = A_L^\WX A_L^\YZ
\int_{\vl_1,\vl_2}g_\WX(\vl_1,\VL)g_\YZ(\vl_2,\VL)\mathcal{T}^\mathrm{typeC}_{\WX,\YZ}(\vl_1,\VL-\vl_1,-\vl_2,\vl_2-\VL),
\end{equation}
where the trispectrum is given by all contributions to  $\la\tilde W\tilde X\tilde Y\tilde Z\ra_c$ that are of type C form $\langle\delta^2W X\delta YZ\rangle_c\sim$~`2010'. There are 8 such terms: 2010, 2001, 0210, 0201, 1020, 0120, 1002 and 0102, where `0' denotes the position of unperturbed fields and `1' and `2' that of first and second order perturbed fields.   Let us denote the integral over the 2010 term by
\begin{align}
  \label{eq:33}
  U_{\WX,\YZ}(L) & \equiv \int_{\vl_1,\vl_2}g_\WX(\vl_1,\VL)g_\YZ(\vl_2,\VL)\mathcal{T}^\mathrm{typeC1}_{\WX,\YZ}(\vl_1,\VL-\vl_1,-\vl_2,\vl_2-\VL)\\
\label{eq:Ualternative}
& = \int_{\vl_1',\vl_2'}g_{\WX}(\VL-\vl_1',\VL)g_{\YZ}(\VL-\vl_2',\VL)\mathcal{T}^\mathrm{typeC1}_{\WX,\YZ}(\VL-\vl_1',\vl_1',\vl_2'-\VL,-\vl_2'),
\end{align}
where we changed integration variables in the second line.
Using \eqq{gSymmetry}, $\mathcal{T}(-\vl_1,-\vl_2,-\vl_3,-\vl_4)=\mathcal{T}(\vl_1,\vl_2,\vl_3,\vl_4)$ and substitution of integration variables, the 8 type C terms contributing to \eqq{N32polStep1} can be written simply by permuting field labels of $U$:
\begin{align}
  \label{eq:N32polarizationInTermsofU}
  N^{(3/2),\mathrm{typeC}}_{\WX,\YZ}(L) = A^\WX_L A^\YZ_L \big[
& U_{\mathit{WX},\mathit{YZ}}(L) + U_{\mathit{WX},\mathit{ZY}}(L)
+ U_{\mathit{XW},\mathit{YZ}}(L) + U_{\mathit{XW},\mathit{ZY}}(L) \nonumber\\
&+U_{\mathit{YZ},\mathit{WX}}(L) + U_{\mathit{YZ},\mathit{XW}}(L)
 +U_{\mathit{ZY},\mathit{WX}}(L) + U_{\mathit{ZY},\mathit{XW}}(L)
\big].
\end{align}

It remains to calculate $U_{\WX,YZ}$.  Extending \eqq{TrispTypeCFinalMainText} to the general polarization case, the connected 4-point function of coupling type C due to the 2010 contraction is
\begin{align}
\nonumber
  \langle\delta^2W_{\vl_1}X_{\vl_2}\delta Y_{\vl_3}Z_{\vl_4}\rangle_c =\, &
-\frac{(2\pi)^2}{2}\delta_D(\vl_1+\vl_2+\vl_3+\vl_4)
C^{\bar W X}_{l_2}C^{\bar Y Z}_{l_4}
h_W(-\vl_2,\vl_1)h_Y(-\vl_4,\vl_3)
\left[(\vl_3+\vl_4)\cdot\vl_4\right]
\\
\nonumber
&\quad\times\int_{\vl'}
(\vl'\cdot\vl_2)\left[(\vl'-\vl_1-\vl_2)\cdot\vl_2\right]
B_\phi(\vl',\vl_1+\vl_2-\vl',\vl_3+\vl_4)
\\
\label{eq:d2WXdYZ}
&+\left(\vl_2\leftrightarrow \vl_4, X\leftrightarrow Z\right),
\end{align}
where the permutation in the last line is obtained by simultaneously replacing every $\vl_2$ by $\vl_4$, every $\vl_4$ by $\vl_2$, every $X$ by $Z$ and every $Z$ by $X$ in the first two lines (in particular, this permutation involves $C^{\bar WZ}_{l_4}C^{\bar Y X}_{l_2}$). We ignore this permutation in the last line of \eqq{d2WXdYZ} from now on because it is expected to lead to more tightly coupled terms that should be subdominant; we call the dominant first two lines 'typeC1'.  For the multipole arguments required for \eqq{Ualternative} we get
\begin{align}
\nonumber
  \langle\delta^2W_{\VL-\vl_1}X_{\vl_1}\delta Y_{\vl_2-\VL}Z_{-\vl_2}\rangle_c^\mathrm{typeC1} =\, &
-\frac{(2\pi)^2}{2}\delta_D(\mathbf{0}) C^{\bar W X}_{l_1} C^{\bar Y Z}_{l_2}h_W(-\vl_1,\VL-\vl_1) h_Y(\vl_2,\vl_2-\VL) (\VL\cdot\vl_2)\\
&\times \int_{\vl'} (\vl'\cdot\vl_1)\left[(\vl'-\VL)\cdot\vl_1\right]
B_\phi(\vl',\VL-\vl',-\VL).
\end{align}
Thus,
\begin{align}
  \label{eq:35}
  U_{\WX,\YZ}(L) &= -\frac{1}{2}\int_{\vl_1,\vl_2}g_\WX(\VL-\vl_1,\VL) g_\YZ(\VL-\vl_2,\VL) 
C^{\bar W X}_{l_1} C^{\bar Y Z}_{l_2}h_W(-\vl_1,\VL-\vl_1) h_Y(\vl_2,\vl_2-\VL) (\VL\cdot\vl_2)\nonumber\\
&\qquad\times \int_{\vl} \left[\vl_1\cdot(\vl-\VL)\right] \left[\vl_1\cdot\vl\right]
B_\phi(\vl,\VL-\vl,-\VL).
\end{align}
The weights in the last integral can be expressed in the separable form of \eqq{N32decompExact}. Then,
\begin{align}
  \label{eq:Ufinal}
  U_{\WX,\YZ}(L) &= -\frac{1}{2} \left[\int_{\vl_2}g_\mathit{ZY}(\vl_2,\VL)h_Y(\vl_2,\vl_2-\VL)(\VL\cdot\vl_2)C_{l_2}^{\bar Y Z}\right]\sum_{n\in\{\parallel,+,\perp\}} R_n^{\WX}(\VL)\beta_n(\VL)
\end{align}
where $\beta_n$ integrals are the same as in Eqs.~\eq{betaparallel},~\eq{betaperp}, and we defined
\begin{eqnarray}
  \label{eq:J0Def}
  R^\WX_\parallel(\VL) &=& \int_{\vl_1} g_\mathit{XW}(\vl_1,\VL) l_1^2\cos^2(\mu_{\vl_1})  h_W(-\vl_1,\VL-\vl_1) C^{\bar W X}_{l_1}\\
  \label{eq:J1Def}
  R^\WX_+(\VL) &=& \int_{\vl_1} g_\mathit{XW}(\vl_1,\VL) l_1^2 \sin(\mu_{\vl_1})\cos(\mu_{\vl_1}) h_W(-\vl_1,\VL-\vl_1) C^{\bar W X}_{l_1}\\
  \label{eq:J2Def}
  R^\WX_\perp(\VL) &=& \int_{\vl_1} g_\mathit{XW}(\vl_1,\VL) l_1^2 \sin^2(\mu_{\vl_1}) h_W(-\vl_1,\VL-\vl_1) C^{\bar W X}_{l_1}.
\end{eqnarray}
When evaluating $h_W$ numerically, the angle $2(\varphi_{\vl_1}-\varphi_{\VL-\vl_1})$ can be obtained brute-force from the components of 2D vectors $\vl_1$ and $\VL$. \footnote{Explicitly, defining angles with respect to the $x$-axis, we have $\varphi_{\vl_1}=\arccos \left[l_{1,x}/ \sqrt{l_{1,x}^2+l_{1,y}^2}\,\right]$ and $\varphi_{\VL-\vl_1}=\arccos\left[(L_x-l_{1,x})/\sqrt{(L_x-l_{1,x})^2+(L_y-l_{1,y})^2}\right]$.}
In the special case of a temperature-only based measurement we recover the previously derived results with $R^{TT}_n=R_n$.
The final $N_{C1}^{(3/2)}$ bias for polarization is obtained by plugging \eqq{Ufinal} into \eqq{N32polarizationInTermsofU}
\begin{align}
  \label{eq:N32PolFinal}
N^{(3/2),\mathrm{typeC1}}_{\WX,\YZ}(L) = -\frac{1}{2} A^\WX_L A^\YZ_L 
S^\mathit{YZ}_L
\sum_{n\in\{\parallel,+,\perp\}} R_n^{\WX}(\VL)\beta_n(\VL)
 \, + \mbox{7 perms in } W,X;Y,Z
\end{align}
where the permutations denote those written out in \eqq{N32polarizationInTermsofU}. 
We also defined 
\begin{align}
  \label{eq:SYZ}
S^\mathit{YZ}_L \equiv \left[\int_{\vl_2}g_\mathit{ZY}(\vl_2,\VL)h_Y(\vl_2,\vl_2-\VL)(\VL\cdot\vl_2)C_{l_2}^{\bar Y Z}\right]  .
\end{align}
One can show that $S^\mathit{YZ}_L+S^\mathit{ZY}_L=A_L^{-1}$ to first order in $C^{\phi\phi}$. This identity also holds for field combinations where one of the $S^\mathit{YZ}_L$ terms is zero (e.g.~$S^\mathit{EB}_L$).

\eqq{N32PolFinal} involves the same integrals $\beta_n$ over the lensing bispectrum $B_\phi$ as the temperature reconstruction bias. The 2D integrals over CMB power spectra $R_n^\WX$ have a similar form as for the temperature-only case, with slightly different weights in the integrands. For $WXYZ=TTTT$ the general bias formula $N^{(3/2),\mathrm{typeC1}}_{\WX,\YZ}(L)$ simplifies to the expression derived for the temperature \eqq{N32decompExact}.
\subsection{C1 bias contribution for $\EB,\EB$ reconstruction}
\label{app:EBEBbias}
The special case of $\EB,\EB$-reconstruction is expected to have relatively high signal-to-noise in comparison with the other polarization-based lensing estimators.  In this case, we have
$W=Y=E$ and $X=Z=B$ so that the $N^{(3/2)}$ bias becomes
\begin{align}
  \label{eq:41}
 N^{(3/2),\mathrm{typeC1}}_{\EB,\EB}(L)
=-(A^\EB_L)^2 
\sum_{n\in\{\parallel,+,\perp\}} \beta_n(\VL)\left[
 S_L^\EB R_n^\EB(\VL)
+S_L^\EB R_n^\BE(\VL)
+S_L^\BE R_n^\EB(\VL)
+S_L^\BE R_n^\BE(\VL)
\right].
\end{align}
We can further simplify \eqq{41} by noting that $R_n^\EB=0$ and $S_L^\EB=0$ (which follows from $C_l^{\EB}=0$) and obtain
\begin{align}
  \label{eq:7}
 N^{(3/2),\mathrm{typeC1}}_{\EB,\EB}(L)
=-(A^\EB_L)^2 S_L^\BE
\sum_{n\in\{\parallel,+,\perp\}} \beta_n(\VL)R_n^\BE(\VL),
\end{align}
where $R_n^\BE$ are integrals over the E-mode power spectrum given by Eqs.~\eq{J0Def}-\eq{J2Def}, and $\beta_n$ are integrated lensing bispectra computed earlier in Eqs.~\eq{betaparallel},~\eq{betaperp}.

\bibliography{marcel_lensing_bisp,LensingBispectrumVanessa}

\label{lastpage}

\end{document}